\documentstyle[12pt]{article}
\newcommand{\ch}{\cosh}
\newcommand{\sh}{\sinh}
\begin{document}
\immediate\write16{<WARNING: FEYNMAN macros work only with
emTeX-dvivers
                    (dviscr.exe, dvihplj.exe, dvidot.exe, etc.) >}
\newdimen\Lengthunit
\newcount\Nhalfperiods
\Lengthunit = 1.5cm
\Nhalfperiods = 9
\catcode`\*=11
\newdimen\L*   \newdimen\d*   \newdimen\d**
\newdimen\dm*  \newdimen\dd*  \newdimen\dt*
\newdimen\a*   \newdimen\b*   \newdimen\c*
\newdimen\a**  \newdimen\b**
\newdimen\xL*  \newdimen\yL*
\newcount\k*   \newcount\l*   \newcount\m*
\newcount\n*   \newcount\dn*  \newcount\r*
\newcount\N*   \newcount\*one \newcount\*two  \*one=1 \*two=2
\newcount\*ths \*ths=1000
\def\GRAPH(hsize=#1)#2{\hbox to #1\Lengthunit{#2\hss}}
\def\Linewidth#1{\special{em:linewidth #1}}
\Linewidth{.4pt}
\def\sm*{\special{em:moveto}}
\def\sl*{\special{em:lineto}}
\newbox\spm*   \newbox\spl*
\setbox\spm*\hbox{\sm*}
\setbox\spl*\hbox{\sl*}
\def\mov#1(#2,#3)#4{\rlap{\L*=#1\Lengthunit\kern#2\L*\raise#3\L*\hbox{#4}}}
\def\smov#1(#2,#3)#4{\rlap{\L*=#1\Lengthunit
\xL*=\xscale\L*\yL*=\yscale\L*\kern#2\xL*\raise#3\yL*\hbox{#4}}}
\def\mov*(#1,#2)#3{\rlap{\kern#1\raise#2\hbox{#3}}}
\def\lin#1(#2,#3){\rlap{\sm*\mov#1(#2,#3){\sl*}}}
\def\arr*(#1,#2,#3){\mov*(#1\dd*,#1\dt*){%
\sm*\mov*(#2\dd*,#2\dt*){\mov*(#3\dt*,-#3\dd*){\sl*}}%
\sm*\mov*(#2\dd*,#2\dt*){\mov*(-#3\dt*,#3\dd*){\sl*}}}}
\def\arrow#1(#2,#3){\rlap{\lin#1(#2,#3)\mov#1(#2,#3){%
\d**=-.012\Lengthunit\dd*=#2\d**\dt*=#3\d**%
\arr*(1,10,4)\arr*(3,8,4)\arr*(4.8,4.2,3)}}}
\def\arrlin#1(#2,#3){\rlap{\L*=#1\Lengthunit\L*=.5\L*%
\lin#1(#2,#3)\mov*(#2\L*,#3\L*){\arrow.1(#2,#3)}}}
\def\dasharrow#1(#2,#3){\rlap{%
{\Lengthunit=0.9\Lengthunit\dashlin#1(#2,#3)\mov#1(#2,#3){\sm*}}%
\mov#1(#2,#3){\sl*\d**=-.012\Lengthunit\dd*=#2\d**\dt*=#3\d**%
\arr*(1,10,4)\arr*(3,8,4)\arr*(4.8,4.2,3)}}}
\def\clap#1{\hbox to 0pt{\hss #1\hss}}
\def\ind(#1,#2)#3{\rlap{%
\d*=.1\Lengthunit\kern#1\d*\raise#2\d*\hbox{\lower2pt\clap{$#3$}}}}
\def\sh*(#1,#2)#3{\rlap{%
\dm*=\the\n*\d**\xL*=\xscale\dm*\yL*=\yscale\dm*
\kern#1\xL*\raise#2\yL*\hbox{#3}}}
\def\calcnum*#1(#2,#3){\a*=1000sp\b*=1000sp\a*=#2\a*\b*=#3\b*%
\ifdim\a*<0pt\a*-\a*\fi\ifdim\b*<0pt\b*-\b*\fi%
\ifdim\a*>\b*\c*=.96\a*\advance\c*.4\b*%
\else\c*=.96\b*\advance\c*.4\a*\fi%
\k*\a*\multiply\k*\k*\l*\b*\multiply\l*\l*%
\m*\k*\advance\m*\l*\n*\c*\r*\n*\multiply\n*\n*%
\dn*\m*\advance\dn*-\n*\divide\dn*2\divide\dn*\r*%
\advance\r*\dn*%
\c*=\the\Nhalfperiods5sp\c*=#1\c*\ifdim\c*<0pt\c*-\c*\fi%
\multiply\c*\r*\N*\c*\divide\N*10000}
\def\dashlin#1(#2,#3){\rlap{\calcnum*#1(#2,#3)%
\d**=#1\Lengthunit\ifdim\d**<0pt\d**-\d**\fi%
\divide\N*2\multiply\N*2\advance\N*1%
\divide\d**\N*\sm*\n*\*one\sh*(#2,#3){\sl*}%
\loop\advance\n*\*one\sh*(#2,#3){\sm*}\advance\n*\*one\sh*(#2,#3){\sl*}%
\ifnum\n*<\N*\repeat}}
\def\dashdotlin#1(#2,#3){\rlap{\calcnum*#1(#2,#3)%
\d**=#1\Lengthunit\ifdim\d**<0pt\d**-\d**\fi%
\divide\N*2\multiply\N*2\advance\N*1\multiply\N*2%
\divide\d**\N*\sm*\n*\*two\sh*(#2,#3){\sl*}\loop%
\advance\n*\*one\sh*(#2,#3){\kern-1.48pt\lower.5pt\hbox{\rm.}}%
\advance\n*\*one\sh*(#2,#3){\sm*}%
\advance\n*\*two\sh*(#2,#3){\sl*}\ifnum\n*<\N*\repeat}}
\def\shl*(#1,#2)#3{\kern#1#3\lower#2#3\hbox{\unhcopy\spl*}}
\def\trianglin#1(#2,#3){\rlap{\toks0={#2}\toks1={#3}\calcnum*#1(#2,#3)%
\dd*=.57\Lengthunit\dd*=#1\dd*\divide\dd*\N*%
\d**=#1\Lengthunit\ifdim\d**<0pt\d**-\d**\fi%
\multiply\N*2\divide\d**\N*\advance\N*-1\sm*\n*\*one\loop%
\shl**{\dd*}\dd*-\dd*\advance\n*2%
\ifnum\n*<\N*\repeat\n*\N*\advance\n*1\shl**{0pt}}}
\def\wavelin#1(#2,#3){\rlap{\toks0={#2}\toks1={#3}\calcnum*#1(#2,#3)%
\dd*=.23\Lengthunit\dd*=#1\dd*\divide\dd*\N*%
\d**=#1\Lengthunit\ifdim\d**<0pt\d**-\d**\fi%
\multiply\N*4\divide\d**\N*\sm*\n*\*one\loop%
\shl**{\dd*}\dt*=1.3\dd*\advance\n*1%
\shl**{\dt*}\advance\n*\*one%
\shl**{\dd*}\advance\n*\*two%
\dd*-\dd*\ifnum\n*<\N*\repeat\n*\N*\shl**{0pt}}}
\def\w*lin(#1,#2){\rlap{\toks0={#1}\toks1={#2}\d**=\Lengthunit\dd*=-.12\d**%
\N*8\divide\d**\N*\sm*\n*\*one\loop%
\shl**{\dd*}\dt*=1.3\dd*\advance\n*\*one%
\shl**{\dt*}\advance\n*\*one%
\shl**{\dd*}\advance\n*\*one%
\shl**{0pt}\dd*-\dd*\advance\n*1\ifnum\n*<\N*\repeat}}
\def\l*arc(#1,#2)[#3][#4]{\rlap{\toks0={#1}\toks1={#2}\d**=\Lengthunit%
\dd*=#3.037\d**\dd*=#4\dd*\dt*=#3.049\d**\dt*=#4\dt*\ifdim\d**>16mm%
\d**=.25\d**\n*\*one\shl**{-\dd*}\n*\*two\shl**{-\dt*}\n*3\relax%
\shl**{-\dd*}\n*4\relax\shl**{0pt}\else\ifdim\d**>5mm%
\d**=.5\d**\n*\*one\shl**{-\dt*}\n*\*two\shl**{0pt}%
\else\n*\*one\shl**{0pt}\fi\fi}}
\def\d*arc(#1,#2)[#3][#4]{\rlap{\toks0={#1}\toks1={#2}\d**=\Lengthunit%
\dd*=#3.037\d**\dd*=#4\dd*\d**=.25\d**\sm*\n*\*one\shl**{-\dd*}%
\n*3\relax\sh*(#1,#2){\xL*=\xscale\dd*\yL*=\yscale\dd*
\kern#2\xL*\lower#1\yL*\hbox{\sm*}}%
\n*4\relax\shl**{0pt}}}
\def\arc#1[#2][#3]{\rlap{\Lengthunit=#1\Lengthunit%
\sm*\l*arc(#2.1914,#3.0381)[#2][#3]%
\smov(#2.1914,#3.0381){\l*arc(#2.1622,#3.1084)[#2][#3]}%
\smov(#2.3536,#3.1465){\l*arc(#2.1084,#3.1622)[#2][#3]}%
\smov(#2.4619,#3.3086){\l*arc(#2.0381,#3.1914)[#2][#3]}}}
\def\dasharc#1[#2][#3]{\rlap{\Lengthunit=#1\Lengthunit%
\d*arc(#2.1914,#3.0381)[#2][#3]%
\smov(#2.1914,#3.0381){\d*arc(#2.1622,#3.1084)[#2][#3]}%
\smov(#2.3536,#3.1465){\d*arc(#2.1084,#3.1622)[#2][#3]}%
\smov(#2.4619,#3.3086){\d*arc(#2.0381,#3.1914)[#2][#3]}}}
\def\wavearc#1[#2][#3]{\rlap{\Lengthunit=#1\Lengthunit%
\w*lin(#2.1914,#3.0381)%
\smov(#2.1914,#3.0381){\w*lin(#2.1622,#3.1084)}%
\smov(#2.3536,#3.1465){\w*lin(#2.1084,#3.1622)}%
\smov(#2.4619,#3.3086){\w*lin(#2.0381,#3.1914)}}}
\def\shl**#1{\c*=\the\n*\d**\d*=#1%
\a*=\the\toks0\c*\b*=\the\toks1\d*\advance\a*-\b*%
\b*=\the\toks1\c*\d*=\the\toks0\d*\advance\b*\d*%
\a*=\xscale\a*\b*=\yscale\b*%
\raise\b*\rlap{\kern\a*\unhcopy\spl*}}
\def\wlin*#1(#2,#3)[#4]{\rlap{\toks0={#2}\toks1={#3}%
\c*=#1\l*\c*\c*=.01\Lengthunit\m*\c*\divide\l*\m*%
\c*=\the\Nhalfperiods5sp\multiply\c*\l*\N*\c*\divide\N*\*ths%
\divide\N*2\multiply\N*2\advance\N*1%
\dd*=.002\Lengthunit\dd*=#4\dd*\multiply\dd*\l*\divide\dd*\N*%
\d**=#1\multiply\N*4\divide\d**\N*\sm*\n*\*one\loop%
\shl**{\dd*}\dt*=1.3\dd*\advance\n*\*one%
\shl**{\dt*}\advance\n*\*one%
\shl**{\dd*}\advance\n*\*two%
\dd*-\dd*\ifnum\n*<\N*\repeat\n*\N*\shl**{0pt}}}
\def\wavebox#1{\setbox0\hbox{#1}%
\a*=\wd0\advance\a*14pt\b*=\ht0\advance\b*\dp0\advance\b*14pt%
\hbox{\kern9pt%
\mov*(0pt,\ht0){\mov*(-7pt,7pt){\wlin*\a*(1,0)[+]\wlin*\b*(0,-1)[-]}}%
\mov*(\wd0,-\dp0){\mov*(7pt,-7pt){\wlin*\a*(-1,0)[+]\wlin*\b*(0,1)[-]}}%
\box0\kern9pt}}
\def\rectangle#1(#2,#3){%
\lin#1(#2,0)\lin#1(0,#3)\mov#1(0,#3){\lin#1(#2,0)}\mov#1(#2,0){\lin#1(0,#3)}}
\def\dashrectangle#1(#2,#3){\dashlin#1(#2,0)\dashlin#1(0,#3)%
\mov#1(0,#3){\dashlin#1(#2,0)}\mov#1(#2,0){\dashlin#1(0,#3)}}
\def\waverectangle#1(#2,#3){\L*=#1\Lengthunit\a*=#2\L*\b*=#3\L*%
\ifdim\a*<0pt\a*-\a*\def\x*{-1}\else\def\x*{1}\fi%
\ifdim\b*<0pt\b*-\b*\def\y*{-1}\else\def\y*{1}\fi%
\wlin*\a*(\x*,0)[-]\wlin*\b*(0,\y*)[+]%
\mov#1(0,#3){\wlin*\a*(\x*,0)[+]}\mov#1(#2,0){\wlin*\b*(0,\y*)[-]}}
\def\calcparab*{%
\ifnum\n*>\m*\k*\N*\advance\k*-\n*\else\k*\n*\fi%
\a*=\the\k* sp\a*=10\a*\b*\dm*\advance\b*-\a*\k*\b*%
\a*=\the\*ths\b*\divide\a*\l*\multiply\a*\k*%
\divide\a*\l*\k*\*ths\r*\a*\advance\k*-\r*%
\dt*=\the\k*\L*}
\def\arcto#1(#2,#3)[#4]{\rlap{\toks0={#2}\toks1={#3}\calcnum*#1(#2,#3)%
\dm*=135sp\dm*=#1\dm*\d**=#1\Lengthunit\ifdim\dm*<0pt\dm*-\dm*\fi%
\multiply\dm*\r*\a*=.3\dm*\a*=#4\a*\ifdim\a*<0pt\a*-\a*\fi%
\advance\dm*\a*\N*\dm*\divide\N*10000%
\divide\N*2\multiply\N*2\advance\N*1%
\L*=-.25\d**\L*=#4\L*\divide\d**\N*\divide\L*\*ths%
\m*\N*\divide\m*2\dm*=\the\m*5sp\l*\dm*%
\sm*\n*\*one\loop\calcparab*\shl**{-\dt*}%
\advance\n*1\ifnum\n*<\N*\repeat}}
\def\arrarcto#1(#2,#3)[#4]{\L*=#1\Lengthunit\L*=.54\L*%
\arcto#1(#2,#3)[#4]\mov*(#2\L*,#3\L*){\d*=.457\L*\d*=#4\d*\d**-\d*%
\mov*(#3\d**,#2\d*){\arrow.02(#2,#3)}}}
\def\dasharcto#1(#2,#3)[#4]{\rlap{\toks0={#2}\toks1={#3}\calcnum*#1(#2,#3)%
\dm*=\the\N*5sp\a*=.3\dm*\a*=#4\a*\ifdim\a*<0pt\a*-\a*\fi%
\advance\dm*\a*\N*\dm*%
\divide\N*20\multiply\N*2\advance\N*1\d**=#1\Lengthunit%
\L*=-.25\d**\L*=#4\L*\divide\d**\N*\divide\L*\*ths%
\m*\N*\divide\m*2\dm*=\the\m*5sp\l*\dm*%
\sm*\n*\*one\loop%
\calcparab*\shl**{-\dt*}\advance\n*1%
\ifnum\n*>\N*\else\calcparab*%
\sh*(#2,#3){\kern#3\dt*\lower#2\dt*\hbox{\sm*}}\fi%
\advance\n*1\ifnum\n*<\N*\repeat}}
\def\*shl*#1{%
\c*=\the\n*\d**\advance\c*#1\a**\d*\dt*\advance\d*#1\b**%
\a*=\the\toks0\c*\b*=\the\toks1\d*\advance\a*-\b*%
\b*=\the\toks1\c*\d*=\the\toks0\d*\advance\b*\d*%
\raise\b*\rlap{\kern\a*\unhcopy\spl*}}
\def\calcnormal*#1{%
\b**=10000sp\a**\b**\k*\n*\advance\k*-\m*%
\multiply\a**\k*\divide\a**\m*\a**=#1\a**\ifdim\a**<0pt\a**-\a**\fi%
\ifdim\a**>\b**\d*=.96\a**\advance\d*.4\b**%
\else\d*=.96\b**\advance\d*.4\a**\fi%
\d*=.01\d*\r*\d*\divide\a**\r*\divide\b**\r*%
\ifnum\k*<0\a**-\a**\fi\d*=#1\d*\ifdim\d*<0pt\b**-\b**\fi%
\k*\a**\a**=\the\k*\dd*\k*\b**\b**=\the\k*\dd*}
\def\wavearcto#1(#2,#3)[#4]{\rlap{\toks0={#2}\toks1={#3}\calcnum*#1(#2,#3)%
\c*=\the\N*5sp\a*=.4\c*\a*=#4\a*\ifdim\a*<0pt\a*-\a*\fi%
\advance\c*\a*\N*\c*\divide\N*20\multiply\N*2\advance\N*-1\multiply\N*4%
\d**=#1\Lengthunit\dd*=.012\d**\ifdim\d**<0pt\d**-\d**\fi\L*=.25\d**%
\divide\d**\N*\divide\dd*\N*\L*=#4\L*\divide\L*\*ths%
\m*\N*\divide\m*2\dm*=\the\m*0sp\l*\dm*%
\sm*\n*\*one\loop\calcnormal*{#4}\calcparab*%
\*shl*{1}\advance\n*\*one\calcparab*%
\*shl*{1.3}\advance\n*\*one\calcparab*%
\*shl*{1}\advance\n*2%
\dd*-\dd*\ifnum\n*<\N*\repeat\n*\N*\shl**{0pt}}}
\def\triangarcto#1(#2,#3)[#4]{\rlap{\toks0={#2}\toks1={#3}\calcnum*#1(#2,#3)%
\c*=\the\N*5sp\a*=.4\c*\a*=#4\a*\ifdim\a*<0pt\a*-\a*\fi%
\advance\c*\a*\N*\c*\divide\N*20\multiply\N*2\advance\N*-1\multiply\N*2%
\d**=#1\Lengthunit\dd*=.012\d**\ifdim\d**<0pt\d**-\d**\fi\L*=.25\d**%
\divide\d**\N*\divide\dd*\N*\L*=#4\L*\divide\L*\*ths%
\m*\N*\divide\m*2\dm*=\the\m*0sp\l*\dm*%
\sm*\n*\*one\loop\calcnormal*{#4}\calcparab*%
\*shl*{1}\advance\n*2%
\dd*-\dd*\ifnum\n*<\N*\repeat\n*\N*\shl**{0pt}}}
\def\hr*#1{\clap{\xL*=\xscale\Lengthunit\vrule width#1\xL*
height.1pt}}
\def\shade#1[#2]{\rlap{\Lengthunit=#1\Lengthunit%
\smov(0,#2.05){\hr*{.994}}\smov(0,#2.1){\hr*{.980}}%
\smov(0,#2.15){\hr*{.953}}\smov(0,#2.2){\hr*{.916}}%
\smov(0,#2.25){\hr*{.867}}\smov(0,#2.3){\hr*{.798}}%
\smov(0,#2.35){\hr*{.715}}\smov(0,#2.4){\hr*{.603}}%
\smov(0,#2.45){\hr*{.435}}}}
\def\dshade#1[#2]{\rlap{%
\Lengthunit=#1\Lengthunit\if#2-\def\t*{+}\else\def\t*{-}\fi%
\smov(0,\t*.025){%
\smov(0,#2.05){\hr*{.995}}\smov(0,#2.1){\hr*{.988}}%
\smov(0,#2.15){\hr*{.969}}\smov(0,#2.2){\hr*{.937}}%
\smov(0,#2.25){\hr*{.893}}\smov(0,#2.3){\hr*{.836}}%
\smov(0,#2.35){\hr*{.760}}\smov(0,#2.4){\hr*{.662}}%
\smov(0,#2.45){\hr*{.531}}\smov(0,#2.5){\hr*{.320}}}}}
\def\vdot{\rlap{\kern-1.9pt\lower1.8pt\hbox{$\scriptstyle\bullet$}}}
\def\vtimes{\rlap{\kern-3pt\lower1.8pt\hbox{$\scriptstyle\times$}}}
\def\vDot{\rlap{\kern-2.3pt\lower2.7pt\hbox{$\bullet$}}}
\def\vTimes{\rlap{\kern-3.6pt\lower2.4pt\hbox{$\times$}}}
\catcode`\*=12
\newcount\CatcodeOfAtSign
\CatcodeOfAtSign=\the\catcode`\@
\catcode`\@=11
\newcount\n@ast
\def\n@ast@#1{\n@ast0\relax\get@ast@#1\end}
\def\get@ast@#1{\ifx#1\end\let\next\relax\else%
\ifx#1*\advance\n@ast1\fi\let\next\get@ast@\fi\next}
\newif\if@up \newif\if@dwn
\def\up@down@#1{\@upfalse\@dwnfalse%
\if#1u\@uptrue\fi\if#1U\@uptrue\fi\if#1+\@uptrue\fi%
\if#1d\@dwntrue\fi\if#1D\@dwntrue\fi\if#1-\@dwntrue\fi}
\def\halfcirc#1(#2)[#3]{{\Lengthunit=#2\Lengthunit\up@down@{#3}%
\if@up\smov(0,.5){\arc[-][-]\arc[+][-]}\fi%
\if@dwn\smov(0,-.5){\arc[-][+]\arc[+][+]}\fi%
\def\lft{\smov(0,.5){\arc[-][-]}\smov(0,-.5){\arc[-][+]}}%
\def\rght{\smov(0,.5){\arc[+][-]}\smov(0,-.5){\arc[+][+]}}%
\if#3l\lft\fi\if#3L\lft\fi\if#3r\rght\fi\if#3R\rght\fi%
\n@ast@{#1}%
\ifnum\n@ast>0\if@up\shade[+]\fi\if@dwn\shade[-]\fi\fi%
\ifnum\n@ast>1\if@up\dshade[+]\fi\if@dwn\dshade[-]\fi\fi}}
\def\halfdashcirc(#1)[#2]{{\Lengthunit=#1\Lengthunit\up@down@{#2}%
\if@up\smov(0,.5){\dasharc[-][-]\dasharc[+][-]}\fi%
\if@dwn\smov(0,-.5){\dasharc[-][+]\dasharc[+][+]}\fi%
\def\lft{\smov(0,.5){\dasharc[-][-]}\smov(0,-.5){\dasharc[-][+]}}%
\def\rght{\smov(0,.5){\dasharc[+][-]}\smov(0,-.5){\dasharc[+][+]}}%
\if#2l\lft\fi\if#2L\lft\fi\if#2r\rght\fi\if#2R\rght\fi}}
\def\halfwavecirc(#1)[#2]{{\Lengthunit=#1\Lengthunit\up@down@{#2}%
\if@up\smov(0,.5){\wavearc[-][-]\wavearc[+][-]}\fi%
\if@dwn\smov(0,-.5){\wavearc[-][+]\wavearc[+][+]}\fi%
\def\lft{\smov(0,.5){\wavearc[-][-]}\smov(0,-.5){\wavearc[-][+]}}%
\def\rght{\smov(0,.5){\wavearc[+][-]}\smov(0,-.5){\wavearc[+][+]}}%
\if#2l\lft\fi\if#2L\lft\fi\if#2r\rght\fi\if#2R\rght\fi}}
\def\Circle#1(#2){\halfcirc#1(#2)[u]\halfcirc#1(#2)[d]\n@ast@{#1}%
\ifnum\n@ast>0\clap{%
\dimen0=\xscale\Lengthunit\vrule width#2\dimen0 height.1pt}\fi}
\def\wavecirc(#1){\halfwavecirc(#1)[u]\halfwavecirc(#1)[d]}
\def\dashcirc(#1){\halfdashcirc(#1)[u]\halfdashcirc(#1)[d]}
%
\def\xscale{1}
\def\yscale{1}
\def\Ellipse#1(#2)[#3,#4]{\def\xscale{#3}\def\yscale{#4}%
\Circle#1(#2)\def\xscale{1}\def\yscale{1}}
\def\dashEllipse(#1)[#2,#3]{\def\xscale{#2}\def\yscale{#3}%
\dashcirc(#1)\def\xscale{1}\def\yscale{1}}
\def\waveEllipse(#1)[#2,#3]{\def\xscale{#2}\def\yscale{#3}%
\wavecirc(#1)\def\xscale{1}\def\yscale{1}}
\def\halfEllipse#1(#2)[#3][#4,#5]{\def\xscale{#4}\def\yscale{#5}%
\halfcirc#1(#2)[#3]\def\xscale{1}\def\yscale{1}}
\def\halfdashEllipse(#1)[#2][#3,#4]{\def\xscale{#3}\def\yscale{#4}%
\halfdashcirc(#1)[#2]\def\xscale{1}\def\yscale{1}}
\def\halfwaveEllipse(#1)[#2][#3,#4]{\def\xscale{#3}\def\yscale{#4}%
\halfwavecirc(#1)[#2]\def\xscale{1}\def\yscale{1}}
\catcode`\@=\the\CatcodeOfAtSign
\hfill
\begin{minipage}{35mm}
UPR-0849-T\\
hep-th/9906141
\end{minipage}

\begin{center}
{\LARGE One-loop effective potential of N=1 supersymmetric theory
and decoupling effects} \end{center}

\centerline{\large I.L. Buchbinder$^{*}$, M.
Cveti\v{c}$^{+}$ and A.Yu. Petrov$^{*}$}
\begin{center}
$^{*}${\small\it
Department
of Theoretical Physics, Tomsk State Pedagogical University}\\
 {\small\it 634041
Tomsk, Russia}

\vspace*{2mm}

 $^{+}${\small\it Department of Physics and Astronomy, University of
Pennsylvania}\\ {\small\it Philadelphia, PA 19104--6396, USA}
\end{center}
\begin{abstract}
We study  the  decoupling effects in  $N=1$ (global) supersymmetric
theories with
chiral superfields at the one-loop
level. Examples of gauge neutral chiral superfields
with
minimal (renormalizable) as well as  non-minimal
(non-renormalizable)
couplings are considered, and decoupling in gauge theories with
$U(1)$
gauge superfields that couple to heavy chiral matter is studied.
We calculate the one-loop corrected effective Lagrangians that
involve light fields
and heavy fields with mass  of order  $M$. Elimination of
heavy
fields by equations of motion  leads to decoupling effects with
terms
that grow logarithmically with $M$.
These  corrections
renormalize light  fields  and couplings in the theory (in
accordance with the ``decoupling theorem'').
 When the field theory is an effective theory of the underlying
fundamental theory, like superstring theory,  where the couplings
are
calculable, such decoupling effects
 modify the low energy predictions for the
effective couplings of light
fields.
In particular, for the class of string vacua with an
``anomalous'' $U(1)$, the vacuum restabilization  triggers
decoupling effects, which can significantly modify the low energy
predictions for couplings of the surviving light fields.
We also demonstrate that quantum corrections to the chiral
potential depending on massive background superfields and
corresponding to
supergraphs with internal massless lines and external massive
lines can also
arise at the two-loop level.
 \end{abstract}
\tableofcontents
\section{Introduction}
\renewcommand{\theequation}{\arabic{section}.\arabic{equation}}
This paper is devoted to the calculation of the one-loop effective
action for
several
models of the global $N=1$ supersymmetric  theory  with
 chiral superfields and a  subsequent study  of some of their
phenomenologically interesting aspects. In particular, we
investigate in detail the decoupling effects due to
the couplings of heavy and light chiral superfields in the theory
and
subsequent implications for the low energy  effective action of
light
superfields.

In principle the  decoupling  effects of heavy  fields in   field
theory are  well understood.  According to  the decoupling theorem
\cite{AppleCat,Syma} (for additional references see, e.g.,
\cite{Collins}) in
the field
theory of interacting light (with masses $m$) and heavy fields
(with masses $M$)
the heavy fields decouple;
the effective Lagrangian of the light fields can be written in terms
of the
original
classical Lagrangian of   light  fields with
 loop effects  of heavy fields
absorbed into redefinitions of
 new light fields, masses and couplings, and the only
new terms in this effective  Lagrangian
 are non- renormalizable, proportional to inverse powers of $M$
(both at tree-
 and loop-levels).

In a field theory as an effective description of phenomena at
certain energies,
the rescaling of the fields and couplings due to heavy
 fields  does not affect the structure of the  couplings,
since those are  { free parameters} whose
values are determined by experiments.  On the other hand if the
field theory is describing an effective theory of an underlying
fundamental theory, like superstring theory, where the couplings at
the
string scale are calculable, the  decoupling effects of the heavy
field
can be  important and can significantly  affect the low energy
predictions  for the couplings of light fields at low energies.
Therefore the quantitative study of decoupling effects
 at the  loop-level  in  effective
  supersymmetric theories is important; it  should
    improve our understanding of such effects for the
effective  Lagrangians from superstring theory and provide us with
 calculable corrections for the  low energy predictions of the
theory. We also note that as the decoupling theorem is based on
finite
renormalization of fields and parameters as all parameters in the
effective
theory (fields, masses, couplings) are determined from the
corresponding
string theory and hence cannot be renormalized. Therefore we will
use consistence
with the decoupling theorem only to check the results.

Effective theories of $N=1$ supersymmetric  four-dimensional
perturbative
 string vacua
can be obtained by employing techniques
of  two-dimensional conformal field theory~\cite{Shenkeretal}.
In particular the k\"ahlerian and  the chiral (super-) potential
 can be calculated  explicitly at the tree level.
 While the chiral potential terms   calculated at the
  string tree-level  are protected from higher genus
 corrections (for a
 representative work on the
 subject see, e.g., \cite{bailin,GSW}, and references
therein), the k\"ahlerian  potential is not.  Such higher
  genus corrections to the
 k\"ahlerian potential could be
 significant; however, their structure has not been studied very
much.
In this paper we shall not address these issues and assume that the
string
theory calculation provides us with a (reliable, calculable)  form
of the
effective theory at $M_{string}$, which would in turn serve as a
starting point of our study.

One of the compelling motivations for a detailed study of decoupling
effects
 is  the phenomenon of
vacuum restabilization~\cite{dsw}  for  a class of  four-dimensional
(quasi-realistic)    heterotic superstring vacua with an
``anomalous'' $U(1)$. (On the open Type I string side these effects
are
closely related to the blowing-up procedure of Type I orientifolds
and
were recently studied in \cite{CELW}.)
 For  such  string vacua of perturbative heterotic  string
theory, the
Fayet-Iliopoulos  (FI) $D$- term is generated at
genus-one~\cite{atick}, thus triggering certain  fields to acquire
vacuum
expectation values (VEV's) of order $M_{String}\sim g_{gauge}
M_{Planck}\sim 5 \times 10^{17}
 \,$ GeV
along $D$- and $F$- flat directions of the effective $N=1$
supersymmetric
theory.
(Here $g_{gauge}$ is the gauge
coupling and $M_{Planck}$ the Planck scale.) Due to these large
string-scale VEV's  a number of additional fields obtain large
string-scale masses. Some of them in turn couple through
(renormalizable) interactions to the remaining light fields, and
thus through
decoupling effects
affect the effective theory of  light fields
at low energies. (For the study of the effective Lagrangians and
their
phenomenological implications for a  class  of
such four-dimensional string vacua see, e.g., \cite{cceel2}
--\cite{Cv4} and references therein.)

 The  tree level decoupling effects  within $N=1$ supersymmetric
theories,
 were studied within an effective string theory  in \cite{Burgess}.
In  a
related
 work \cite{katehou} it was  shown that 
the leading order corrections  of order $1\over M$ are to the
effective chiral potential, however there are also important next
order effects
in  the  effective chiral potential \cite{cew}.
In addition, in \cite{cew} the nonrenormalizable modifications of
the
k\"{a}hlerian potential  (as was also pointed out in \cite{dine})
were
systematically studied.
These tree level  decoupling effects (as triggered by, e.g., vacuum
restabilization for a class of string vacua)
 lead to new nonrenormalizable interactions which are competitive
with the nonrenormalizable  terms that are calculated
 directly in the superstring  theory.

In this paper  we  consider one-loop decoupling effects in
$N=1$ supersymmetric theory. We
study both the effects on chiral (gauge neutral)  superfields and
on the effects of gauge superfields. (In another context see
\cite{Rattari}.)
It turns out that an
essential modification of low energy predictions takes place not
only for
chiral superfields \cite{new} but also for gauge superfields.
As stated earlier such effective Lagrangians arise  naturally due to
the vacuum
restabilization  for  a class of supersymmetric string vacua and
trigger couplings between heavy
fields with mass scale $M\sim 10^{17}$ GeV and the light (massless)
fields \cite{Cleaver:1999cj}.
(Note however, that we do  not include supergravity effects which
could
also be significant.)

As a result we find that the one-loop effective action after a
redefinition of fields, masses and couplings coincides with the
one-loop
effective action of the corresponding theory where heavy
superfields are completely
absent, in accordance with the decoupling theorem. However, since
the masses and the couplings of the fields are calculable in string
theory
(at the mass scale $M_{string}$), the decoupling effects add
additional
corrections to the effective action of the  light superfields. These
corrections grow logarithmically
with $M$ (mass of the heavy superfields)
and modify the effective couplings in an
essential way,  which for a class of string vacua under
consideration can be significant.

Another interesting result presented in this paper pertains to the
chiral
effective potential. When the chiral potential
depends  on massive
superfields, quantum corrections due to these fields
appear earlier than in the case
when one considers light fields only.

This paper is organized as follows. In Section 2 the general
structure of the
effective action studied   is given and
the general approach to addressing the decoupling effects is
presented.
Section 3 is devoted to the study of the  effective action for
the ``minimal''  model with one heavy and one light
(gauge neutral) chiral superfield.  In Section
4 the leading order decoupling  corrections to the effective action
for
non-minimal models (with more  general couplings)
are considered.  Section 5 is devoted to the investigation of the
one-loop decoupling effects in $N=1$
supersymmetric theory with $U(1)$ gauge vector superfields and
chiral superfields charged under $U(1)$.
A summary and  discussion of the obtained results are given
in Section 6.
In Appendix A  details
of the calculation of the one-loop k\" ahlerian effective potential
for
the minimal model  are presented,  in Appendix B the calculation
of  the one-loop
k\"{a}hlerian effective action via diagram technique for the
minimal model
is described, and in Appendix C details of the calculation for the
effective action of  non-minimal models are given.

\section{Effective action in the model of interacting light and
heavy
superfields}
\setcounter{equation}{0}
\renewcommand{\theequation}{\arabic{section}.\arabic{equation}}
\subsection{General structure of the effective action}

$N=1$ supersymmetric  actions  with chiral supermultiplets
arise as  a subsector of an effective theory of $N=1$ supersymmetric
string vacua. Such calculations are carried out for perturbative
string vacua primarily  by employing conformal field theory
techniques.
(Though less powerful techniques, e.g., sigma-model approach,  in
which
the integration over
massive string modes is carried out in the  the background of the
ten-dimensional manifold  with  the structure
$M^4\times K$ where $M^4$ is a four-dimensional Minkowski space and
$K$ is a suitable six-dimensional compact  (Calabi-Yau) manifold,
can also
be employed.)
The
resulting effective theories contain as an ingredient $N=1$
chiral superfields $\Phi^i$ with action
\begin{equation}
\label{act1}
S[\Phi,\bar{\Phi}]=\int d^8 z K(\bar{\Phi}^i,\Phi^i)+
(\int d^6 z W(\Phi^i)+h.c.)
\end{equation}
Here $\Phi^i=\Phi^i(z)$, $z^A\equiv(x^a,\theta_{\alpha},
\bar{\theta}_{\dot{\alpha}})$; $a=0,1,2,3$; $\alpha=1,2$,
$\dot{\alpha}=\dot{1},
\dot{2}$, $d^8 z= d^4 x d^2\theta d^2\bar{\theta}$. Real function
$K(\bar{\Phi}^i,\Phi^i)$ is called the k\"ahlerian potential, the
holomorphic function
$W(\Phi^i)$ is called the chiral potential \cite{BK0}. Expression
(\ref{act1})  represents the  most general action of gauge neutral
chiral
superfields which does not
contain higher derivatives at a component level \cite{BK0}. We refer
to this action as the  chiral superfield model of  a general form.
In a
special case
$K(\bar{\Phi}^i,\Phi^i)=\Phi\bar{\Phi}$, $W(\Phi^i)\sim\Phi^3$ we
obtain
the well-known Wess-Zumino model. For  $W(\Phi^i)=0$ the present
theory
represents
itself as a $N=1$ supersymmetric four-dimensional sigma-model (see,
e.g.,\cite{GSW}).
The action (\ref{act1}),  which originates  from  superstring
theory,
can be treated as a classical effective action of the fundamental
theory,
suitable for
description of phenomena at energies
much less than the Planck scale.
Such   models of chiral superfields
are widely used for the study
of possible phenomenological implications of superstring theories
(see,
e.g., recent papers \cite{CELW,cceel2,Cvet,Cv4,new} and references
therein). One of the most
important aspects of the study of these models pertains to the
investigation of the decoupling effects, which is
the main subject of the present paper.

The starting point in the study of the decoupling effects is
the  model with the classical
action (\ref{act1})  and, for the sake of simplicity,
two chiral superfields: a light one,
$\phi$, and a heavy one, $\Phi$, i.e. $\Phi^i=\{\Phi,\phi\}$.
The aim is to  to calculate
the low-energy effective action  in the one-loop approximation
and to compute the  one-loop corrected effective action of light
superfield, only.

We refer to the model in which the k\"ahlerian potential is of the
canonical (minimal) form:
\begin{equation}
\label{kmin}
K(\Phi,\bar{\Phi},\phi,\bar{\phi})=\Phi\bar{\Phi}+\phi\bar{\phi}
\end{equation}
as the minimal model,
and the model in which
\begin{equation}
\label{knon}
K(\Phi,\bar{\Phi},\phi,\bar{\phi})=\Phi\bar{\Phi}+\phi\bar{\phi}
+\tilde{K}(\Phi,\bar{\Phi},\phi,\bar{\phi})
\end{equation}
with $\tilde{K}\neq 0$  --
as the non-minimal one (in analogy with \cite{cceel2}). We assume
that the function $\tilde{K}(\Phi,\bar{\Phi},\phi,\bar{\phi})$ can
be expanded into power series in superfields
$\Phi,\bar{\Phi},\phi,\bar{\phi}$
where the leading order term is at least of the third order in
the chiral  superfields (and thus proportional to  at least one
inverse power of $M$)
\begin{equation}
\tilde{K}(\Phi,\bar{\Phi},\phi,\bar{\phi})=
\frac{\phi\bar{\Phi}^2}{M}+\frac{\Phi\bar{\phi}^2}{M}\ldots
\end{equation}

The chiral potential  $M$ is taken to be of the form:
\begin{equation}
W=\frac{M}{2}\Phi^2+\tilde{W}(\Phi,\phi)
\label{W}
\end{equation}
where $\tilde{W}$ is also at least of the third order in
$\{\Phi,\phi\}$
superfields. The functions
$\tilde{W}$ could have the following structure
\begin{equation}
\label{tW}
\tilde{W}(\Phi,\phi)
=
\phi\Phi^2+\Phi\phi^2+ \phi^3+ \Phi^3+
\frac{\phi\Phi^3}{M}+\ldots
\end{equation}
with $M$ as  a massive parameter.
Hence the possible vertices of interaction of superfields have the
form
$$
\phi\Phi^2, \Phi\phi^2, \frac{\Phi\bar{\phi}^2}{M}, \phi^3, \Phi^3,
\frac{\phi\Phi^3}{M}\ldots $$

The effective action $\Gamma [\Phi,\bar{\Phi},\phi,\bar{\phi}]$ is
defined as
the Legendre transform from  the generating functional of connected
Green functions \cite{BO} $W[J,\bar{J}]$:
 \begin{eqnarray}
 \label{Green}
 \exp(\frac{i}{\hbar}W[J,\bar{J},j,\bar{j}]) & = &
\int {\cal D} \varphi
{\cal D} \bar{\varphi} {\cal D} \bar{\Psi} {\cal D} \Psi
      \exp(\frac{i}{\hbar}(S[\Psi,\bar{\Psi},\varphi,\bar{\varphi}]+
\nonumber\\&+&(\int d^6 z
      (J\Psi+j\varphi )+h.c.)))\\
\Gamma[\Phi,\bar{\Phi},\phi,\bar{\phi}] & = &
 W[J,\bar{J}]-(\int d^6 z (J\Phi+j\phi )+h.c.)\nonumber
\end{eqnarray}
$\Gamma [\Phi, \bar{\Phi},\phi,\bar{\phi}]$ can be calculated
using
the loop-expansion method.  This method employs
the splitting of all  the chiral superfields
into a sum of the background superfields $\Phi,\phi$ and  the
quantum ones
$\Phi_q,\phi_q$, using the rule
\begin{eqnarray}
\Phi&\to&\Phi+\sqrt{\hbar}\Phi_q\\
\phi&\to&\phi+\sqrt{\hbar}\phi_q\nonumber
\end{eqnarray}
As a result the action (\ref{act1}) after such changes can be
written as
\begin{eqnarray}
\label{actq}
S_q&=&\int d^8 z
K(\Phi+\sqrt{\hbar}\Phi_q,\bar{\Phi}+\sqrt{\hbar}\bar{\Phi}_q,
\phi+\sqrt{\hbar}\phi_q,\bar{\phi}+\sqrt{\hbar}\bar{\phi}_q)
+\nonumber\\
&+&[\int d^6 z W(\Phi+\sqrt{\hbar}\Phi_q,
\phi+\sqrt{\hbar}\phi_q)+h.c.]
\end{eqnarray}
and the effective action takes the form:
\begin{eqnarray}
\label{Green1}
 \exp(\frac{i}{\hbar}\Gamma[\Phi,\bar{\Phi},\phi,\bar{\phi}]) &=&
\int {\cal D} \phi_q
{\cal D} \bar{\phi}_q {\cal D} \bar{\Phi}_q {\cal D} \Phi_q
 \exp\big(\frac{i}{\hbar}
S[\Phi+\sqrt{\hbar}\Phi_q,\bar{\Phi}+\sqrt{\hbar}\bar{\Phi}_q,
\nonumber\\& &
\phi+\sqrt{\hbar}\phi_q,\bar{\phi}+\sqrt{\hbar}\bar{\phi}_q
] -\nonumber\\&-&
\sqrt{\hbar}(\int d^6 z
(\frac{\delta\Gamma}{\delta\Phi(z)}\Phi_q(z)+
\frac{\delta\Gamma}{\delta\phi(z)}\phi_q(z))
+h.c. )
\big)
\end{eqnarray}
(for details see  \cite{BK0,BO}).
The effective action (\ref{Green1}) can be cast  in the form
$\Gamma[\Phi,\bar{\Phi},\phi,\bar{\phi}]=
S[\Phi,\bar{\Phi},\phi,\bar{\phi}]+
\tilde{\Gamma}[\Phi,\bar{\Phi},\phi,\bar{\phi}]$.
Here $\tilde{\Gamma}[\Phi,\bar{\Phi},\phi,\bar{\phi}]$ is a
quantum correction in effective action which can be expanded into
power series
in $\hbar$ as
\begin{equation}
\label{Gamma}
 \tilde{\Gamma}[\Phi,\bar{\Phi},\phi,\bar{\phi}] =
\sum_{n=1}^{\infty}\hbar^n
\Gamma^{(n)} [\Phi,\bar{\Phi},\phi,\bar{\phi}]
\end{equation}

The one-loop quantum correction $\Gamma^{(1)}$
to the effective action is defined through the following expression
\cite{BO}:
\begin{equation}
e^{i\Gamma^{(1)}}=\int
{\cal D}\Phi_q {\cal D} \bar{\Phi}_q {\cal D}\phi_q {\cal D}
\bar{\phi}_q
\exp (iS^{(2)}_q)
\end{equation}
Here $S^{(2)}_q$  corresponds to the  part of $S_q$
(\ref{actq})  which is  quadratic in  quantum
superfields. It is of the form
\begin{eqnarray}
\label{actq2}
S^{(2)}_q&=&\int d^8 z (K_{\Phi\bar{\Phi}}\Phi_q\bar{\Phi}_q+
K_{\phi\bar{\Phi}}\phi_q\bar{\Phi}_q+K_{\Phi\bar{\phi}}\Phi_q\bar{\phi}_q+
K_{\phi\bar{\phi}}\phi_q\bar{\phi}_q)+\nonumber\\
&+& [\int d^6 z W_{\Phi\Phi}\Phi^2_q+W_{\phi\Phi}\Phi_q\phi_q+
W_{\phi\phi}\phi^2_q+h.c.]
\end{eqnarray}
As a result we arrive at the one-loop effective action of the form
\begin{eqnarray}
\label{eact}
\Gamma[\Phi,\bar{\Phi},\phi,\bar{\phi}] &=&S+\hbar\Gamma^{(1)}=\int
d^8 z
K(\Phi,\bar{\Phi},\phi,\bar{\phi}) + [\int d^6 z W(\Phi,\phi)+h.c.]+
\nonumber\\&+&
\hbar\big(\int d^8 z K^{(1)}(\Phi,\bar{\Phi},\phi,\bar{\phi})
+(\int d^6 z W^{(1)}(\Phi,\phi)+h.c.)\big)
\end{eqnarray}
Here we suppose that the one-loop correction in the effective action
$\Gamma^{(1)}$ can be represented in the form
\begin{equation}
\label{eact1}
\Gamma^{(1)}=\int d^8 z K^{(1)}(\Phi,\bar{\Phi},\phi,\bar{\phi}) +
(\int d^6 z W^{(1)}(\Phi,\phi)+h.c.)+\ldots
\end{equation}
Dots denote terms that depend on the supercovariant derivatives of
the chiral superfields.

The loop corrected effective action 
 has the
following structure
\begin{eqnarray}
& &\Gamma[\Phi,\bar{\Phi},\phi,\bar{\phi}] = \int d^8z
L_{eff}(\Phi,D_A\Phi,D_A D_B\Phi,
\bar{\Phi},D_A\bar{\Phi},D_AD_B\bar{\Phi},
\nonumber\\
& &\phi,D_A\phi,D_A D_B\phi,
\bar{\phi},D_A\bar{\phi},D_AD_B\bar{\phi}) +
(\int d^6z L^{(c)}_{eff}(\Phi,\phi) + h.c.)
+\ldots
\end{eqnarray}
Here $D_A$ are supercovariant derivatives,
$D_A=(\partial_a, D_{\alpha}, \bar{D}_{\dot{\alpha}})$.
$L_{eff}$ is the effective super-Lagrangian
that we write in the form\begin{eqnarray}
 L_{eff} & = & K_{eff}(\Phi,\bar{\Phi},\phi,\bar{\phi}) +\ldots
\nonumber \\
 K & = & K(\Phi,\bar{\Phi},\phi,\bar{\phi}) + \sum_{n=1}^\infty
\hbar^n K^{(n)}
\end{eqnarray}
and $L^{(c)}$ is the effective chiral Lagrangian
\begin{equation}
L^{(c)}=W_{eff}(\Phi,\phi)+\ldots
\end{equation}
$K_{eff}$ is the k\"ahlerian effective potential that depends
only on the chiral superfields
$\Phi$, $\bar{\Phi},\phi,\bar{\phi}$  but not on their
(covariant) derivatives.
$W_{eff}$ is the chiral effective potential that depends on
on (holomorphic) chiral superfields
$\{ \Phi$, $\phi\}$, only.
Dots denote the terms that depend on the the space-time
derivatives of chiral superfields only.
Furthermore, one can  prove that the
one-loop correction to the chiral potential  is zero
(for  the $N=1$ supersymmetric theory which does not
include gauge superfields). However,
higher corrections can exist
(cf. \cite{Buch1}--\cite{West2}), i.e.
\begin{equation}
W_{eff}(\Phi,\phi)=W(\Phi,\bar{\phi})+\sum_{n=2}^\infty \hbar^n
W^{(n)}(\Phi,\phi)
\end{equation}
Here $K^{(n)}$ and $W^{(n)}$ are loop corrections to the
k\"{a}hlerian
and chiral potential, respectively.

Since in this paper we concentrate on the one-loop corrected
effective
action only,  we are mainly interested in the correction to the
k\"ahlerian potential which is the leading term in the
one-loop
corrected low-energy effective
action. (At low energies ($E\ll M$)   higher derivative terms are
suppressed.)

Our ultimate goal is
to obtain  the effective action for light superfields, only. For
that
purpose one  must eliminate  heavy  superfields from the one-loop
effective action
$\Gamma[\Phi,\bar{\Phi},\phi,\bar{\phi}]$ (\ref{eact}) by means of
the effective
equations of motion.
These equations  can
be solved by an iterative method up to
a certain order in  the inverse mass  $M$ of heavy superfield.
Substituting a solution of these
equations into the effective action (\ref{eact}) we then obtain the
one-loop
corrected effective action of light superfields only.
In the following subsection we shall describe the procedure in
detail.

\subsection{The effective equations of motion}
The effective equations of motion for heavy superfields in the model
with the
effective action (\ref{eact},\ref{eact1}) are of the form
\begin{eqnarray}
\label{eqm}
\frac{\delta \Gamma}{\delta\Phi}=0
&: &-\frac{1}{4}\bar{D}^2(\frac{\partial K}{\partial \Phi}+
\frac{\partial K^{(1)}}{\partial \Phi})+
\frac{\partial W}{\partial \Phi}
=0\nonumber\\
\frac{\delta \Gamma}{\delta\bar{\Phi}}=0
&: &-\frac{1}{4}\bar{D}^2(\frac{\partial K}{\partial \bar{\Phi}}+
\frac{\partial K^{(1)}}{\partial \bar{\Phi}})+
\frac{\partial \bar{W}}{\partial \bar{\Phi}}
=0
\end{eqnarray}
The effective equations of motion for light superfields have an
analogous
form. We  consider the case when the  interactions with the  gauge
superfields are absent (see however Section 5)
and $W^{(1)}=0$ (which is absent at one-loop level (cf., discussion
above)).

The equations (\ref{eqm}) can be solved via an iterative method,
described
below.
We can represent the heavy superfield $\Phi$ in the form
\begin{equation}
\label{ex}
\Phi=\Phi_0+\Phi_1+\ldots+\Phi_n+\ldots
\end{equation}
where $\Phi_0$ is zeroth-order approximation, $\Phi_1$ is
first-order one,
etc..
We assume that
$|D^2\Phi|\ll M\bar{\Phi}$ since the superfield $\Phi$ is heavy, and
thus
the assumption is valid.
The zeroth-order approximation $\Phi_0$ can be found from the
condition
$$
\frac{\partial W}{\partial \Phi}|_{\Phi=\Phi_0}=0
$$
After a substitution of  the expansion (\ref{ex}) into
equations (\ref{eqm})
we arrive at the following equation for the $(n+1)$-th-order
solution for $\bar{\Phi}_{n+1}$
\begin{eqnarray}
\label{ee}
& &(\frac{\partial \bar{W}}{\partial \bar{\Phi}}
)|_{\Phi=\Phi_0+\ldots+\Phi_{n+1}}-
(\frac{\partial \bar{W}}{\partial \bar{\Phi}}
)|_{\Phi=\Phi_0+\ldots+\Phi_{n}}=\\
&=&\frac{\bar{D}^2}{4}(
\frac{\partial K}{\partial \Phi}|_{\Phi=\Phi_0+\ldots+\Phi_n}-
\frac{\partial K}{\partial \Phi}|_{\Phi=\Phi_0+\ldots+\Phi_{n-1}}
+
\frac{\partial K^{(1)}}{\partial \Phi}|_{\Phi=\Phi_0+\ldots+\Phi_n}-
\frac{\partial K^{(1)}}{\partial
\Phi}|_{\Phi=\Phi_0+\ldots+\Phi_{n-1}})
\nonumber
\label{rec}
\end{eqnarray}
and an analogous  equation for  $\Phi_{n+1}$.  These sets of
equations
can be used
to calculate $\Phi_n$ for any  $n\geq 1$.
Since the chiral potential is of the form
$W=\frac{M}{2}\Phi^2+\tilde{W}$  (see eqs. (\ref{W}-\ref{tW}),
eq. (\ref{rec}))  can be rewritten in the form
\begin{eqnarray}
\label{ee1}
& &M\bar{\Phi}_{n+1}+(\frac{\partial \bar{\tilde{W}}}{\partial
\bar{\Phi}}
)|
_{\Phi=\Phi_0+\ldots+\Phi_{n+1}}-
(\frac{\partial \bar{\tilde{W}}}{\partial \bar{\Phi}}
)
|_{\Phi=\Phi_0+\ldots+\Phi_{n}}=
\\
&=&\frac{\bar{D}^2}{4}(
\frac{\partial K}{\partial \Phi}|_{\Phi=\Phi_0+\ldots+\Phi_n}-
\frac{\partial K}{\partial \Phi}|_{\Phi=\Phi_0+\ldots+\Phi_{n-1}}
+
\frac{\partial K^{(1)}}{\partial \Phi}|_{\Phi=\Phi_0+\ldots+\Phi_n}-
\frac{\partial K^{(1)}}{\partial
\Phi}|_{\Phi=\Phi_0+\ldots+\Phi_{n-1}})
\nonumber
\end{eqnarray}
with an analogous equation for $\Phi_{n+1}$.
As a result we can find $\Phi$, $\bar{\Phi}$ in principle to any
order.

Thus, a substitution of the heavy superfields $\Phi$, $\bar{\Phi}$
expressed
in terms of light superfields by means of equations
(\ref{ee}-\ref{ee1})
allows one to obtain the  one-loop corrected effective action of
light
superfields only.

\subsection{Calculation of the one-loop k\"ahlerian effective
potential}
Let us now consider the method to calculate the one-loop
effective action.   The most effective
method  consists of  expressing the quantities in terms  of matrices
(matrix approach). First we write the
heavy quantum superfield $\Phi_q$ and light one $\phi_q$  as a
column vector:
$\psi_q=\left(\begin{array}{c}\Phi_q\\ \phi_q\end{array}\right)$.
As a result, the part of the classical action that is  quadratic in
quantum superfields
(\ref{actq2}) takes the form
\begin{equation}
\label{spsi}
S^{(2)}[\psi_q,\bar{\psi}_q]=\int d^8 z
K_{\psi\bar{\psi}}\psi_q\bar{\psi}_q+
(\int d^6 z W^{''}\psi^2_q+h.c.)
\end{equation}
Here $K_{\psi\bar{\psi}}$ and $W^{''}$ are matrices of the form
\begin{eqnarray}
\label{wpp}
K_{\psi\bar{\psi}}=
\left(\begin{array}{cc}
K_{\phi\bar{\phi}}&K_{\phi\bar{\Phi}}\\
K_{\Phi\bar{\phi}}&K_{\Phi\bar{\Phi}}
\end{array}\right)
\
W^{''}=
\left(\begin{array}{cc}
W_{\phi\phi}&W_{\phi\Phi}\\
W_{\phi\Phi}&W_{\Phi\Phi}
\end{array}\right)
\end{eqnarray}
The one-loop effective action
 \cite{BO} (\ref{Gamma1}) is of the form
\begin{equation}
\label{gam}
exp(i\Gamma^{(1)})=\int D\psi_q D\bar{\psi}_q
S^{(2)}[\psi_q,\bar{\psi}_q]
\end{equation}
where $S^{(2)}[\psi_q,\bar{\psi}_q]$ is given by (\ref{spsi}),
$D\psi_q$ denotes
the integration  over  both $\phi_q$ and $\Phi_q$.

To perform the integration we employ the  technique analogous to
that
used for the study of  the Wess-Zumino model \cite{Buch1,Buch2}.
First we
consider a theory of a real
scalar superfield $q$ which represents itself as a column
$q=\left(\begin{array}{c}u\\v\end{array}\right)$. The action for $q$
reads as
\begin{equation}
S^0_q=-\frac{1}{16}\int d^8 z q^T D^{\alpha}\bar{D}^2 D_{\alpha}q
\end{equation}
where $q^T$ is a string (row)  of the form $q^T=(u\ v)$.
The effective action $U_q$ for the superfield $q$ can be
obtained by the
Faddeev-Popov procedure.
We choose the gauge fixing functions
$\chi[q]=\frac{1}{4}\bar{D}^2 q-\psi$ and
$\bar{\chi}[q]=\frac{1}{4}D^2 q-\bar{\psi}$, hence
the effective action $U_q$ is defined with
the following integral:
\begin{eqnarray}
\label{ev}
e^{iU_q}&=&\int Dq \delta(\frac{1}{4}D^2 q-\bar{\psi})
\delta(\frac{1}{4}\bar{D}^2 q-\psi)\exp (iS_v) det M_0
\end{eqnarray}
Here $M_0$ is a Faddeev-Popov matrix
\begin{equation}
M_0=\left(
\begin{array}{cc}
0 & -1_2\frac{1}{4}\bar{D}^2\\
-1_2\frac{1}{4}D^2 & 0
\end{array}
\right)
\end{equation}
where
$1_2$ is an unit $2\times 2$ matrix.
By definition,  $U_q$ is  constant: since the theory of $q$ is
gauge invariant the corresponding effective action  does not depend
on
$\psi$
(cf. \cite{Buch1,Buch2}).

By  multiplying  the corresponding left-hand sides and right-hand
sides
of eqs. (\ref{gam}) and (\ref{ev}) we obtain
\begin{eqnarray}
\exp(i \Gamma^{(1)}[\Phi,\bar{\Phi}]+iU_q) &=&
\int {\cal D} \psi {\cal D} \bar{\psi}
Du Dv \delta(\frac{1}{4}D^2 q-\bar{\psi})
\delta(\frac{1}{4}\bar{D}^2 q-\psi)\exp (iS_q)
\times\nonumber\\&\times&
det M_0
\exp (iS^{(2)}[\psi,\bar{\psi}])
\end{eqnarray}
Then we integrate over $\psi, \bar{\psi}$
by means of the delta functions (cf. \cite{Buch1,Buch2,BK0}) and
since $e^{iU_v}$ and $det M_0$ are constants we
arrive at the following expression for the one-loop effective action
for the superfields $\psi$, $\bar{\psi}$:
\begin{equation}
\label{gm1}
\exp(i\Gamma^{(1)})=\int Dq \exp(i S[q])
\end{equation}
which leads to
\begin{equation}
\label{tr}
\Gamma^{(1)}=\frac{i}{2}Tr\log\Delta
\end{equation}
Here $Tr$ is a functional supertrace, and
\begin{eqnarray}
S[q]&=&\frac{1}{2}q\Delta q\\
\Delta&=&\Box 1_2
-\frac{1}{16}(K_{\psi\bar{\psi}}-1)\{D^2,\bar{D}^2\}
-\frac{1}{4}W^{''}\bar{D}^2-\frac{1}{4}\bar{W}^{''}D^2
\end{eqnarray}
The terms proportional to the  supercovariant derivatives of
$K_{\psi\bar{\psi}}$,
$W^{''}$ and $\bar{W}^{''}$ are omitted since the one-loop k\"
ahlerian
effective potential by definition does not depend on the derivatives
of
superfields.

In order to determine  $Tr\log\Delta$ we use the Schwinger
representation
$$
Tr \log\Delta =tr \int \frac{ds}{s} \exp (is\tilde{\Delta})
\exp(is\Box)
$$
Here $tr$ denotes a trace of the matrix.
Since $\tilde{\Delta}$ is a matrix operator we turn to the problem
of
calculating the  exponent of the matrix. Let  $\Omega=\exp
(is\tilde{\Delta})$.
Then the matrix equation for  $\Omega$ reads
\begin{equation}
\label{omega}
-i\frac{\partial}{\partial s}{\Omega}=\Omega\tilde{\Delta}
\end{equation}
Here $\tilde{\Delta}$ is a matrix operator of the form
\begin{equation}
\tilde{\Delta}=
-\frac{1}{16}(K_{\psi\bar{\psi}}-1)\{D^2,\bar{D}^2\}
-\frac{1}{4}W^{''}\bar{D}^2-\frac{1}{4}\bar{W}^{''}D^2
\end{equation}

In order to solve the equation (\ref{omega}) we represent $\Omega$
in the
form of an expansion in terms of a  spinor supercovariant
derivatives
\begin{eqnarray}
\label{omega1}
 \Omega(s) &=& 1_2 + \frac{1}{16}A(s)D^2\bar {D}^2 +
 \frac{1}{16}\tilde{A}(s)\bar{D}^2D^2+
\frac{1}{8}B^{\alpha}(s)D_{\alpha}\bar{D}^2+
\frac{1}{8}\tilde{B}_{\dot{\alpha}}(s)D^{\dot{\alpha}}D^2+\nonumber\\
&+&\frac{1}{4}C(s)D^2 + \frac{1}{4}\tilde{C}(s)\bar{D}^2
\end{eqnarray}
The equation (\ref{omega}) leads to the following system of
equations
for the coefficients $A,B,C$,
\begin{eqnarray}
\frac{1}{i}\dot{A}&=&F+AF\Box-CW^{''}\\
\frac{1}{i}\dot{B}^{\alpha}&=&\tilde{B}_{\dot{\alpha}}W^{''}
\partial^{\alpha\dot{\alpha}}+B^{\alpha}F\Box\nonumber\\
\frac{1}{i}\dot{C}&=&-\bar{W}^{''}-A\bar{W}^{''}\Box+CF\Box\nonumber
\end{eqnarray}
and an analogous system  of equations for
$\tilde{A},\tilde{B},\tilde{C}$,
which can be obtained from this one by changing $W^{''}$ into
$\bar{W}^{''}$
and vice versa.
Here $F=1-K_{\psi\bar{\psi}}$.
Since the initial
condition for $\Omega$ is
$\Omega|_{s=0}=1$ the
initial conditions for $A, \tilde{A}, B, \tilde{B},C, \tilde{C}$)
are
$A|_{s=0}=\tilde{A}|_{s=0}=C|_{s=0}=\tilde{C}|_{s=0}=0$.
The solution for $B^{\alpha},\tilde{B}^{\dot{\alpha}}$ evidently has
the form
$B^{\alpha}=\tilde{B}^{\dot{\alpha}}=0$.
The manifest form of the matrices $A,\tilde{A}$
$C,\tilde{C}$,
necessary for exact calculations, is of the form
\begin{equation}
A=\left(
\begin{array}{cc}
A_{11} & A_{12} \\
A_{21} & A_{22}
\end{array}\right);
\
C=\left(
\begin{array}{cc}
C_{11} & C_{12} \\
C_{21} & C_{22}
\end{array}\right)
\end{equation}
Here index 1 denotes the sector of heavy superfield $\Phi$ and 2 the
sector of
light superfield $\phi$.
Now let us solve the system for matrices $A,C$. The solution for
$\tilde{A}$,
$\tilde{C}$ can be easily obtained in an analogous way since the
system
with $B^{\alpha}=\tilde{B}^{\dot{\alpha}}=0$ is invariant
under the change $A\to\tilde{A}$, $C\to\tilde{C}$.

Let us study the solution for $A,C$ which  should be
chosen in the form
\begin{eqnarray}
A&=&A_i+A_0\\
C&=&C_i+C_0\nonumber
\end{eqnarray}
Here $A_i$, $C_i$ is a partial solution of the inhomogeneous system,
and
$A_0$,
$C_0$ is a general solution of the homogeneous system. It is
straightforward to see that
$A_i=-1 \Box^{-1}$, $C_i=0$. And $A_0$, $C_0$ should satisfy the
system of
equations
\begin{eqnarray}
\frac{1}{i}\dot{A}_0&=&A_0F\Box+C_0W^{''}\\
\frac{1}{i}\dot{C}_0&=&A_0\bar{G}\Box+C_0F\Box\nonumber
\end{eqnarray}
 $A_0$, $C_0$ should be chosen to be of  the form
$A_0=a_0 \exp(i\omega s)$, $C_0=c_0\exp(i\omega s)$ where
$a_0,c_0,\omega$
are some functions of the background superfields and the
d'Alembertian
operator, but are  independent of $s$.
As a result we arrive at the equations for  $a_0$, $c_0$:
\begin{eqnarray}
\label{sys}
& &a_0(\omega 1-F\Box)+c_0 W^{''}=0\\
& &c_0(\omega 1-F\Box)+a_0\bar{W}^{''}\Box=0\nonumber
\end{eqnarray}
This system of equations has a non-trivial solution at
\begin{equation}
\label{seq}
\det\left(\begin{array}{cc}\omega 1_2-F\Box&W^{''}\\
\bar{W}^{''}\Box&\omega 1_2-F\Box
\end{array}\right)=0
\end{equation}
In principle, parameters $\omega$ can be found from this equation.
Their exact
form is determined by the structure of the matrix $W^{''}$ and $F$.
It
turns out that
for the specific cases  studied in detail the subsequent sections
(minimal
(Section 3)
and non-minimal (Section 4)  cases)
these parameters are different.
As a result the final solution can be cast in the form:
\begin{eqnarray}
A&=&\sum_{k}a_{0k}\exp(i\omega_k s)-\frac{1}{\Box}\\
C&=&\sum_{k}c_{0k}\exp(i\omega_k s)
\end{eqnarray}
where $\omega_k$ are all the roots of the equation (\ref{seq}) and
$a_{0k}$, $c_{0k}$ are
values of $a_0$, $c_0$ which can be found from (\ref{sys}) for
the corresponding
root $\omega_k$.  $a_{0k}$, $c_{0k}$ are fixed by means of initial
conditions.

Since only the coefficients $A$ and $\tilde{A}$ in (\ref{omega})
contribute to the
trace $Tr\log \Delta$  the one-loop k\"ahlerian contribution to the
effective action (\ref{tr}) can be written in the form
\begin{equation}
K^{(1)}=-\frac{i}{2}\int_{0}^{\infty}\frac{ds}{s}tr(A+\tilde{A})
U(x,x';s)|_{x=x'}
\end{equation}
where $U(x,x';s)=\exp(is\Box)\delta^4(x-x')$ (see
\cite{Buch1,Buch2}) and $tr$
denotes a trace of the matrix.  The operator $U(x,x';s)$ satisfies
the equation
\begin{equation}
\Box U=\frac{1}{i}\frac{\partial}{\partial s}U
\end{equation}
and $U(x,x';s)|_{x=x'}=\frac{1}{16\pi^2{(is)}^2}$.  $A$,  and
$\tilde{A}$
are
functions of $\Box$.
Hence in order to calculate the one-loop k\"{a}hlerian effective
potential
it is necessary to
find $A$ and $\tilde{A}$ and to expand them into a  power series in
$\Box$.

In this section we addressed in detail the techniques needed to
calculate
the one-loop corrected effective action, to eliminate the heavy
fields
and to obtain the effective action of light fields only.
In the  subsequent sections these techniques
will  be applied to obtain the explicit form of the one-loop
corrected
actions for   specific models.
In Section 5 we shall also include interactions with the $U(1)$
vector
superfields and modify the procedure accordingly.

\section{One-loop effective action for minimal models}
\setcounter{equation}{0}
\renewcommand{\theequation}{\arabic{section}.\arabic{equation}}
In this section we study decoupling effects for the model with
minimal k\"{a}hlerian potential (\ref{kmin})
$K=\Phi\bar{\Phi}+\phi\bar{\phi}$.
The first part  consists of calculating the one-loop correction
to the k\"{a}hlerian effective potential. In the second part we
solve
the effective equations of motion for the heavy superfields. As a
result
we arrive at the effective action of the
light superfields.
\subsection{Calculation  of the effective action}
\subsubsection{The one-loop k\"{a}hlerian effective potential}
Here we are going to calculate the one-loop contribution to
k\"{a}hlerian
effective potential by means of the effective equations of motion.
We study the minimal model with the chiral potential
in the form
\begin{eqnarray}
\label{spot}
W=\frac{1}{2}(M\Phi^2+\lambda\Phi\phi^2)+\frac{1}{3!}g\phi^3
\end{eqnarray}
with the corresponding  functions  in $W^{''}$  (see eq.(\ref{wpp}))
\begin{eqnarray}
W_{\phi\phi}&=&\lambda\Phi+g\phi\nonumber\\
W_{\phi\Phi}&=&\lambda\phi\nonumber\\
W_{\Phi\Phi}&=&M
\end{eqnarray}
The total classical action with the chiral potential
(\ref{spot}) is of the form
\begin{equation}
\label{stot}
S=\int d^8 z (\phi\bar{\phi}+\Phi\bar{\Phi})+[\int d^6 z
(\frac{1}{2}(M\Phi^2+\lambda\Phi\phi^2)+\frac{1}{3!}g\phi^3)+ h.c.]
\end{equation}
Note that the chiral potential used  is of the ``minimal'' form: it
involves the renormalizable terms only and the renormalizable
coupling
between the light and heavy superfields is {\it linear} in the heavy
fields, which yields a dominant contribution  in the study of the
decoupling effects. These types of  couplings are typical  for  a
class of
effective string models after the vacuum restabilization was taken
into
account, and thus this minimal model provides a prototype example
for the
study of decoupling effects  in $N=1$ supersymmetric theories.
(The results for this model and the physics consequences were
presented in
\cite{new}. For the sake of completeness we present here the
intermediate
steps in the derivation.)

In order to find  the one-loop k\"{a}hlerian effective potential we
should
determine the operator $\Omega(s)$ that satisfies the equation
(\ref{omega}). For the case of this minimal model this equation
leads to the following
system of equations for matrices $A,C$:
\begin{eqnarray}
\label{syst}
\frac{1}{i}\dot{A}&=&-CW^{''}\\
\frac{1}{i}\dot{C}&=&-\bar{W}^{''}-A\bar{W}^{''}\Box\nonumber
\end{eqnarray}
with the analogous equations for $\tilde{A},\tilde{C}$. Initial
conditions are
$A|_{s=0}=\tilde{A}|_{s=0}=C|_{s=0}=\tilde{C}|_{s=0}=0$.
Calculations described in Appendix A show that the one-loop
contribution to
k\"{a}hlerian effective potential is of the form:
\begin{eqnarray}
\label{kahl1}
& &K^{(1)}=
-\frac{1}{32\pi^2}
\Big(
(
|\lambda\Phi+g\phi|^2 +2 \lambda^2\phi \bar{\phi}+M^2+
\\&+&
\sqrt
{(|\lambda\Phi+g\phi|^2-M^2)^2+4
|\lambda^2\Phi\bar{\phi}+\lambda M\phi +\lambda g |\phi|^2)|^2
})
\times\nonumber\\&\times&
\log\frac{
(
|\lambda\Phi+g\phi|^2 +2 \lambda^2\phi \bar{\phi}+M^2+
\sqrt
{(|\lambda\Phi+g\phi|^2-M^2)^2+4
|\lambda^2\Phi\bar{\phi}+\lambda M\phi +\lambda g |\phi|^2|^2
})
}{\mu^2}
+\nonumber\\&+&
(
|\lambda\Phi+g\phi|^2 +2 \lambda^2\phi \bar{\phi}+M^2-
\sqrt
{(|\lambda\Phi+g\phi|^2-M^2)^2+4
|\lambda^2\Phi\bar{\phi}+\lambda M\phi +\lambda g |\phi|^2|^2
})
\times\nonumber\\&\times&
\log\frac{
(
|\lambda\Phi+g\phi|^2 +2 \lambda^2\phi \bar{\phi}+M^2-
\sqrt
{(|\lambda\Phi+g\phi|^2-M^2)^2+4
|\lambda^2\Phi\bar{\phi}+\lambda M\phi +\lambda g |\phi|^2|^2
})
}{\mu^2}\Big)\nonumber
\end{eqnarray}
and the effective action in the one-loop approximation is of the
form
\begin{eqnarray}
\label{ea}
& &\Gamma^{(1)}=\int d^8 z \big(\phi\bar{\phi}+\Phi\bar{\Phi}\big) +
\big(\int d^6 z(\frac{1}{2}M\Phi^2+\Phi\phi^2)+ h.c.\big)
-\frac{1}{32\pi^2}
\Big(
(
|\lambda\Phi+g\phi|^2 +\nonumber\\&+&
2 \lambda^2\phi \bar{\phi}+M^2+
\sqrt
{(|\lambda\Phi+g\phi|^2-M^2)^2+4
|\lambda^2\Phi\bar{\phi}+\lambda M\phi +\lambda g |\phi|^2)|^2
})
\times\nonumber\\&\times&
\log\frac{
(
|\lambda\Phi+g\phi|^2 +2 \lambda^2\phi \bar{\phi}+M^2+
\sqrt
{(|\lambda\Phi+g\phi|^2-M^2)^2+4
|\lambda^2\Phi\bar{\phi}+\lambda M\phi +\lambda g |\phi|^2|^2
})
}{\mu^2}
+\nonumber\\&+&
(
|\lambda\Phi+g\phi|^2 +2 \lambda^2\phi \bar{\phi}+M^2-
\sqrt
{(|\lambda\Phi+g\phi|^2-M^2)^2+4
|\lambda^2\Phi\bar{\phi}+\lambda M\phi +\lambda g |\phi|^2|^2
})
\times\nonumber\\&\times&
\log\frac{
(
|\lambda\Phi+g\phi|^2 +2 \lambda^2\phi \bar{\phi}+M^2-
\sqrt
{(|\lambda\Phi+g\phi|^2-M^2)^2+4
|\lambda^2\Phi\bar{\phi}+\lambda M\phi +\lambda g |\phi|^2|^2
})
}{\mu^2}\Big)\nonumber
\end{eqnarray}
Here parameter $\mu$ is is the renormalization scale parameter (see
Appendix A). This is the focal result of this section and is
further used for the
study of decoupling effects.
Note also that the same result can be obtained via a diagrammatic
approach, which is
described in Appendix B.

\subsubsection{Corrections to the chiral potential}
Let us now address loop corrections to the
chiral effective potential.
By definition  the chiral effective potential depends only on chiral
background
superfields, i.e. in order to consider corrections to the chiral
effective potential one
must set
$\bar{\Phi}=\bar{\phi}=0$.
Possible vertices contributing to one-loop effective potential
should be
quadratic in quantum superfields \cite{BO}. They have the form
\begin{eqnarray}
\label{vert}
& &K_{\bar{\phi}\bar{\phi}}\bar{\phi}^2,\ K_{\phi\phi}\phi^2,
\ (K_{\phi\bar{\phi}}-1)\phi\bar{\phi},\
\frac{1}{2}W_{\phi\phi}\phi^2,\nonumber\\
& &K_{\bar{\Phi}\bar{\phi}}\bar{\phi}\bar{\Phi},
\ K_{\Phi\phi}\Phi\phi,
\ (K_{\Phi\bar{\phi}})\Phi\bar{\phi},\ \frac{1}{2}W_{\Phi\phi}
\Phi\phi,\nonumber\\
& &K_{\bar{\Phi}\bar{\Phi}}\bar{\Phi}^2,\ K_{\Phi\Phi}\Phi^2,
\ (K_{\Phi\bar{\Phi}}-1)\Phi\bar{\Phi},\
\frac{1}{2}W_{\Phi\Phi}\Phi^2\nonumber
\end{eqnarray}
Here all derivatives of $K$ and $W$ are taken at
$\bar{\Phi}=\bar{\phi}=0$.
The standard method
to address these corrections
(\cite{West2}--\cite{Buch4}) is the following.
Corrections contributing to the chiral potential should be of the
form:
\begin{equation}
\label{chr}
\int d^8 z f(\Phi)\big(-\frac{D^2}{4\Box}\big)g(\Phi)
\end{equation}
Namely, after a transformation to an integral over the  chiral
superspace by the rule\\
$\int d^8 z F(\Phi,\bar{\Phi})=\int d^6 z
\big(-\frac{\bar{D}^2}{4}\big) F(\Phi,\bar{\Phi})$ these corrections
take
the form
\begin{equation}
\label{chr1}
\int d^6 z f(\Phi)g(\Phi)
\end{equation}
Here $f(\Phi),g(\Phi)$ are  functions of the chiral superfield
$\Phi$, only.

However, factor $\Box^{-1}$ in (\ref{chr})
can arise only from propagators of massless superfields as
propagators of massive superfields are proportional to
${(\Box-m^2)}^{-1}$.  Namely, in the case of the massive  chiral
superfields,
instead of
(\ref{chr}) we must consider the expression
\begin{equation}
\int d^8 z f(\Phi)\big(-\frac{D^2}{4(\Box-m^2)}\big)g(\Phi)
\end{equation}
A transformation to the form of an integral over the chiral
superspace leads
to
\begin{equation}
\int d^6 z f(\Phi)\big(\frac{\Box}{\Box-m^2}\big)g(\Phi)
\end{equation}
which after Fourier transformation is of the form
\begin{equation}
\int\frac{d^4
p}{{(2\pi)}^4}f(\Phi)\big(\frac{p^2}{p^2+m^2}\big)g(\Phi)
\end{equation}
This expression evidently vanishes for superfields slowly varying in
space-time,
($p^2\to 0$) and $m\neq 0$. Hence supergraphs contributing to the
chiral
effective
potential cannot contain massive propagators.

Let us now carry out a dimensional analysis of possible one-loop
diagrams
contributing to the chiral effective potential.
Each loop corresponds to a contribution with a scaling dimension 2
since
each loop includes an integration over
momenta which contribute a scaling  dimension  4, however  each
contraction
of the loop to the point by
the rule $\delta_{12} D^2_1\bar{D}^2_1\delta_{12}=16\delta_{12}$
decreases a number of $D,\bar{D}$-factors by 4 and the corresponding
scaling dimension by  2.
Each propagator of a massless superfield gives no contribution
(scaling
dimension 0) since it
has the form (cf. \cite{BK0})
\begin{equation}
\label{prop}
G(z_1,z_2)=-\frac{D^2_1\bar{D}^2_2}{16\Box}\delta^8(z_1-z_2)
\end{equation}
(It is more convenient for dimensional analysis to associate
$D$-factors with   propagators instead of  vertices.)
Each vertex proportional to an external chiral superfield
corresponds
to one
factor $\bar{D}^2$ instead of two since one of them is used to
transform
a vertex to a form of an integral over
$d^4\theta$. As a result such a vertex decreases a scaling dimension
by 1.
One of the factors $D^2$ does not contribute to the scaling
dimension
since it
is used to
transform the expression (\ref{chr}) to the form (\ref{chr1}).
Then, during such a transformation one more square of external
momenta
arises.
Therefore the scaling dimension due to  contributions from a
one-loop supergraph has the
form $2L+1-n_{W^{''}}$,
where $L$ is the number of loops (in our case $L=1$), and
$n_{W^{''}}$
is the number of external chiral lines
$W^{''}$. By definition the effective potential is an effective
Lagrangian
in the low energy (infrared) limit
$p^2\to 0$. Hence a non-trivial one-loop chiral effective Lagrangian
can
arise
only for
\begin{equation}
\label{ind}
3-n_{W^{''}}=0
\end{equation}
Otherwise the  contribution would either vanish or be singular
in the infrared limit. We also note that the vertices proportional
to
derivatives
of
$K(\Phi,\bar{\Phi})$ do not contribute to the scaling dimension.

Furthermore, chiral potential contributions should contain one
factor
$D^2$ more than $\bar{D}^2$ after a transformation of all vertices
to the
form of
integrals over $d^4\theta$. The reason for this  is the following.
After each pair
$\bar{D}^2 D^2$ is transformed to a d'Alembertian operator, one
factor
$D^2$
should be employed to transform the integral over $d^4\theta$
into an  integral over
$d^2\theta$ (see (\ref{chr})-(\ref{chr1})).

It is easy to see that one of two factors $\bar{D}^2$ associated
with
the vertex
proportional to $K_{\Phi\Phi}$ can be transported only to an
external
line
$K_{\Phi\Phi}$ during the ``$D$-algebra'' transformations. However,
all
the derivatives of
$K$ and $W$ are considered at $\bar{\Phi}=0$,
and therefore $\bar{D}^2K_{\Phi\Phi}=0$. As a result we can put
$K_{\Phi\Phi}=0$.

Now let us consider the possible one-loop diagrams contributing to
the
chiral  effective potential. At the one-loop level
$n_{W^{''}}=3$ (see above) and each  external line $W^{''}$
corresponds
to $\bar{D}^2$.
Hence the supergraph should contain four $D^2$-factors, i.e. two
vertices
proportional to $K_{\bar{\Phi}\bar{\Phi}}$. However, a
straightforward
construction shows that such a diagram must contain the line
proportional
to $\frac{\bar{D}^2_1\bar{D}^2_2}{16}\delta_{12}=0$.
Hence a contribution of such a diagram is equal to zero, and a
one-loop
contribution to the chiral effective potential is absent:
$W^{(1)}(\Phi)=0$.
We note that this situation is analogous to the general model of one
chiral
superfield  studied in \cite{Buch5}.

However, higher order (loop) corrections to the chiral effective
potential
can
arise not only for
diagrams with external massless lines but also for those  with heavy
external lines, in spite of the fact that it was commonly
believed that  quantum corrections to
the chiral effective potential for massive superfields are absent.
For example, consider the supergraph

\hspace{4.5cm}
\Lengthunit=1.5cm
\GRAPH(hsize=3){
\mov(.5,0){\Circle(2)\mov(-1,0){\lin(2,0)}
\ind(-2,10){|}
\ind(-2,-3){\bar{D}^2}\ind(-2,0){|}\ind(-2,-10){|}\ind(-2,-13){\bar{D}^2}
\ind(-2,7){\bar{D}^2}
\ind(-9,2){D^2} \ind(-10,-2){D^2} \ind(8,2){D^2} \ind(8,-2){D^2}
\ind(-18,2){-} \ind(-18,-2){-} \ind(0,2){-} \ind(0,-2){-}
\Linewidth{0.3pt}
\mov(-1,1){\lin(-.7,.7)}\mov(-1.1,1){\lin(-.7,.7)}
\mov(-1,-1){\lin(.7,-.7)}\mov(-1.1,-1){\lin(.7,-.7)}
\mov(-1,0){\lin(-.7,.7)}\mov(-1.1,0){\lin(-.7,.7)}}
}

Here a double  line denotes the external superfield $\Phi$, and
a single line corresponds to the
propagator $<\phi\bar{\phi}>$ of the massless superfield $\phi$.
A contribution of such a supergraph is of the form
\begin{eqnarray}
\label{I1}
I&=&\int \frac{d^4p_1 d^4p_2}{{(2\pi)}^8}\frac{d^4k
d^4l}{{(2\pi)}^8}
\int d^4\theta_1 d^4\theta_2 d^4\theta_3 d^4\theta_4 d^4\theta_5
{\big(\frac{g}{3!}\big)}^2\lambda^3
\Phi(-p_1,\theta_3)\Phi(-p_2,\theta_4)\Phi(p_1+p_2,\theta_5)
\times\nonumber\\&\times&
\frac{1}{k^2 l^2 {(k+p_1)}^2{(l+p_2)}^2{(l+k)}^2{(l+k+p_1+p_2)}^2}
\times\nonumber\\&\times&
\delta_{13}\frac{\bar{D}^2_3}{4}\delta_{32}
\frac{D^2_1 \bar{D}^2_4}{16}\delta_{14}\delta_{42}
\frac{D^2_1 \bar{D}^2_5}{16}\delta_{15}\delta_{52}
\end{eqnarray}
After a tedious calculation analogous to that  carried out in
\cite{Buch5} we
arrive at the expression:
\begin{eqnarray}
\label{app}
I&=&{\big(\frac{g}{3!}\big)}^2\lambda^3
\int \frac{d^4p_1 d^4p_2}{{(2\pi)}^8}\frac{d^4k d^4l}{{(2\pi)}^8}
\int d^2\theta
\Phi(-p_1,\theta)
\Phi(-p_2,\theta)
\Phi(p_1+p_2,\theta)
\times\nonumber\\&\times&
\frac{k^2 p_1^2+ l^2 p_2^2 +2 (k l)(p_1 p_2)}
{k^2 l^2 {(k+p_1)}^2{(l+p_2)}^2{(l+k)}^2{(l+k+p_1+p_2)}^2}
\end{eqnarray}
which can be cast in the form
\begin{eqnarray}
\label{cont}
I&=&{\big(\frac{g}{3!}\big)}^2\lambda^3\int d^2\theta \int
\frac{d^4p_1 d^4p_2}{{(2\pi)}^8}
{\bar{W}}^{'''2}
\Phi(-p_1,\theta)\Phi(-p_2,\theta)\Phi(p_1+p_2,\theta)
S(p_1, p_2)
\end{eqnarray}
Here $p_1, p_2$ are external momenta. The expression for $S(p_1,
p_2)$
is of the form
$$ \int\frac{d^4 k d^4 l}{{(2\pi)}^8}\frac{k^2 p_1^2+ l^2
p_2^2 +2 (kl)(p_1 p_2)} {k^2 l^2
{(k+p_1)}^2{(l+p_2)}^2{(l+k)}^2{(l+k+p_1+p_2)}^2} $$
After Fourier transformation, eq. (\ref{cont}) has the form
\begin{eqnarray}
\label{cont0}
I&=&{\big(\frac{g}{3!}\big)}^2\lambda^3\int d^2\theta \int d^4 x_1
d^4x_2 d^4 x_3
\int\frac{d^4p_1 d^4p_2}{{(2\pi)}^8}
\Phi(x_1,\theta)\Phi(x_2,\theta)\Phi(x_3,\theta)
\times\nonumber\\&\times&
\exp[i(-p_1 x_1- p_2 x_2+(p_1+p_2)x_3)]
S(p_1, p_2)
\end{eqnarray}
The effective potential is obtained
from the above expression by taking superfields that
slowly vary in space-time, i.e. we take
$
\Phi(x_1,\theta)
\Phi(x_2,\theta)
\Phi(x_3,\theta)
\simeq \Phi^3(x_1,\theta)
$.
As a result one obtains
\begin{eqnarray}
I&=&{\big(\frac{g}{3!}\big)}^2\lambda^3
\int d^2\theta \int d^4 x_1 d^4 x_2 d^4 x_3
\int\frac{d^4p_1 d^4p_2}{{(2\pi)}^8}
\Phi^3(x_1,\theta)
\times\nonumber\\&\times&
\exp[i(-p_1 x_1- p_2 x_2+(p_1+p_2)x_3)]
S(p_1, p_2)
\end{eqnarray}
The integration over $d^4 x_2 d^4 x_3$ leads to delta-functions
$\delta(p_2)\delta(p_1+p_2)$ and  hence the eq. (\ref{cont0}) takes
the
form
\begin{equation}
I={\big(\frac{g}{3!}\big)}^2\lambda^3
\int d^2\theta \int d^4 x_1 \Phi^3(x_1,\theta)
S(p_1,p_2)|_{p_1,p_2=0}
\end{equation}
Therefore the result for  the two-loop contribution to  the chiral
effective
potential due to this  diagram is of the form
\begin{equation}
\label{l2c}
\frac{6}{{(16\pi^2)}^2}\zeta(3) {\big(\frac{g}{3!}\big)}^2\lambda^3
\Phi^3
\end{equation}
As the last step we took into account that
$$
\int \frac{d^4k d^4l}{{(2\pi)}^8}
\frac{k^2 p_1^2+ l^2 p_2^2 +2 (k_1 k_2)(p_1 p_2)}
{k^2 l^2
{(k+p_1)}^2{(l+p_2)}^2{(l+k)}^2{(l+k+p_1+p_2)}^2}|_{p_1=p_2=0}
=\frac{6}{{(4\pi)}^4}\zeta(3)
$$

We have therefore  demonstrated that  chiral  corrections to the
effective action that depend on massive superfields can also arise.
However, such effects can appear at the two-loop level at most, and
after
solving the effective equations of motion for heavy fields such
corrections are  proportional to
$1 \over M^3$. Therefore, they are highly suppressed compared to
the decoupling effects due to the one-loop corrected k\"ahlerian
effective potential. In the following we shall concentrate on the
latter effects only.

By analogous reasoning,  corrections to the chiral
effective
potential  that involve a coupling of the heavy
superfield to the
light ones, are of the form
$\frac{6}{{(16\pi^2)}^2}\zeta(3)
{\big(\frac{g}{3!}\big)}^2\lambda^3\Phi\phi^2$
which is of the same  form as the corresponding  classical
chiral
effective potential term. They are described by supergraphs
analogous to those
given above, but with one heavy external line and two massless ones.
At the two-loop level $K_{eff}=K+K^{(1)}+K^{(2)}$, and
$W_{eff}=W+W^{(2)}$ where $K^{(2)}$ and $W^{(2)}$ are two-loop
corrections to the k\"{a}hlerian potential and chiral potential,
 respectively. $W^{(2)}$ depends non-trivially on $\Phi$.
Hence the presence of such a correction to $W$ should modify the
effective
equations of motion (\ref{ee},\ref{ee1}). After substitution of the
solution of effective equations of motion into the effective action,
$W^{(2)}$ turns out to be
suppressed by $1\over M$, i.e.  it can have the form
$\sim\frac{\phi^4}{M}$.
 Hence the effective dynamics of the two-loop corrected theory
 allow us to conclude that a two-loop contribution to the
 effective superpotential can yield corrections of order $1\over
 M$, i.e.
it can be competitive with the leading tree-level (classical)
decoupling
effects.



\subsection{The effective action for light superfields}
Now we solve the effective equations of motion
for this minimal model that allows us to
eliminate heavy superfields from effective action up to some
order
in the inverse powers of the  mass parameters, and as  a consequence
obtain the one-loop effective action of light superfields only.
The one-loop corrected effective action is given by (\ref{ea}).
The effective
equations of motion for the heavy superfield with  the
chiral potential
(\ref{spot}) have the form
\begin{eqnarray}
\label{eqmh}
\frac{\delta\Gamma^{(1)}}{\delta\Phi}=0:
& &-\frac{1}{4}\bar{D}^2(\bar{\Phi}+
\frac{\partial K^{(1)}}{\partial\Phi})+
M\Phi+\frac{\lambda}{2}\phi^2=0\nonumber\\
\frac{\delta\Gamma^{(1)}}{\delta\bar{\Phi}}=0:
& &-\frac{1}{4}D^2(\Phi+\frac{\partial K^{(1)}}{\partial
\bar{\Phi}})+M\bar{\Phi}+\frac{\lambda}{2}\bar{\phi}^2=0
\end{eqnarray}
As described in the previous Section these equations
can be solved for the heavy superfield via an iterative method and a
subsequent substitution
of this solution into the one-loop corrected  effective action.
(Analogous equations for $\phi$ are not
essential since our aim is to minimize  the effective action with
respect to heavy superfields.)

Now let us solve the equations (\ref{eqmh}) via the iterative
method.
Using  $|D^2\Phi|\ll M\Phi$,
since we consider the case with  large $M\gg p$,
one finds that the lowest
(zeroth) order  solution of these equations is
\begin{equation}
\label{zero}
\Phi_0=-\lambda\frac{\phi^2}{2M}
\end{equation}
and with the corresponding value for $\bar{\Phi}_0$.

The first order correction  
(cf.
(\ref{ee})-(\ref{ee1}))
is obtained from  the
equation
\begin{eqnarray}
\Phi_1=
\frac{1}{4M}\bar{D}^2(\bar{\Phi}_0+
\frac{\partial K^{(1)}}{\partial \Phi}|_{\Phi=\Phi_0})
\end{eqnarray}
where $\Phi_0$ is given by (\ref{zero}).
As for higher order corrections,  they can be found by
continuing the iterative procedure.

It turns out to be that for the case with  $g\neq 0$,  $\Phi_1$
is  of order  $\frac{1}{M}$  (as  $\Phi_0$), and
it  is of the
form
\begin{equation}
\label{app1}
\Phi_1=-\frac{\bar{D}^2}{64\pi^2 M}\Big(\lambda g\bar{\phi}
(1+\log\frac{2g^2|\phi|^2}{\mu^2})\Big)+O(\frac{1}{M^2})
\end{equation}
On the other hand for the case with  $g=0$ (i.e. the
self-interactions of the light superfield in the classical chiral
potential are absent),
$\Phi_1$ is of the order $\frac{1}{M^2}$ and takes
the form
\begin{eqnarray}
\label{app2}
\Phi_1&=&\frac{\lambda \bar{D}^2(\bar{\phi}^2)}{8 M^2}+
\frac{1}{32\pi^2}\frac{\lambda^3}{2 M^2}\bar{D}^2\Big\{|\phi|^2
\big(1+\log\frac{2 M^2}{\mu^2})+\nonumber\\&+&
(\frac{\bar{\phi}^2}{2}+|\phi|^2\big)
\big(1+\log[5\lambda^4\frac{|\phi|^4}{M^4}-\lambda^4\frac
{|\phi|^2(\phi^2+\bar{\phi}^2)}{2M^4}]\big)\Big\}+
O(\frac{1}{M^3})
\nonumber
\end{eqnarray}
As for the  second order correction  $\Phi_2$,
it is proportional
to $\frac{1}{M^2}$ for $g\neq 0$ and to $\frac{1}{M^3}$ for $g=0$.
Note, the cases with $g=0$ and $g\neq 0$ lead to essentially
different expressions
for the heavy superfield $\Phi$ in terms of the light one $\phi$.

For the  calculation of the effective action for the  light
superfield it
is convenient to
expand $K^{(1)}$ (\ref{kahl1}) into a power series in  $\frac{1}{M}$
(up to
the order $\frac{1}{M^3}$):
\begin{eqnarray}
\label{kahlt}
K^{(1)}&=&-\frac{1}{32\pi^2}\Big\{
\big(4\lambda^2|\phi|^2+2\frac{\lambda^2}{M}(\lambda\Phi+g\phi+h.c.)
|\phi|^2
+\frac{|\phi|^4}{M^2}(g^4+2\lambda^2 g^2)\big)
(1+\log\frac{2 M^2}{\mu^2})
+\nonumber\\&+&
2\lambda^4\frac{|\phi|^4}{M^2}(1-\log\frac{2 M^2}{\mu^2})+
\nonumber\\&+&
\big[2|\lambda\Phi+g\phi|^2-2\frac{\lambda^2}{M}(\lambda\Phi+g\phi)|\phi|^2
+\frac{|\phi|^4}{M^2}(2\lambda^4-g^4-2\lambda^2 g^2)
\big]\times\nonumber\\&\times&
\Big(\log\big\{\frac{2|\lambda\Phi+g\phi|^2}{\mu^2}-
\frac{2\lambda^2}{M\mu^2}(\lambda\Phi+g\phi+h.c.)|\phi|^2
+\nonumber\\&+&\frac{|\phi|^4}{M^2\mu^2}(2\lambda^4-g^4-2\lambda^2
g^2)\big\}
\Big)
\Big\}+O(\frac{1}{M^3})
\end{eqnarray}
Here we took into account that $\Phi$ is suppressed by at least one
inverse power of $M$.
In the following we shall substitute the solutions of effective
equations of motion
into the expression (\ref{kahlt}).
\subsubsection{Contribution of  the self-interaction of the light
superfield}
First we consider the case $g\neq 0$.
In this case we can write $\Phi=\Phi_0+\Phi_1$ in the form (cf.
(\ref{app1}))
\begin{equation}
\label{app3}
\Phi=-\lambda\frac{\phi^2}{2M}-
\frac{\bar{D}^2}{64\pi^2 M}
\Big(\lambda g\bar{\phi}
(1+\log\frac{2g^2|\phi|^2}{\mu^2})\Big)+O(\frac{1}{M^2})
\end{equation}
Its substitution into (\ref{kahlt}) leads to the expression
\begin{eqnarray}
\label{kahll}
K^{(1)}&=&-\frac{1}{32\pi^2}2 g^2|\phi|^2\Big\{
\log\frac{g^2|\phi|^2}{\mu^2}+\frac{\lambda}{g|\phi|^2}
\big(\bar{\phi}(-\frac{\lambda\phi^2}{2M}+\lambda
g\frac{\bar{D}^2}{64\pi^2 M}
[\bar{\phi}(1+\nonumber\\&+&
\log\frac{g^2|\phi|^2}{\mu^2})]+h.c.
)\big)-\frac{\lambda^2}{M}\frac{\phi+\bar{\phi}}{g}
\Big\}-\nonumber\\
&-&2\frac{\lambda^2}{M}g(\phi+\bar{\phi})|\phi|^2
\log\frac{g^2|\phi|^2}{\mu^2}
+O(\frac{1}{M^2})
\end{eqnarray}
Here a factor $e=\exp(1)$ is introduced to cancel superfluous factor
$-1$.

As for the chiral potential which has the form (\ref{spot}) it can
be written as
(cf. (\ref{app3})):
\begin{equation}
\label{sp}
V=\frac{g}{3!}\phi^3-\frac{1}{8}\frac{\lambda^2\phi^4}{M}+
\frac{\lambda^2 g^2}{2M}
{\Big\{\frac{\bar{D}^2}{64\pi^2}
\big(\bar{\phi}
(1+\log\frac{2g^2|\phi|^2}{\mu^2})\big)\Big\}}^2+O(\frac{1}{M^2})
\end{equation}
The effective action for light superfields can now be written
up to the order $\frac{1}{M^2}$ in the form
\begin{equation}
\label{e0}
S_{eff}[\phi,\bar{\phi}]=S_{eff}^{cl}[\phi,\bar{\phi}]+
S_{eff}^{q}[\phi,\bar{\phi}]
\end{equation}
where $S_{eff}^{cl}[\phi,\bar{\phi}]$ is the classical part:
 \begin{eqnarray}
\label{ecl}
S_{eff}^{cl}[\phi,\bar{\phi}]&=&
\int d^8z \phi\bar{\phi}+
\Big\{\int d^6 z\Big(
\frac{g}{3!}\phi^3-\frac{1}{8}\frac{\lambda^2\phi^4}{M}+
\frac{\lambda^2 g^2}{2M}
{\Big\{\frac{\bar{D}^2}{64\pi^2}
\big(\bar{\phi}
(1+\log\frac{2g^2|\phi|^2}{\mu^2})\big)\Big\}}^2
\Big)+\nonumber\\&+&h.c.
\Big\}
+
O(\frac{1}{M^2})
\end{eqnarray}
and $S_{eff}^{q}[\phi,\bar{\phi}]$ is the  quantum part:
\begin{eqnarray}
\label{eqt}
S_{eff}^{q}[\phi,\bar{\phi}]&=& -
\frac{1}{32\pi^2}\int d^8 z\Big[
2 g^2|\phi|^2\Big\{ \log\frac{g^2|\phi|^2}{\mu^2}+\nonumber\\&+&
\frac{\lambda}{g|\phi|^2}
\big(\bar{\phi}(-\frac{\lambda\phi^2}{2M}+\lambda
g\frac{\bar{D}^2}{64\pi^2 M}
[\bar{\phi}(1+
\log\frac{g^2|\phi|^2}{\mu^2})]+h.c.
)\big)-\frac{\lambda^2}{M}\frac{\phi+\bar{\phi}}{g}
\Big\}-\nonumber\\
&-&2\frac{\lambda^2}{M}
g(\phi+\bar{\phi})|\phi|^2\log\frac{g^2|\phi|^2}{\mu^2}\Big]
+O(\frac{1}{M^2})
\end{eqnarray}
The main feature of the expressions (\ref{e0})-(\ref{eqt}) is the
fact that
with increasing $M$ the leading contribution of  the decoupling
effects grow
logarithmically with $M$.
Let us   fix $\mu$ by imposing the condition
$\log\frac{2M^2}{\mu^2}=-1$. This
 renormalization condition eliminates
from the one-loop effective action the terms that are proportional
to
$\lambda$, i.e., the coupling  of the heavy and light
superfields in the classical action. As a consequence,  $S^q_{cl}$
in eq.
(\ref{eqt})  is of the
form:
$$-\frac{1}{32\pi^2}2 g^2|\phi|^2\log\frac{g^2|\phi|^2}{e M^2}
$$
This term {\it increases} as $M\to\infty$ and becomes leading in
comparison with the classical contribution (\ref{ecl}) which in the
limit
$M\to\infty$ is of the form:
 \begin{eqnarray}
S_{eff}^{cl}[\phi,\bar{\phi}]&=&
\int d^8z \phi\bar{\phi}+
\Big\{\int d^6 z
\frac{g}{3!}\phi^3
+h.c.\Big\}
+O(\frac{1}{M})\nonumber
\end{eqnarray}

Alternatively, we do not fix $\mu$  and address  the effective
action  (\ref{e0})-(\ref{eqt}) at energy scales specified by the
values of $\mu$.  In the leading (zeroth) order  in ${1\over M}$
expansion
the quantum effective action  (\ref{eqt})
in the form
\begin{eqnarray}
\label{two}
S^q_{eff}[\phi,\bar{\phi}]&=&-\frac{1}{32\pi^2}
\Big\{4\lambda^2|\phi|^2(1+\log\frac{2 M^2}{\mu^2})
+2|g\phi|^2\Big(\log\frac{2|g\phi|^2}{\mu^2}\Big)
\Big\}+O(\frac{1}{M})
\end{eqnarray}
As $\mu$ is  taken as a free parameter, describing the effective
theory
at energy scales $\mu$,  we see that the second term of this
expression (proportional to
$\log\frac{2M^2}{\mu^2}$) can become
competitive with  those of the
classical action as
 $\mu\ll M\to\infty$.
Therefore from both perspectives, i.e. when we fix $\mu$ and when
we treat $\mu$ as a free parameter,
the decoupling effects grow logarithmically with
$M$ and  as  $M\to\infty$ they become dominant.
Hence the appearance  of such terms
is not an effect caused by a choice of normalization parameter $\mu$
but
it is a genuine  property
of the model itself.

The total one-loop effective action can be finally written as
\begin{eqnarray}
\label{three}
S_{eff}[\phi,\bar{\phi}]&=&\int d^8 z (|\phi|^2-
\frac{\hbar}{32\pi^2}
\Big\{4\lambda^2|\phi|^2(1+\log\frac{2 M^2}{\mu^2})
+2|g\phi|^2\Big(\log\frac{2|g\phi|^2}{\mu^2}\Big)
\Big\})+\nonumber\\
&+&\big[\frac{1}{3!}\int d^6 z g\phi^3+h.c.]+O(\frac{1}{M})
\end{eqnarray}
Here we took into account that all the terms that depend on the
heavy
superfield $\Phi$,
after the substitution of the solution of the effective equations of
motion,  are
at least of the first order in the inverse mass $M$.

According to the decoupling theorem when we eliminate heavy
superfields
via effective equations of motion we can split any quantum
correction
to the effective action into a sum of two terms. The first term
is a
quantum  correction in the corresponding theory
describing dynamics of light superfields only, with heavy
superfields
put
to zero, i.e. the standard Coleman-Weinberg potential associated
with the
interaction of the light fields only. The second one corresponds to
a sum of
 terms that are proportional to
at least one power of ${1\over M}$.

To convince ourselves that  our results are in accordance with
the decoupling theorem, we
introduce the following field  and coupling redefinitions.
First, we see that the kinetic energy term (proportional to
$|\phi|^2$)
of the one-loop effective action is of the form
$$
|\phi|^2
[1-\frac{\hbar}{32\pi^2}\Big\{4\lambda^2(1+\log\frac{2 M^2}{\mu^2})
\Big\}]
$$
We can redefine the kinetic energy term by the rule
\begin{equation}
\label{re}
\tilde{\phi}=Z\phi
\end{equation}
where $\tilde{\phi}$ is a new chiral superfield, and $Z$ is a
finite renormalization
constant equal to
\begin{equation}
\label{z}
Z=
{[1-\frac{\hbar}{32\pi^2}
4\lambda^2(1+\log\frac{2 M^2}{\mu^2})]}^{1/2}
\end{equation}
Then, if we also redefine coupling $g$ by the rule
$\tilde{g}=Z^{-3}g$ we
see that the chiral potential $g\phi^3$ remains  invariant.
We note that  the coupling $\lambda$ corresponds to the interaction
between light and heavy superfields, and after a redefinition of the
light
superfield (\ref{re}) it will present itself only in terms that are
at
least  first
order  in the inverse mass $M$.
(Note also that analogous redefinitions of the superfield and
couplings could be carried out at higher orders of the effective
action.)
As a result
we arrive at the one-loop effective action of the form
\begin{eqnarray}
\label{eac}
S_{eff}[\phi,\bar{\phi}]&=&\int d^8 z \Big\{|\tilde{\phi}|^2-
\frac{\hbar}{32\pi^2}
2 Z^4\tilde{g}^2|\tilde{\phi}|^2\big(\log
\frac{2Z^4\tilde{g}^2|\tilde{\phi}|^2}{\mu^2}\big)\Big\}+\nonumber\\
&+&\big[\int d^6
z\frac{1}{3!}\tilde{g}\tilde{\phi}^3+h.c.]+O(\frac{1}{M})
\end{eqnarray}
However, we can expand $Z$ (\ref{z}) into a power series in $\hbar$
as
$$
Z=
1-\frac{\hbar}{64\pi^2}
4\lambda^2(1+\log\frac{2 M^2}{\mu^2})+O(\hbar^2)
$$
Substituting this expansion into (\ref{eac}) and taking into account
only terms of zeroth and first order in $\hbar$ we arrive at
\begin{eqnarray}
S_{eff}[\phi,\bar{\phi}]&=&\int d^8 z \Big\{|\tilde{\phi}|^2-
\frac{\hbar}{32\pi^2}
2 \tilde{g}^2|\tilde{\phi}|^2\big(\log
\frac{2\tilde{g}^2|\tilde{\phi}|^2}{\mu^2}\big)\Big\}+\nonumber\\&+&
\big[\int d^6
z\frac{1}{3!}\tilde{g}\tilde{\phi}^3+h.c.]+O(\frac{1}{M})
\end{eqnarray}
 This effective action coincides at the zeroth order with  the form
of the one-loop effective action for the ``pure light'' theory
where quantum
corrections
originate from the couplings of the light fields only.
Therefore the result for the effective action of light superfields
given by
expressions (\ref{e0}--\ref{eqt}) is in formal agreement with the
decoupling theorem.

Note, however,  that the parameters of the theory, i.e. fields,
masses and
 couplings,
are determined from the string theory;  they are fixed-calculable
at $M_{string}$ and therefore
they cannot be adjusted  via redefinitions performed above. Thus,
the
corrections due to decoupling effects correct in a quantitative
manner the
predictions for the low-energy values of the couplings. For a class
of
perturbative string vacua with an anomalous $U(1)$
after the vacuum  restabilization, the effective theory  has
couplings of the
heavy and light fields of the type discussed in this section.
Typical values of
the couplings  $g$ and $\lambda$  (at $M_{string}\sim 10^{17}$ GeV)
are of the order of
the gauge coupling $\sim 0.8$ and the mass parameter of the heavy
fields $M\sim
M_{string}$.  The decoupling effects then  modify the low-energy
 ($\mu \sim 1$ TeV) predictions for the corresponding tri-linear
couplings
  of the light fields  by an amount of order
 ${{g^2}\over {16\pi^2}}\log ({{M^2}\over \mu^2})\sim 0.25$.
corrections, and
Thus for such classes of  string vacua  decoupling effects could
correct the
 predictions for  the couplings
by  10\% -- 50\%.

\subsubsection{Absence of the self-interaction of the light
superfield}
Now let us put $g=0$ in (\ref{spot}), i.e.  a self-interaction of
light
superfields is absent.  In this case $\Phi_1$ is of the order
$\frac{1}{M^2}$ and $\Phi=\Phi_0+\Phi_1$ takes the form
\begin{eqnarray}
\Phi&=&-\lambda\frac{\phi^2}{2M}+\frac{\lambda
\bar{D}^2(\bar{\phi}^2)}{8 M^2} +\nonumber\\&+&
\frac{1}{32\pi^2}\frac{\lambda^3}{2 M^2}\bar{D}^2\Big\{
(\frac{\bar{\phi}^2}{2}+|\phi|^2\big)
\big(2+\log[5\lambda^4\frac{|\phi|^4}{M^2\mu^2}-\lambda^4\frac
{|\phi|^2(\phi^2+\bar{\phi}^2)}{M^2\mu^2}]\big)\Big\}+
O(\frac{1}{M^3})
\nonumber
\end{eqnarray}

The one-loop k\"ahlerian effective potential is of the form
\begin{eqnarray}
K^{(1)}&=&
-\frac{1}{32\pi^2}\int d^8z \Big\{
4\lambda^2|\phi|^2(1+\log\frac{2M^2}{\mu^2})+
4\lambda^4\frac{|\phi|^4}{M^2}+
\big[
\frac{5}{2}\lambda^4\frac{|\phi|^4}{M^2}+2\lambda^4\frac{|\phi|^2}{M^2}
(\phi^2+\bar{\phi}^2)
\big]
\times\nonumber\\&\times&
\Big(
\log\big[
\frac{5}{2}\lambda^4\frac{|\phi|^4}{M^2\mu^2}-
\lambda^4\frac{|\phi|^2}{M^2\mu^2}
(\phi^2+\bar{\phi}^2)
\big]
\Big)
\Big\}+O(\frac{1}{M^3})
\end{eqnarray}
As a result, in this expression the leading order correction
is proportional to
$\frac{1}{M^2}$.
The classical Lagrangian takes the form
$$S_0=\phi\bar{\phi}+\Phi\bar{\Phi}
$$
which for this case looks like
\begin{equation}
S_0=\phi\bar{\phi}+\frac{\lambda^2 |\phi|^4}{4 M^2}
\end{equation}
and the chiral potential, which is of the  special form:
$$W=\frac{1}{2}(\lambda\Phi\phi^2+M\Phi^2)
$$
can be written as
\begin{eqnarray}
W&=&
-\frac{1}{8}\frac{\lambda^2\phi^4}{M}+\frac{1}{4}\lambda\phi^2
\Big\{\frac{\lambda\bar{D}^2(\bar{\phi}^2)}{8M^2}+\frac{\lambda^3}{64\pi^2
M^2}
\bar{D}^2\Big((\frac{\bar{\phi}^2}{2}+|\phi|^2)
(2+\nonumber\\&+&
\log\big[
\frac{5}{2}\lambda^4\frac{|\phi|^4}{M^2\mu^2}-
\lambda^4\frac{|\phi|^2}{M^2\mu^2}
(\phi^2+\bar{\phi}^2)
\big])
\Big)
\Big\}+O(\frac{1}{M^3})
\end{eqnarray}
As a result the total effective action of light superfields for this
case
can be cast in the form
\begin{equation}
S_{eff}[\phi,\bar{\phi}]=S_{eff}^{cl}[\phi,\bar{\phi}]+
S_{eff}^{q}[\phi,\bar{\phi}]
\end{equation}
where $S_{eff}^{cl}[\phi,\bar{\phi}]$ is a classical part of the
effective action which is of the form
\begin{eqnarray}
\label{cl1}
S_{eff}^{cl}[\phi,\bar{\phi}]&=&\int d^8 z\Big\{
\phi\bar{\phi}+\frac{\lambda^2 |\phi|^4}{4 M^2}\Big\}-\nonumber\\
&+&
\Big(\int d^6 z\big(
-\frac{1}{8}\frac{\lambda^2\phi^4}{M}+\frac{1}{4}\lambda\phi^2
\Big\{\frac{\lambda\bar{D}^2(\bar{\phi}^2)}{8M^2}+\frac{\lambda^3}{64\pi^2
M^2}
\bar{D}^2\Big((\frac{\bar{\phi}^2}{2}+|\phi|^2)
(2+\nonumber\\&+&
\log\big[
\frac{5}{2}\lambda^4\frac{|\phi|^4}{M^2\mu^2}-
\lambda^4\frac{|\phi|^2}{M^2\mu^2}
(\phi^2+\bar{\phi}^2)
\big])
\Big)
\Big\}\big)+h.c.
\Big)
+O(\frac{1}{M^3})
\end{eqnarray}
\begin{eqnarray}
\label{q1}
S_{eff}^{q}&=&
-\frac{1}{32\pi^2}\int d^8z \Big\{
4\lambda^2|\phi|^2(1+\log\frac{2M^2}{\mu^2})+
4\lambda^4\frac{|\phi|^4}{M^2}+
\big[
\frac{5}{2}\lambda^4\frac{|\phi|^4}{M^2}+2\lambda^4\frac{|\phi|^2}{M^2}
(\phi^2+\bar{\phi}^2)
\big]
\times\nonumber\\&\times&
\Big(
\log\big[
\frac{5}{2}\lambda^4\frac{|\phi|^4}{M^2\mu^2}-
\lambda^4\frac{|\phi|^2}{M^2\mu^2}
(\phi^2+\bar{\phi}^2)
\big]
\Big)
\Big\}+O(\frac{1}{M^3})
\end{eqnarray}
To convince ourselves that the results are consistent with the
decoupling
theorem, we can again carry out the field redefinition (\ref{re})
with the finite renormalization
constant (\ref{z}). As a result we arrive at the one-loop
effective action of light superfields which is a sum of (\ref{cl1})
and
(\ref{q1}) and has the form
\begin{equation}
S_{eff}[\phi,\bar{\phi}]=\int d^8 z\phi\bar{\phi}+O(\frac{1}{M})
\end{equation}
Therefore we find that in the  absence of self-interaction
of the light superfield the effective action for $\phi,\bar{\phi}$,
after
a suitable choice of $\mu$ and redefinitions of fields and couplings
becomes  a sum of the classical action of the light superfield and
terms at the one loop level that are
at least  second order in the
inverse mass parameter. Hence the result (\ref{cl1})-(\ref{q1})
for the effective action
of the light superfield is again consistent with the decoupling
theorem.
However, note again, since the parameters of the theory (fields,
masses,
couplings) are determined from  string theory, the one-loop
decoupling effects again modify in a quantitative manner the
low energy predictions for the couplings.

\section{One-loop effective action for non-minimal models}
\setcounter{equation}{0}
\renewcommand{\theequation}{\arabic{section}.\arabic{equation}}
In this section we generalize the discussion to the case  of
non-minimal
models.
We consider examples with the k\"ahlerian potential of the form
(\ref{knon}).
However, as it turns out an exact computation
of the one-loop
correction in the effective action for a general form of
$\tilde{K}$
leads to essential difficulties. Therefore we consider two  specific
examples.

\subsection{The model with heavy quantum
superfields and  external light superfields}
\subsubsection{Calculation of the effective action}
We consider for example the theory with the  action
\begin{eqnarray}
\label{acnm0}
S&=&\int d^8
z\Big(\phi\bar{\phi}+(1+\frac{\alpha}{M}(\phi+\bar{\phi}))
\Phi\bar{\Phi}
\Big)+\nonumber\\&+&\Big(\int d^6 z
\big[\frac{1}{2}(M+c\phi)\Phi^2+\frac{\lambda}{2}\Phi\phi^2\big]+h.c.\Big)
\end{eqnarray}
which corresponds to the functions $K$ and $W$ of the form
\begin{eqnarray}
\label{nm0}
W&=&\frac{1}{2}(M+c\phi)\Phi^2+\frac{\lambda}{2}\Phi\phi^2\nonumber\\
K&=&\Phi\bar{\Phi}+\frac{\alpha}{M}(\phi+\bar{\phi})\Phi\bar{\Phi}
\end{eqnarray}
In principle it is possible to consider the effective action for the
general case when both light and heavy superfields are split into
a sum
of background and quantum parts. However, calculation in this case
is
very complicated; hence in order
to simplify the computing we put light
superfields $\phi$, $\bar{\phi}$ to be pure background ones. Heavy
superfields $\Phi$, $\bar{\Phi}$ are split into a sum of background
superfields $\Phi_0$, $\bar{\Phi}_0$ and quantum ones $\Phi$,
$\bar{\Phi}$ by a standard way $$\Phi\to\Phi_0+\Phi,\
\bar{\Phi}\to\bar{\Phi}_0+\bar{\Phi} $$ As a result the quadratic
action of quantum superfields looks like
\begin{eqnarray}
S&=&\int d^8z\Big([1+\frac{\alpha}{M}(\phi+\bar{\phi})]
\Phi\bar{\Phi}\Big)+\Big(\int d^6 z
\frac{1}{2}(M+c\phi)\Phi^2+h.c.\Big)
\end{eqnarray}
and does not depend on heavy background superfields. Therefore
the one-loop quantum correction in the effective action is also
$\Phi_0,\bar{\Phi}_0$ independent due to the special (quadratic)
structure of action (\ref{acnm0}) in heavy superfields.
In this
case the column vector $\psi$ (see Section 2.3) consists of  the
heavy
superfield $\Phi$ only, and all the derivatives with respect to
$\psi$
are equivalent to those with respect to $\Phi$ only.

$\Phi$,

The one-loop correction in the effective action $\Gamma^{(1)}$
can be defined in this case by the equation
\begin{equation}
\label{Gamma1}
e^{i\Gamma^{(1)}}=\int D\Phi D\bar{\Phi} \exp (iS^{(2)})
\end{equation}
where $S_2$ is the  part of the action  quadratic  in quantum
superfields  $\Phi,\bar{\Phi}$ (see Section 2.3)
\begin{eqnarray}
\label{qua}
S_2&=&\int d^8 z (K_{\Phi\bar{\Phi}}\Phi\bar{\Phi} +
\frac{1}{2}K_{\bar{\Phi}\bar{\Phi}}\bar{\Phi}\bar{\Phi}+
\frac{1}{2}K_{\Phi\Phi}\Phi\Phi)+\nonumber\\
&+&[\int d^6 z \frac{1}{2}W_{\Phi\Phi}\Phi\Phi+h.c.]
\end{eqnarray}
and $K_{\Phi\bar{\Phi}}$, $K_{\bar{\Phi}\bar{\Phi}}$, $K_{\Phi\Phi}$
are
functions of light background superfields $\phi$, $\bar\phi$, only.

Using (\ref{Gamma1}) one can write $\Gamma^{(1)}$ as
\begin{equation}
\label{gr1}
\Gamma^{(1)}=\frac{i}{2} Tr \log G^{(\psi)}
\end{equation}
The straightforward calculation of the effective potential based on
(\ref{gr1}) is very complicated because elements of the matrix
$G^{(\psi)}$
are defined in different chiral superspaces hence  their chiralities
are mixed.
However, in this model it is possible to  employ a technique
 analogous to that
one used in \cite{Buch1,Buch2}.
Namely, let us consider a theory of  a real scalar superfield with
the standard
action
\begin{equation}
\label{gt}
S_v=-\frac{1}{16}\int d^8 z v D^{\alpha}\bar{D}^2 D_{\alpha}v
\end{equation}
This theory can be quantized  using the  Faddeev-Popov procedure
(cf.
\cite{Buch1,Buch2}), and
the effective action $W_v$ corresponding to this theory can be
determined by
the following integral:
\begin{eqnarray}
\label{ev1}
e^{iW_v}=\int Dv \delta(\frac{1}{4}D^2 v-\bar{\Phi})
\delta(\frac{1}{4}\bar{D}^2 v-\Phi)\exp (iS_v) det M_0
\end{eqnarray}
Here $M_0$ is a Faddeev-Popov matrix,
\begin{equation}
M_0=\left(
\begin{array}{cc}
0 & -\frac{1}{4}\bar{D}^2\\
-\frac{1}{4}D^2 & 0
\end{array}
\right)
\end{equation}
It is evident that $W_v$ is a constant: since  the model (\ref{gt})
is gauge
invariant its  effective action does not depend on $\Phi$
(cf. \cite{Buch1,Buch2}).

If we  multiply  the corresponding left-hand sides and right-hand
sides
of (\ref{Gamma1}) and (\ref{ev1}) we obtain
\begin{eqnarray}
\exp(i \Gamma^{(1)}[\phi,\bar{\phi}]+iW_v) &=&
\int {\cal D} \Phi {\cal D} \bar{\Phi}
Dv \delta(\frac{1}{4}D^2 v-\bar{\Phi})
\delta(\frac{1}{4}\bar{D}^2 v-\Phi)\exp (iS_v) det M_0
\times\nonumber\\&\times&
\exp (i\{
\int d^8 z (K_{\Phi\bar{\Phi}}\Phi\bar{\Phi} +
\frac{1}{2}K_{\bar{\Phi}\bar{\Phi}}\bar{\Phi}\bar{\Phi}+
\frac{1}{2}K_{\Phi\Phi}\Phi\Phi)+\nonumber\\
&+&[\int d^6 z \frac{1}{2}W_{\Phi\Phi}\Phi\Phi+h.c.]\}
)
\end{eqnarray}
Here $S_v$ is given by (\ref{gt}).
Then we integrate over $\Phi, \bar{\Phi}$
by means of delta functions (cf. \cite{Buch1,Buch2,BK0}) and
since $e^{iW_v}$ and $det M_0$ are constants we
arrive at the following expression for the one-loop effective action
for the superfields $\phi$, $\bar{\phi}$:
\begin{equation}
\label{g1}
\exp(i\Gamma^{(1)})=\int Dv \exp(\frac{i}{2}v\Delta v);\
\Gamma^{(1)}=
\frac{i}{2}Tr\log\Delta
\end{equation}
where
\begin{eqnarray}
\label{delta}
\Delta&=&\Box+\frac{1}{16}(K_{\Phi\bar{\Phi}}-1)\{D^2,\bar{D}^2\}+
\frac{1}{4}W_{\Phi\Phi}\bar{D}^2+
\frac{1}{4}\bar{W}_{\bar{\Phi}\bar{\Phi}}D^2
+\nonumber\\&+&
\frac{1}{4}(\bar{D}^2 K_{\Phi\Phi})\bar{D}^2+
\frac{1}{4}(D^2 K_{\bar{\Phi}\bar{\Phi}})D^2
\end{eqnarray}
This expression can be used  to calculate the explicit form of the
one-loop effective action.
We note that the last two  terms in this expression depend on
derivatives
of the background superfields and therefore do not contribute to
the k\"{a}hlerian
effective potential.

To find the one-loop effective action from (\ref{g1}) we use the
Schwinger
representation
\begin{equation}
Tr \log\Delta=Tr\int \frac{ds}{s} e^{is\Delta}
\end{equation}
Since the k\"ahlerian effective action  (by definition) does not
depend on
supercovariant derivatives of superfields, one
 calculates the  k\"{a}hlerian effective potential via proper-time
method
where  one can
omit  the last two terms
in (\ref{delta}) and obtain
(cf. \cite{Buch1,Buch2})
\begin{eqnarray}
K^{(1)}&=&\frac{i}{2}Tr \int \frac{ds}{s} e^{is\Delta'}\nonumber\\
&=&\frac{i}{2}\int d^8 z_1 d^8 z_2
\delta^8(z_1-z_2) \int \frac{ds}{s}e^{is\tilde{\Delta'}} e^{is\Box}
\delta^8(z_1-z_2)
\end{eqnarray}
where
\begin{eqnarray}
\label{dlt2}
\Delta'&=&
\Box+\frac{1}{16}(K_{\Phi\bar{\Phi}}-1)\{D^2,\bar{D}^2\}+
\frac{1}{4}W_{\Phi\Phi}\bar{D}^2+
\frac{1}{4}\bar{W}_{\bar{\Phi}\bar{\Phi}}D^2
\equiv\nonumber\\&\equiv&
\Box+H\{D^2,\bar{D}^2\}+
\frac{1}{4}W''\bar{D}^2+\frac{1}{4}\bar{W}''D^2
\end{eqnarray}
Here $W''=W_{\Phi\Phi}$, $H=\frac{1}{16}(K_{\Phi\bar{\Phi}}-1)$.
After calculations described in Appendix C we arrive at the
renormalized one-loop correction in the k\"ahlerian effective
potential
 in the form
\begin{eqnarray}
\label{res0}
K^{(1)}&=&
-\frac{1}{32\pi^2}\int d^8 z
\frac{W''\bar{W}''}{K_{\Phi\bar{\Phi}}^2}
\Big[\log\Big\{\frac{W''\bar{W}''}{\mu^2 K_{\Phi\bar{\Phi}}^2}\Big\}
\Big]
\end{eqnarray}
In the case when $K$ and $W$ are given by (\ref{nm0}), and
we  can write this correction as
\begin{eqnarray}
\label{res}
K^{(1)}&=&
-\frac{1}{32\pi^2}\int d^8 z
\frac{|(M+c\phi)|^2}{{(1+\frac{\alpha}{M}(\phi+\bar{\phi}))}^2}
\Big[\log\Big\{\frac{|(M+c\phi)|^2}{\mu^2{(1+\frac{\alpha}{M}
(\phi+\bar{\phi}))}^2}
\Big\}
\Big]
\end{eqnarray}
We note that the result for the one-loop contribution to
k\"{a}hlerian
effective potential does not depend on $\lambda$. This result,
however,
is natural since the vertex $\lambda\Phi\phi^2$ is linear
in the quantum superfield $\Phi$ and hence it can lead to
one-particle-reducible diagrams, only. The same follows from
(\ref{dlt2})
with $K$ and $W$ given by (\ref{nm0}).

To study the effective action of light superfields in the model
one can
expand the $K^{(1)}$ into a  power series in the inverse mass.


\subsubsection{Solving the effective equations of motion}

Let us consider decoupling effects for the non-minimal model with
the
one-loop k\"ahlerian effective
potential  given by (\ref{res}).
The result (\ref{res}) can be represented as a power series in
$\frac{1}{M}$.
The leading order  terms in this expansion have the form
\begin{eqnarray}
\label{g1t}
K^{(1)}&=&
-\frac{1}{32\pi^2}
\Big\{
\phi\bar{\phi}\big[2(c^2-3\alpha c+2\alpha^2)+(c^2-4\alpha
c+6\alpha^2)
\log\frac{M^2}{\mu^2}\big]
+\nonumber\\&+&
\frac{\phi\bar{\phi}(\phi+\bar{\phi})}{M}\big[
\log\frac{M^2}{\mu^2}(-12\alpha^3+9c\alpha^2-2\alpha
c^2)+\nonumber\\&+&
(3\alpha^2-2\alpha c)(c-\alpha)+(6\alpha^2-4\alpha c+c^2)(c-\alpha)+
\nonumber\\&+&
\frac{1}{2}(c-2\alpha)(\alpha^2-c^2)+\alpha^2(c-\alpha)
\big]+O(\frac{1}{M^2})\Big\}
\end{eqnarray}

For  $K$, $W$ of the form (\ref{nm0}) the effective equations of
motion (\ref{ee},\ref{ee1})
take  the form:
\begin{eqnarray}
& &
-\frac{1}{4}\bar{D}^2(\bar{\Phi}(1+\alpha\frac{\phi+\bar{\phi}}{M}))+
(M+c\phi)\Phi+\frac{\lambda}{2}\phi^2
=0\nonumber\\
& &
-\frac{1}{4}D^2(\Phi(1+\alpha\frac{\phi+\bar{\phi}}{M}))+
(M+c\bar{\phi})\bar{\Phi}+\frac{\lambda}{2}\bar{\phi}^2
=0
\end{eqnarray}
The zeroth order solution is
$$\Phi_0=-\frac{\lambda}{2(M+c\phi)}\phi^2=
-\frac{\lambda}{2M}\phi^2+O(\frac{1}{M^2})
$$
Due to (\ref{eqmh}) the $n$th order equation
has the form
\begin{eqnarray}
-\frac{1}{4}\bar{D}^2\big(\Phi_{n-1}(1+\alpha\frac{\phi+\bar{\phi}}{M})
\big)+(M+c\phi)\Phi_n=0
\end{eqnarray}
and the analogous equation for $\bar{\Phi}_n$. Hence the first order
approximation
can be written as
$$
\Phi_1=
\frac{1}{4(M+c\phi)}\bar{D}^2
\big(-\frac{\lambda}{2(M+c\bar{\phi})}\bar{\phi}^2
(1+\alpha\frac{\phi+\bar{\phi}}{M})
\big)=-\frac{\lambda}{16
M^2}\bar{D}^2(\bar{\phi}^2)+O(\frac{1}{M^3})
$$
As a result we can substitute  $\Phi=\Phi_0+\Phi_1$ into the
one-loop
effective action 
and
write the effective action for light superfields in the form
\begin{equation}
S_{eff}[\phi,\bar{\phi}]=S_{eff}^{cl}[\phi,\bar{\phi}]+
S_{eff}^{q}[\phi,\bar{\phi}]
\end{equation}
where $S_{eff}^{cl}[\phi,\bar{\phi}]$ is a classical part of the
effective action which is of the form
\begin{eqnarray}
\label{cl2}
S_{eff}^{cl}[\phi,\bar{\phi}]&=&\int d^8z\phi\bar{\phi}
-
\Big\{\int d^6 z\frac{3\lambda^2\phi^4}{8M}
+h.c.
\Big\}
+O(\frac{1}{M^2})
\end{eqnarray}
and $S^{q}_{eff}[\phi,\bar{\phi}]$ is a quantum part of the
effective action
looking like
\begin{eqnarray}
\label{q2}
S^{q}_{eff}[\phi,\bar{\phi}]&=&-\frac{\hbar}{32\pi^2}\int d^8
z\Big\{
\phi\bar{\phi}
\big[2(c^2-3\alpha c+2\alpha^2)+(c^2-4\alpha c+6\alpha^2)
\log\frac{M^2}{\mu^2}\big]
+\nonumber\\&+&
\big\{\frac{\phi\bar{\phi}(\phi+\bar{\phi})}{M}\big[
-\frac{c^2-3\alpha c+2\alpha^2}{c^2-4\alpha c+6\alpha^2}
(-12\alpha^3+9c\alpha^2-2\alpha c^2)+\nonumber\\&+&
(3\alpha^2-2\alpha c)(c-\alpha)+(6\alpha^2-4\alpha
c+c^2)(c-\alpha)+\nonumber\\
&+&\frac{1}{2}(c-2\alpha)(\alpha^2-c^2)+\alpha^2(c-\alpha)
\big]+\nonumber\\&+&
O(\frac{1}{M^2})
\big\}\Big\}
\end{eqnarray}
To check that the result corresponds to the decoupling theorem
we can renormalize the superfield $\phi$ according to  (\ref{re})
with the
finite renormalization constant $Z$  of the form
\begin{equation}
Z={\Big(1-\frac{h}{32\pi^2}
\big[2(c^2-3\alpha c+2\alpha^2)+(c^2-4\alpha c+6\alpha^2)
\log\frac{M^2}{\mu^2}\big]
\Big)}^{1/2}
\end{equation}
As a result the effective action being the sum of the  classical
part
(\ref{cl2}) and the quantum one (\ref{q2}) takes the form
\begin{equation}
S_{eff}[\phi,\bar{\phi}]=\int d^8 z \tilde{\phi}\tilde{\bar{\phi}}+
O(\frac{1}{M})
\end{equation}
which is thus formally consistent with the decoupling theorem.
  (However, note  again that
fields and couplings $\alpha, c$ are determined from string theory,
and thus
the  classical action gets essential one-loop corrections.)

\subsection{The model with light and heavy quantum superfields and
light
and heavy external superfields}
\subsubsection{Calculation of the  effective action}
It turns out that for non-minimal models it
 is possible obtain explicit expressions for the
 the one-loop corrected action  in
the  leading orders in
$\frac{1}{M}$   when both light and heavy
superfields have background and quantum parts, by employing the
 diagram approach.

 Let
us consider a specific  non-minimal model with action
\begin{eqnarray}
\label{nonmin}
S&=&\int d^8 z \big[\phi\bar{\phi}+\Phi\bar{\Phi}
+\frac{\lambda}{M}(W_1(\phi)\bar{\Phi}+\bar{W}_1(\bar{\phi})\Phi)\big]+
\nonumber\\&+&\big[
\int d^6 z(\frac{g}{3!}\phi^3+\frac{M}{2}\Phi^2)+h.c.\big]
\end{eqnarray}
The propagators of the model have the standard form (\ref{props}).
The zeroth order  in $\frac{1}{M}$ the expansion
of the one-loop k\"ahlerian effective potential
corresponds to the case where vertices, proportional to
$\frac{1}{M}$, are
absent. The corresponding supergraphs consist of
$<\phi\bar{\phi}>$-propagators with alternating background  fields
$\phi$  and
$\bar{\phi}$ (cf. \cite{PW,GR}):

\vspace*{3mm}

\hspace{1.5cm}
\Lengthunit=.5cm
\GRAPH(hsize=3){
\Circle(2)\mov(-1,0){\lin(-1,0)}\mov(1,0){\lin(1,0)}
\mov(8,0){\GRAPH(hsize=3){
\mov(.5,0){\Circle(2)\mov(-1,0){\lin(-1,0)}
\mov(1,0){\lin(1,0)}
\mov(-.2,1){\lin(0,1)}\mov(-.2,-1){\lin(0,-1)} }}
\mov(4,0){\GRAPH(hsize=3){
\mov(.5,0){\Circle(2)\mov(-1,.3){\lin(-.8,.3)}
\mov(.8,.3){\lin(.8,.3)}
\mov(-.2,1){\lin(0,1)}\mov(-.2,-1){\lin(0,-1)}
\mov(-1.7,-.3){\lin(-.8,-.3)}
\mov(0,-.3){\lin(.8,-.3)}
}
\mov(5,0){\ldots}}}}}

\vspace*{3mm}

\noindent Here the external legs correspond to alternating $\phi$
and $\bar{\phi}$.

The result is given by a sum
of all such supergraphs.  After calculations  that  are completely
parallel
with  those  carried out above we arrive at the leading order
 correction $K^{(1)}_0$ displayed in eq.
 (\ref{cr0}). (Cf. \cite{GR,PW}. Other methods of calculating the
k\"{a}hlerian effective potential were given in \cite{BP}.)

To study corrections that are  higher orders in the $1\over M$
expansion,
it is more convenient
to split superfields $\phi$, $\Phi$ into a sum of background $\phi$,
$\Phi$
and quantum $\phi_q$, $\Phi_q$ superfields. As a result we arrive at
the following
quadratic action of the quantum superfields $\phi_q$, $\Phi_q$ in
external
superfields $\phi$, $\Phi$
\begin{eqnarray}
\label{nonmin1}
S&=&\int d^8 z \big[\phi_q\bar{\phi}_q+\Phi_q\bar{\Phi}_q
+\frac{\lambda}{M}(W'_1(\phi)\phi_q\bar{\Phi}_q+W^{''}_1(\phi)\Phi\phi^2_q
+h.c.)\big]+
\nonumber\\&+&\big[
\int d^6 z(\frac{g}{2}\phi\phi^2_q+\frac{M}{2}\Phi^2_q)+h.c.\big]
\end{eqnarray}
Let us study  corrections at different
 orders in $1\over M$ expansion. It is evident
 that  the presence
of corrections at higher orders  is due to the  presence of vertices
proportional to $\lambda$ in the supergraphs.
One can assume  that corrections at  higher  orders in the $1\over
M$
expansion
can arise in two cases.  In the first case the
diagram contains   the background heavy superfield
$\Phi$
and the vertex is of the form  $\int d^8 z
\bar{W}^{''}\Phi\bar{\phi}^2_q$.
Note  also that in  the diagrams   each factor $D^2$ corresponds
to a quantum
superfield $\phi_q$.  Therefore a contribution of such vertices is
of the form

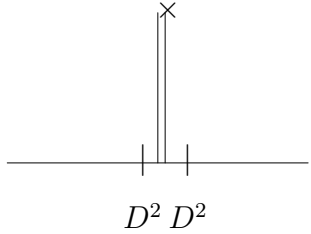
\vspace{2mm}

\Lengthunit=2cm
\hspace{2cm}
\GRAPH(hsize=3){
\mov(.5,0){
\mov(1,0){\lin(2,0)}\mov(2,0){\lin(0,1)}\mov(2.05,0){\lin(0,1)}
\ind(20,10){\times}
}
\ind(27,-4){D^2}\ind(27,0){|}\ind(24,-4){D^2}\ind(24,0){|}
}

\vspace{2mm}

\noindent Here the double external line with sign $\times$
corresponds
to $\bar{W}^{''}\Phi$ (an analogous fragment of a diagram arises
for the
contribution proportional to $W^{''}\bar{\Phi}$). Then a
transportation  of
any of the two $D^2$-factors during $D$-algebra transformations
leads to
 the factor $D^2 \Phi$, i.e. diagrams containing such
vertices do not contribute to the  k\" ahlerian effective potential.
Therefore corrections at the  first and higher orders in the
inverse mass
expansion can appear
only in  supergraphs with  light external field lines only.

Let us study  the possible diagrams of that type.
We must pay attention to the fact
that the propagator of the light superfield
$\phi$ depends on the  background superfield. We denote it with a
 bold line  which has the following  diagrammatic representation

\vspace{2mm}

\Lengthunit=.8cm
\Linewidth{1.5pt}
\hspace{1.5cm}
\GRAPH(hsize=3){\mov(.5,0){
\mov(.5,0){\lin(1,0)}\ind(21,0){=}
\Linewidth{.4pt}\mov(2.5,0){\lin(1,0)}
\ind(37,0){+}\mov(4,0){\lin(3,0)}\mov(5,0){\lin(0,1)}
\mov(6,0){\lin(0,1)}
\ind(74,0){+}\ind(79,0){\ldots}\ind(93,0){(}
\mov(8.1,0){\lin(1,0)}\mov(9.2,0){\lin(0,1)\lin(2,0)}
\mov(10.2,0){\lin(0,1)} 
\ind(113,0){)^n}\ind(118,0){+}\ind(120,0){\ldots}
}}

\vspace{2mm}

\noindent Here  an external line denotes the
background superfields $\phi$, $\bar{\phi}$.  Summation of this
chain
of diagrams allows one to show that the  total propagator of  the
light
superfield is of the form
\begin{eqnarray}
<\phi_q(z_1)\bar{\phi}_q(z_2)>=-\frac{\bar{D}^2_1
D^2_2}{16(\Box-g^2|\phi|^2)}
\delta_{12}
\end{eqnarray}
The propagator of the  heavy superfield $\Phi$ is given by
(\ref{props}).
One can easily see that in addition to the
 diagrams discussed above,
which do not contain vertices proportional to $\frac{\lambda}{M}$,
there are two other types
of supergraphs. The first one  consists of  an arbitrary number of
 repeating links
of the form $<\Phi\bar{\Phi}><\phi\bar{\phi}>$, i.e.

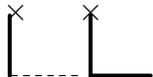
\vspace{2mm}

\Lengthunit=.8cm
\hspace{2cm}
\GRAPH(hsize=3){\Linewidth{.4pt}
\mov(.5,0){\mov(.2,0){\dashlin(1.1,0)}
\Linewidth{1.5pt}\lin(0,1)\ind(1,10){\times}\ind(13.5,10){\times}
\mov(1,0){
\lin(1,0)\lin(0,1)}
}}

\vspace{2mm}

\noindent where  bold external lines with sign $\times$ correspond
to
external $\frac{\lambda}{M}W'_1(\phi)$ term, and   dashed internal
ones
to $<\Phi\bar{\Phi}>$ propagators.  The second  type of diagrams
consist of  an arbitrary number of repeating links of the form\\ $
<\phi\bar{\phi}>_t<\Phi\bar{\Phi}><\phi\bar{\phi}>_t<\Phi\Phi>
<\phi\bar{\phi}>_t<\bar{\Phi}\Phi><\bar{\phi}\phi>_t<\bar{\Phi}\bar{\Phi}>
$
which can be diagrammatically represented as

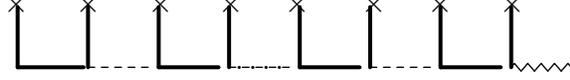
\vspace{2mm}

\Lengthunit=.8cm
\hspace*{2.5cm}
\GRAPH(hsize=3){\Linewidth{1.5pt}\lin(0,1)\lin(1.1,0)\Linewidth{0.5pt}
\mov(1,0){\dashlin(1,0)\Linewidth{1.5pt}\lin(0,1)\Linewidth{0.5pt}}
\mov(2,0){\Linewidth{1.5pt}\lin(1,0)\lin(0,1)\Linewidth{0.5pt}}
\mov(3,0){\dashdotlin(1,0)\Linewidth{1.5pt}\lin(0,1)\Linewidth{0.5pt}}
\mov(4,0){\Linewidth{1.5pt}\lin(1,0)\lin(0,1)\Linewidth{0.5pt}}
\mov(5,0){\dashlin(1,0)\Linewidth{1.5pt}\lin(0,1)\Linewidth{0.5pt}}
\mov(6,0){\Linewidth{1.5pt}\lin(1,0)\lin(0,1)\Linewidth{0.5pt}}
\mov(7,0){\trianglin(1,0)\Linewidth{1.5pt}\lin(0,1)\Linewidth{0.5pt}}
\ind(-14,10){\times}\ind(-2,10){\times}\ind(10,10){\times}
\ind(20,10){\times}\ind(31,10){\times}
\ind(42,10){\times}\ind(53,10){\times}\ind(65,10){\times}
}

\vspace{2mm}

\noindent Here  the dashed-and-dotted line denotes $<\Phi\Phi>$  and
the jagged line denotes $<\bar{\Phi}\bar{\Phi}>$.

The contribution of  a diagram of the first type, which consists of
$n$ such
links,  is of the form
\begin{eqnarray}
\label{cont1}
I_n=\frac{1}{n}\Big\{\frac{\lambda^2}{M^2}{|W'_1|}^2
<\phi\bar{\phi}><\Phi\bar{\Phi}>\Big\}^n
\end{eqnarray}
and  the contribution of a  diagram of  the second type  that
consists
 of $n$ chains is of the form
\begin{eqnarray}
\label{cont2}
J_n=\frac{1}{n}\Big\{\frac{\lambda^8}{M^8}{|W'_1|}^8
<\phi\bar{\phi}><\Phi\bar{\Phi}><\phi\bar{\phi}><\Phi\Phi>
<\phi\bar{\phi}><\bar{\Phi}\Phi><\bar{\phi}\phi><\bar{\Phi}\bar{\Phi}>
\Big\}^n
\end{eqnarray}
Using  the exact form of the propagator given by (\ref{props}) and
carrying out
$D$-algebra transformations we find that  (\ref{cont1}) and
(\ref{cont2})  can be cast
in the form
\begin{equation}
I_n=\frac{1}{\Box}\frac{1}{n}{\Big(\frac{\Box^2}{(\Box-M^2)
(\Box-g^2|\phi^2|)}
\Big)}^n\frac{\lambda^{2n}}{M^{2n}}{|W'_1|}^{2n}
\end{equation}
and
\begin{equation}
J_n=\frac{1}{\Box}\frac{1}{n}{\Big(\frac{\Box^7}{{(\Box-M^2)}^4
{(\Box-g^2|\phi^2|)}^4}
\Big)}^n\frac{\lambda^{2n}}{M^{6n}}{|W'_1|}^{8n}
\end{equation}
 respectively.

We can now  find  the leading terms in  the inverse mass  expansion
for the
above  contributions.
It is easy to see that as $n$ grows  the  order in the  inverse mass
contribution increases.
 $J_n$  is proportional to $M^{-6n}\Box^{-n-1}$. After Fourier
transformation and  the integration over
momenta the leading term will be  of the order $M^{2-8n}$, i.e.  it
will
involve at least six powers of the inverse mass parameter.
As for $I_n$ it is  proportional to
$M^{-2n}\Box^{-1}$; after  the integration the
leading term is of the order $ M^{2-2n}$.
Therefore
the zeroth and  the second order are given by $I_1$ and $I_2$.
A straightforward calculation carried out in the framework of
dimensional regularization yields
\begin{eqnarray}
\label{cr1}
I_1&=&\lambda^2\frac{{|W'_1|}^2}{32\pi^2}
[\frac{2}{\epsilon}+\log\frac{M^2}{\mu^2}+1]
+\frac{\lambda^2}{32\pi^2 M^2}{|W'_1|}^2 g^2|\phi|^2
[\frac{2}{\epsilon}+\log\frac{M^2}{\mu^2}]\nonumber\\
I_2&=&
-\frac{\lambda^4{|W'_1|}^4}{32\pi^2 M^2}(\frac{2}{\epsilon}+
\log\frac{M^2}{\mu^2}+1)
\end{eqnarray}
Therefore the leading  term for the  one-loop correction
to the effective action of the
 non-minimal model with  the classical action (\ref{nonmin}) is
the sum of
$I_1$ and $I_2$ (eq.(\ref{cr1}))  as well as  the correction
(\ref{cr0})
and it takes  the  form:
\begin{eqnarray}
\label{nmef0}
K^{(1)}&=&-\frac{1}{32\pi^2}g^2|\phi|^2
(\frac{2}{\epsilon}+\log\frac{g^2|\phi|^2}{\mu^2})+
\lambda^2\frac{{|W'_1|}^2}{32\pi^2}[\frac{2}{\epsilon}+\log\frac{M^2}{\mu^2}
+1]-
\\&-&\frac{1}{32\pi^2 M^2}\Big[
\lambda^4{|W'_1|}^4(\frac{2}{\epsilon}+\log\frac{M^2}{\mu^2}+1)-
\lambda^2{|W'_1|}^2 g^2|\phi|^2\{
\frac{2}{\epsilon}+\log\frac{M^2}{\mu^2}\}
\Big]+O(\frac{1}{M^3})\nonumber
\end{eqnarray}
To cancel the one-loop divergences we must add a counterterm of the
form
\begin{equation}
S^{(1)}_{ct}=-\frac{1}{16\pi^2\epsilon}
\Big(-g^2|\phi|^2+\lambda^2{|W'_1|}^2-\frac{1}{M^2}
\big[\lambda^2 g^2{|W'_1|}^2 |\phi|^2-\lambda^4{|W'_1|}^4\big]
\Big)
\end{equation}
As a result we arrive at the final form of the renormalized one-loop
 correction in the  k\"{a}hlerian
effective potential:
\begin{eqnarray}
\label{nmef}
K^{(1)}&=&-\frac{1}{32\pi^2}g^2|\phi|^2
\log\frac{g^2|\phi|^2}{\mu^2}+
\lambda^2\frac{{|W'_1|}^2}{32\pi^2}[\log\frac{M^2}{\mu^2}+1]-
\\&-&\frac{1}{32\pi^2 M^2}\Big[
\lambda^4{|W'_1|}^4(\log\frac{M^2}{\mu^2}+1)-
\lambda^2{|W'_1|}^2 g^2|\phi|^2
\log\frac{M^2}{\mu^2}
\Big]+O(\frac{1}{M^3})\nonumber
\end{eqnarray}
The essential feature of this  correction is that  terms that would
depend on
the  background heavy superfield $\Phi$ are absent. This fact
is due to the form of the vertex $\int d^8 z W(\phi)\bar{\Phi}$
whose
presence necessarily leads to terms which depend on supercovariant
derivatives of $\Phi$.


\subsubsection{Solution of the effective equations of motion}
Now let us consider the effective action (\ref{nmef})
corresponding to the model
with classical action (\ref{nonmin}).
The effective equation of motion corresponding to this model have
the form
\begin{eqnarray}
& &M\Phi-\frac{1}{4}\bar{D}^2(\bar{\Phi}+\frac{\lambda}{M}
\bar{W}_1(\bar{\phi}))=0\\
&
&M\bar{\Phi}-\frac{1}{4}D^2(\Phi+\frac{\lambda}{M}W_1(\phi))=0\nonumber
\end{eqnarray}
We note that  quantum corrections in this model do not depend on
$\Phi$.
The zeroth order  approximation for the  superfield $\Phi$ is
$\Phi_0=0$ (since
$\frac{\partial W}{\partial \Phi}=0$ only at $\Phi=0$) (see
(\ref{eqm})).
The first order  approximation (\ref{eqmh}) looks like
\begin{equation}
\Phi_1=\frac{\lambda}{4}\frac{\bar{D}^2}{M^2}\bar{W}_1(\bar{\phi})
\end{equation}
It follows from (\ref{eqmh}) that for this model  the $n$-th  order
approximation
$\Phi_n$ is of the order $\frac{1}{M^{n+1}}$.
Therefore the final result for  the
effective action of light superfields reads as
\begin{eqnarray}
\label{gnm}
\Gamma^{(1)}&=&\int d^8 z \Big(\phi\bar{\phi}
-\frac{\hbar}{32\pi^2}
\big\{g^2|\phi|^2\log\frac{g^2|\phi|^2}{\mu^2}+
\lambda^2{|W'_1|}^2[1+\log\frac{M^2}{\mu^2}]-
\nonumber\\&-&\frac{1}{M^2}\big[
\lambda^4{|W'_1|}^4(1+\log\frac{M^2}{\mu^2})-
\lambda^2{|W'_1|}^2 g^2|\phi|^2\log\frac{M^2}{\mu^2}
\big]\big\}
\Big)+\nonumber\\&+&
(\int d^6 z \frac{g}{3!}\phi^3 +h.c.)+O(\frac{1}{M^3})
\end{eqnarray}
This is a final result for the one-loop corrected effective action
of light
superfields and applies to a general choice of $W_1$.

We note that the result (\ref{gnm}) is consistent with the
decoupling theorem.
If we choose for example $W_1=\phi^2$, $W'_1=2\phi$ (the minimal
model),
we can carry out the redefinition of the  superfield $\phi$ by the
rule
(\ref{re})
 with
\begin{eqnarray}
Z=\Big(1-\frac{\hbar}{32\pi^2}[4\lambda^2(1+\log\frac{M^2}{\mu^2})]
\Big)^{1/2}
\end{eqnarray}
Coupling $g$ is redefined by the rule $\tilde{g}=Z^{-3}g$.
As a result the effective action of light superfields takes the form
\begin{equation}
\Gamma^{(1)}=\int d^8 z (\tilde{\phi}\tilde{\bar{\phi}}
-\frac{\hbar}{32\pi^2}g^2|\tilde{\phi}|^2\log\frac{g^2|\tilde{\phi}|^2}{\mu^2})+
(\int d^6 z \frac{\tilde{g}}{3!}\tilde{\phi}^3
+h.c.)+O(\frac{1}{M^2})
\end{equation}
The result is consistent with the decoupling theorem and
  coincides in form with the case that corresponds to the minimal
model discussed in section 3. However, we note again that since
parameters of the theory (fields,
masses, couplings) are fixed from string theory, they cannot be
redefined.
Therefore the classical action
is modified by one-loop corrections. The final result for the
effective action of light superfields is given by (\ref{gnm}).


\section{Quantum corrections to the effective action in gauge
theories}
\subsection{Gauge invariant model of massive chiral
superfields}
\setcounter{equation}{0}
\renewcommand{\theequation}{\arabic{section}.\arabic{equation}}
For the sake of completeness we would like to address the decoupling
effects in
the $N=1$ supersymmetric gauge theory  with
 chiral superfields that
 interact   with the corresponding gauge superfields
 (i.e. chiral superfields that  are charged under the gauge
symmetry)\footnote{M.C. would like to thank L. Everett for a
discussion
on this topic.}.
To simplify the consideration we choose the Abelian $U(1)$  gauge
theory with the
chiral superfields charged under $U(1)$.
 The simplest (minimal)  model describing  gauge invariant
interactions
 between massive and light chiral superfields, and the $U(1)$  gauge
superfield
 is of the form:
\begin{eqnarray}
\label{actg}
S&=&\int d^8 z (\bar{\Phi}_1 e^{2gV}\Phi_1+\bar{\Phi}_2
e^{-2gV}\Phi_2+
\bar{\phi} e^{-gV}\phi)+\nonumber\\
&+&\frac{1}{64}\int d^6 z W^2 +
\big(\int d^6 z (M\Phi_1\Phi_2+\lambda\Phi_1\phi^2)+h.c.\big)
\end{eqnarray}
Here $V$ is a real scalar superfield, $\Phi_1,\Phi_2,\phi$ are heavy
chiral
superfields,
$W_{\alpha}=\bar{D}^2 D_{\alpha}V$ is a superfield strength.
The model is invariant under  the following gauge transformations
with $\Lambda$ is a chiral superfield, $\bar{\Lambda}$ is an
antichiral
one
 (c.f. \cite{BK0}):~ 
 denotes
\begin{eqnarray}
V&\to&V+i(\bar{\Lambda}-\Lambda)\nonumber\\
\Phi_1&\to& e^{2ig\Lambda}\Phi_1\nonumber\\
\Phi_2&\to& e^{-2ig\Lambda}\Phi_2\nonumber\\
\phi&\to& e^{-ig\Lambda}\phi
\end{eqnarray}

For the sake of concreteness we chose the  specific $U(1)$ charges
for the
chiral superfields such that there is an allowed  tri-linear
interaction term between
$\Phi_i$ and $\phi$.  For the sake of simplicity we did  not include
self-interactions of the  light
fields in the chiral potential. In addition, we did  not include
explicitly the terms of possible  other light chiral
superfields which  we chose not to interact with  $\Phi_i$ and
$\phi$
superfields; these
additional chiral superfields  however should be present, in order
to cancel
chiral anomalies.

 The gauge-fixing term  is of the form
\begin{equation}
\label{gf}
S_{gf}=-\frac{1}{16\xi}\int d^8 z D^2 V\bar{D}^2 V
\end{equation}
where $\xi$ is a gauge parameter.
The propagators  for the fields of the model
have the form (see, e.g., \cite{BK0})
\begin{eqnarray}
\label{prq}
<\Phi_1\bar{\Phi}_1>&=&-\frac{\bar{D}^2_1
D^2_2}{16(\Box-M^2)}\delta^8(z_1-z_2)
\nonumber\\
<\Phi_2\bar{\Phi}_2>&=&
-\frac{\bar{D}^2_1 D^2_2}{16(\Box-M^2)}\delta^8(z_1-z_2)
\nonumber\\
<\Phi_1\Phi_2>&=&-\frac{M\bar{D}^2_1 }{4(\Box-M^2)}\delta^8(z_1-z_2)
\nonumber\\
<\phi\bar{\phi}>&=&-\frac{\bar{D}^2_1
D^2_2}{16\Box}\delta^8(z_1-z_2)
\nonumber\\
<vv>&=&(-\frac{D^{\alpha}\bar{D}^2 D_{\alpha}}{8\Box^2}+
\xi\frac{\{D^2,\bar{D}^2\}}{16\Box^2})\delta^8(z_1-z_2)
\end{eqnarray}

Possible vertices associated with  the action (\ref{actg}) are
\begin{equation}
\label{vertg}
g^n \bar{\phi} v^n \phi,\  g^l \bar{\Phi}_1 v^l \Phi_1,\
g^m \bar{\Phi}_2 v^m \Phi_2,\ \lambda\Phi_1\phi^2,\
\lambda\bar{\Phi}_1\bar{\phi}^2
\end{equation}
The effective action can again be studied in  the framework of  the
loop expansion.
To carry out the  calculations we  split  all  the superfields
into  a sum of
background
 $\Phi_1,\Phi_2,\phi,V$ and quantum
$\Phi_{1q},\Phi_{2q},\phi_q,v$  superfields by the rule
\begin{eqnarray}
\Phi_1&\to&\Phi_1+\Phi_{1q},\ \Phi_2\to\Phi_2+\Phi_{2q}\nonumber\\
\phi&\to&\phi+\phi_q,\ v\to V+v
\end{eqnarray}
As a result we arrive at the quadratic action of quantum superfields
\begin{eqnarray}
\label{q2g}
S_q&=&\int d^8 z
\Big\{\frac{1}{2}v\big[\Box+(\frac{1}{\xi}-1)\frac{1}{16}\{D^2,\bar{D}^2\}+
g^2(4\bar{\Phi}_1e^{2gV}\Phi_1+4\bar{\Phi}_2e^{-2gV}\Phi_2+
\bar{\phi}e^{-gV}\phi)
\big]v+\nonumber\\&+&
\bar{\Phi}_{1q}e^{2gV}\Phi_{1q}+\bar{\Phi}_{2q}e^{-2gV}\Phi_{2q}+
\phi_q e^{-gV}\bar{\phi}_q+\nonumber\\&+&
2g(\bar{\Phi}_1 e^{2gV} v\Phi_{1q}+\bar{\Phi}_{1q} e^{2gV} v\Phi_1)
-2g(\bar{\Phi}_2 e^{-2gV} v\Phi_{2q}+\bar{\Phi}_{2q} e^{-2gV}
v\Phi_2)
-\nonumber\\&-&
g(\bar{\phi} e^{-gV} v\phi_q+\bar{\phi}_q e^{-gV} v\phi)
\Big\}+\nonumber\\&+&
\Big(\int d^6 z(M\Phi_{1q}\Phi_{2q}+\lambda\Phi_1\phi^2_q+
2\lambda\phi\Phi_{1q}\phi_q)+h.c.\Big)
\end{eqnarray}
In the following we employ this action to determine the one-loop
corrections.

The one-loop contribution to  the effective action $\Gamma^{(1)}$
is defined in the usual way
(see \cite{BO}):
\begin{eqnarray}
\label{oncor}
\exp (i\Gamma^{(1)}[\Phi_1,\Phi_2,\phi,V])=\int D\Phi_{1q}
D\Phi_{2q} D\phi_q Dv
\exp (iS_q[\Phi_{1q},\Phi_{2q},\phi_q,v;\Phi_1,\Phi_2,\phi,V])
\end{eqnarray}
The low-energy leading terms in the effective
action  are the k\"{a}hlerian effective
potential that depends only on
$\Phi_1,\Phi_2,\phi,\bar{\Phi}_1,\bar{\Phi}_2,
\bar{\phi},V$ but not on their supercovariant derivatives,
the chiral effective potential
that is a
holomorphic function of  chiral superfields $\Phi_1,\Phi_2,\phi$,
and the
 field depending effective gauge coupling
$f(\Phi_1,\Phi_2,\phi)$ holomorphic function of superfields
that multiplies the term  proportional to
$W^2$.
 They completely specify the low-energy effective action of
 $N=1$ supersymmetric gauge theory
(see, e.g.,  \cite{GR}).

\subsection{One-loop k\"{a}hlerian potential in supersymmetric
gauge theory}
To study decoupling  effects in the model (\ref{actg})  we
determine the one-loop
k\"{a}hlerian effective potential and
  consider the effective equations of motion for the heavy
superfields. We carry out the calculation of one-loop quantum
contribution
to k\"{a}hlerian effective potential on the base of the expression
(\ref{oncor}) with quadratic action of quantum superfields
(\ref{q2g}).
The perturbative approach for the action (\ref{q2g}) can be
developed in the
following way. We consider
\begin{eqnarray}
\label{q2g0}
S_q&=&\int d^8 z
\Big\{\frac{1}{2}v\big[\Box+(\frac{1}{\xi}-1)\frac{1}{16}\{D^2,\bar{D}^2\}
\big]v+
\bar{\Phi}_{1q}\Phi_{1q}+\bar{\Phi}_{2q}\Phi_{2q}+
\phi_q \bar{\phi}_q+\nonumber\\&+&
\Big(\int d^6 z(M\Phi_{1q}\Phi_{2q}+h.c.\Big)
\end{eqnarray}
as a quadratic part which leads just to the propagators (\ref{prq})
and all other terms in (\ref{q2g}) are treated as vertices of
interaction.

Let us consider possible one-loop supergraphs. They can contain
both $<vv>$ and
$<\phi\bar{\phi}>$, $<\Phi_1\bar{\Phi}_1>$, $<\Phi_2\bar{\Phi}_2>$,
$<\Phi_1\Phi_2>$ propagators. We will use the Landau gauge
$\xi=0$. The
propagator $<vv>$ is proportional to the projection operator
$\Pi_1=-\frac{D^{\alpha}\bar{D}^2 D_{\alpha}}{8\Box}$ in this
gauge, and propagators $<\phi\bar{\phi}>$,
$<\Phi_1\bar{\Phi}_1>$, $<\Phi_2\bar{\Phi}_2>$ are proportional to
the
projection operator $\Pi_0=\frac{\bar{D}^2 D^2}{16\Box}$ (see
e.g.(\ref{prq})).  The projection operators $\Pi_0$ and $\Pi_1$ are
orthogonal to each other \cite{BK0}. As a result each supergraph
containing propagators of both chiral superfields and gauge
superfields has zero contribution (it is easy to see that the
analogous
 situation takes place for supergraphs containing the propagator
$<\Phi_1\Phi_2>$ due to the identity $\bar{D}^2\Pi_0=0$).
Therefore there
are no supergraphs containing both propagators of chiral and gauge
superfields.  As a result, in the Landau gauge there can exist
only two
types of supergraphs: the first of them contains propagators of
chiral
superfields only, and the second one contains propagators of gauge
superfields
only.  Hence in order to calculate the one-loop k\"{a}hlerian
effective
potential we can omit in action (\ref{q2g}) the cross  terms, i.e.
those
that include both the chiral and vector quantum superfields
because such terms will lead to supergraphs containing both
propagators
of chiral and gauge superfields.  As a result one can use the
expression for the quadratic action \begin{eqnarray} S_q&=&\int
d^8 z
\Big\{\frac{1}{2}v\big[\Box+(\frac{1}{\xi}-1)\frac{1}{16}\{D^2,\bar{D}^2\}+
g^2(4\bar{\Phi}_1 e^{2gV}\Phi_1+4\bar{\Phi}_2
e^{-2gV}\Phi_2+\bar{\phi}
e^{-gV}
\phi)
\big]v+\nonumber\\&+&
\bar{\Phi}_{1q}e^{2gV}\Phi_{1q}+4\bar{\Phi}_{2q}e^{-2gV}\Phi_{2q}+
\bar{\phi}_q e^{-gV}\phi_q
\Big\}+\nonumber\\&+&
\Big(\int d^6 z(M\Phi_{1q}\Phi_{2q}+\lambda\Phi_1\phi^2_q+
2\lambda\phi\Phi_{1q}\phi_q)+h.c.\Big)
\end{eqnarray}
instead of (\ref{q2g}). It leads to the same supergraphs as
(\ref{q2g}).
At the end
of the calculations the parameter $\xi$ must be put to zero.

To study the one-loop contribution to k\"{a}hlerian effective
potential it
is convenient to introduce background superfields
$\tilde{\phi},\tilde{\Phi}_1,\tilde{\Phi}_2$ and conjugated ones
in the following way
\begin{eqnarray}
\label{trs}
& &\tilde{\phi}=e^{-gV/2}\phi,\
\tilde{\bar{\phi}}=\bar{\phi}e^{-gV/2}\nonumber\\
& &\tilde{\Phi}_1=e^{gV}\Phi_1,\ \tilde{\bar{\Phi}}_1=\bar{\Phi}_1
e^{gV}\nonumber\\
& &\tilde{\Phi}_2=e^{-gV}\Phi_2,\
\tilde{\bar{\Phi}}_2=\bar{\Phi}_2 e^{-gV}
\end{eqnarray}
The quantum superfields are redefined by the same way.
Such redefinitions represent themselves as a sort of introduction of
background covariantly chiral superfields \cite{BK0}.
We note that background superfields
$\tilde{\phi},\tilde{\Phi}_1,\tilde{\Phi}_2$
can be treated as constants.
The quadratic action of quantum superfields takes the form
\begin{eqnarray}
S_q&=&\int d^8 z
\Big\{\frac{1}{2}v\big[\Box+(\frac{1}{\xi}-1)\frac{1}{16}\{D^2,\bar{D}^2\}+
g^2(4\tilde{\bar{\Phi}}_1\tilde{\Phi}_1
+4\tilde{\bar{\Phi}}_2\Phi_2 +
\tilde{\bar{\phi}}\tilde{\phi} )
\big]v+\nonumber\\&+&
\tilde{\bar{\Phi}}_{1q}\tilde{\Phi}_{1q}+
\tilde{\bar{\Phi}}_{2q}\tilde{\Phi}_{2q}+
\tilde{\phi}_q \tilde{\bar{\phi}}_q
\Big\}+\nonumber\\&+&
\Big(\int d^6
z(M\tilde{\Phi}_{1q}\tilde{\Phi}_{2q}+\lambda\tilde{\Phi}_1
\tilde{\phi}^2_q+
2\lambda\tilde{\phi}\tilde{\Phi}_{1q}\tilde{\phi}_q)+h.c.\Big)
\end{eqnarray}
The propagators $<\tilde{\phi}\tilde{\bar{\phi}}>,
<\tilde{\Phi}_1\tilde{\bar{\Phi}}_1>$ and other ones have the
standard form
(\ref{prq}).

As a result we find the  one-loop correction
to the effective action to be of the form
\begin{equation}
\Gamma^{(1)}=\Gamma^{(1)}_1+\Gamma^{(1)}_2
\end{equation}
where
\begin{eqnarray}
\label{gam3}
\exp(i\Gamma^{(1)}_1)&=&\int D\tilde{\Phi}_{1q} D\tilde{\Phi}_{2q}
D\tilde{\phi}_q
\exp[
\int d^8z \{\tilde{\bar{\Phi}}_{1q}\tilde{\Phi}_{1q}+
\tilde{\bar{\Phi}}_{2q}\tilde{\Phi}_{2q}+
\tilde{\bar{\phi}}_q \tilde{\phi}_q
\Big\}+\nonumber\\&+&
\Big(\int d^6
z(M\tilde{\Phi}_{1q}\tilde{\Phi}_{2q}+\lambda\tilde{\Phi}_1
\tilde{\phi}^2_q+
2\lambda\tilde{\phi}\tilde{\Phi}_{1q}\tilde{\phi}_q)+h.c.\Big)
]\\
\exp(i\Gamma^{(1)}_2)&=&\int Dv\exp\Big(i
\int d^8 z
\frac{1}{2}v\big[\Box+(\frac{1}{\xi}-1)\frac{1}{16}\{D^2,\bar{D}^2\}+
g^2(4\tilde{\bar{\Phi}}_1\tilde{\Phi}_1+\nonumber\\&+&4
\tilde{\bar{\Phi}}_2\tilde{\Phi}_2+\tilde{\bar{\phi}}\tilde{\phi} )
\big]v\Big)\nonumber
\end{eqnarray}
We note that since transformations of the quantum superfields
(\ref{trs}) are local their Jacobian is equal to 1.
The contributions $\Gamma^{(1)}_1$, $\Gamma^{(1)}_2$ can be
calculated in a
straightforward way.
Correction $\Gamma^{(1)}_1$ is given in terms of supergraphs of
two types.
Diagrams of the first type consist only of propagators
$<\tilde{\phi}\tilde{\bar{\phi}}>$,
i.e. they can be represented as  rings consisting of different
numbers of
repeating links

\vspace{3mm}

\hspace*{4cm}\GRAPH(hsize=3){\Linewidth{.5pt}\mov(.1,0){\lin(2,0)}
\lin(0,1)\mov(.1,0){\lin(0,1)}\mov(1,0){\lin(0,1)}\mov(1.1,0){\lin(0,1)}
}

\vspace{3mm}

\noindent Each link has the form
\begin{equation}
K=\frac{\lambda^2|\tilde{\Phi}_1|^2 D^2\bar{D}^2}{16k^4}\delta_{12}
\end{equation}
The contribution of  a diagram consisting of $n$ such links has the
form
\begin{equation}
J_n=\int d^4\theta\int\frac{d^4k}{{(2\pi)}^4}K^n
\end{equation}
The total contribution of all such diagrams is then of
 the form $J=\sum_{n=1}^{\infty}J_n$.
After calculations analogous to those  carried out in Section 4 we
arrive
at the explicit expression
\begin{equation}
J=-\frac{1}{32\pi^2}\int d^4\theta
\lambda^2|\tilde{\Phi}_1|^2\frac{\lambda^2|\tilde{\Phi}_1|^2}{\mu^2}
\end{equation}

The diagrams of  the second type consist of repeating links of the
form

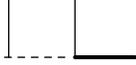
\vspace{2mm}

\Linewidth{.5pt}
\hspace*{2cm}\GRAPH(hsize=3){\mov(.1,0){\dashlin(1,0)}
\lin(0,1)\mov(1.1,0){\lin(0,1)\Linewidth{1.5pt}\lin(1,0)}
}

\vspace{2mm}

\noindent where  the bold line is a total propagator of a light
superfield which depends on the background superfield
$\tilde{\Phi}_1$.
It can be
obtained by a method analogous to the one used in the previous
section
(but with background, $\tilde{\Phi}_1,\tilde{\bar{\Phi}}_1$
instead of
$\phi,\bar{\phi}$)
and has the form
\begin{eqnarray}
\label{prt}
<\tilde{\phi}\tilde{\bar{\phi}}>&=&
\frac{\lambda^2|\tilde{\Phi}_1|^2}{k^2+\lambda^2|\tilde{\Phi}_1|^2}
\frac{D^2\bar{D}^2}{16k^2}
\end{eqnarray}
The dashed line corresponds to
$<\tilde{\Phi}_1\tilde{\bar{\Phi}}_1>$, and the external
line with sign $\times$ denotes the background superfield $\phi$.
The link
has the form
$\lambda^2|\tilde{\phi}|^2<\tilde{\bar{\phi}}\tilde{\phi}>
<\tilde{\Phi}_1\tilde{\bar{\Phi}}_1>$
As a result the contribution of a diagram consisting of $n$ such
links is
of the form
\begin{equation}
I_n=\int
d^4\theta\int\frac{d^4k}{{(2\pi)}^4}L^n
\end{equation}
where
\begin{equation}
L=\frac{\lambda^2|\tilde{\phi}|^2
D^2\bar{D}^2}{16(k^2+M^2)(k^2+\lambda^2|\tilde{\Phi}_1|^2)}
\end{equation}
The total contribution of all these diagrams, after  calculations
analogous to those
carried out in the previous section, has the form
\begin{eqnarray}
I&=&\sum_{n=1}^{\infty}I_n=
-\frac{1}{32\pi^2}\Big\{
\frac{2}{\epsilon}(-\lambda^2|\tilde{\phi}|^2)
+[M^2+\lambda^2(|\tilde{\Phi_1}|^2-|\tilde{\phi}|^2)]
\log\frac{[M^2+\lambda^2(|\tilde{\Phi}_1|^2-|\tilde{\phi}|^2)]}{\mu^2}-
\nonumber\\
&-&\lambda^2|\tilde{\Phi}_1|^2\log\frac{\lambda^2|\tilde{\Phi}_1|^2}{\mu^2}
\Big\}
\end{eqnarray}
After cancellation of a divergence  proportional to
$1\over\epsilon$,
via a  suitable counterterm, we arrive at the
following one-loop contribution to the k\"{a}hlerian effective
potential:
\begin{eqnarray}
K^{(1)}_{1r}&=&I+J=
-\frac{1}{32\pi^2}\Big\{
[M^2+\lambda^2(|\tilde{\Phi}_1|^2-|\tilde{\phi}|^2)]
\log\frac{[M^2+\lambda^2(|\tilde{\Phi}_1|^2-|\tilde{\phi}|^2)]}{\mu^2}
\Big\}
\end{eqnarray}
This expression can be expanded into a  power series in $1/M$ as
\begin{eqnarray}
K^{(1)}_1&=&-\frac{1}{32\pi^2}\Big\{
\lambda^2(|\tilde{\Phi}|^2-|\tilde{\phi}|^2)
(1+\log\frac{M^2}{\mu^2})+\frac{1}{2M^2}\lambda^4{(|\tilde{\Phi}|^2-
|\tilde{\phi}|^2)}^2
\Big\}+O(\frac{1}{M^4})
\end{eqnarray}

Now we turn to the calculation of $\Gamma^{(1)}_2$. Using
(\ref{gam3}) it is easy to
see that
\begin{equation}
K^{(1)}_2=\frac{i}{2}tr\log\Delta
\end{equation}
where
\begin{equation}
\Delta=\Box+(\frac{1}{\xi}-1)\frac{1}{16}\{D^2,\bar{D}^2\}+
g^2(4\tilde{\Phi}_1\tilde{\bar{\Phi}}_1+4\tilde{\Phi}_2\tilde{\bar{\Phi}}_2
+\tilde{\phi}\tilde{\bar{\phi}})
\end{equation}
Using  a proper-time method (see Section 4) one derives
\begin{equation}
\Gamma^{(1)}_2=\frac{i}{2}tr\int_0^{\infty}\frac{ds}{s}
\exp(is\Delta)
\end{equation}
and  a straightforward computation yields
$$
\exp(is\Delta)=\exp(is\Box)\exp(isA)(1+\sum_{n=1}^{\infty}
\frac{{is\Box(\frac{1}{\xi}-1)}^n}{n!}\frac{1}{16\Box}\{D^2,\bar{D}^2\})
$$
where
$A=g^2(4\tilde{\Phi}_1\tilde{\bar{\Phi}}_1+4\tilde{\Phi}_2
\tilde{\bar{\Phi}}_2
+\tilde{\phi}\tilde{\bar{\phi}})$. A
subsequent  calculation can be carried in a completely parallel
fashion
 as that given in  Appendix A.
As a result we arrive at
\begin{equation}
K^{(1)}_2=\frac{1}{32\pi^2}
\int_0^{\infty}\frac{ds}{s^{2-\epsilon}}\exp(isA)=\frac{1-\xi}{16\pi^2}
A^{1-\epsilon}(\frac{2}{\epsilon}-\gamma+1)
\end{equation}
Here we introduced the dimensional regularization parameter
$\epsilon$ since
the integral is divergent. We also put $\xi=0$, i.e. we work in the
Landau gauge.
Expanding this expression into a power series in $\epsilon$ we
arrive at
\begin{eqnarray}
K^{(1)}_2&=&\frac{1}{32\pi^2}
(\frac{2}{\epsilon}(g^2(4\tilde{\Phi}_1\tilde{\bar{\Phi}}_1+
4\tilde{\Phi}_2\tilde{\bar{\Phi}}_2+\tilde{\phi}\tilde{\bar{\phi}}))
-\nonumber\\&-&
\{
g^2(4\tilde{\Phi}_1\tilde{\bar{\Phi}}_1+4\tilde{\Phi}_2\tilde{\bar{\Phi}}_2
+\tilde{\phi}\tilde{\bar{\phi}})
\log\frac{g^2(4\tilde{\Phi}_1\tilde{\bar{\Phi}}_1+4\tilde{\Phi}_2
\tilde{\bar{\Phi}}_2
+\tilde{\phi}\tilde{\bar{\phi}})}{\mu^2}-
\gamma+1\}
\end{eqnarray}
The divergent part can be cancelled via a suitable counterterm which
leads to
the renormalized correction of  the form
\begin{equation}
K^{(1)}_{2r}=-\frac{1}{32\pi^2}
[g^2(4\tilde{\bar{\Phi}}_1\tilde{\Phi}_1+4\tilde{\bar{\Phi}}_2
\tilde{\Phi}_2
+\tilde{\bar{\phi}}\tilde{\phi})]
\log\frac{g^2(4\tilde{\bar{\Phi}}_1\tilde{\Phi}_1+
4\tilde{\bar{\Phi}}_2\tilde{\Phi}_2+\tilde{\bar{\phi}}\tilde{\phi})}{\mu^2}-
\gamma+1)
\end{equation}
Hence the final result for the one-loop k\"{a}hlerian effective
potential rewritten in terms of standard background chiral
superfields
$\Phi_1,\Phi_2,\phi$ and background gauge superfield $V$ is
\begin{eqnarray}
\label{kk}
K^{(1)}&=&K^{(1)}_{1r}+K^{(1)}_{2r}=\nonumber\\
&=&-\frac{1}{32\pi^2}\Big\{
\lambda^2(\bar{\Phi}_1 e^{2gV}\Phi_1-\bar{\phi}e^{-gV}\phi)
(1+\log\frac{M^2}{\mu^2})+\frac{1}{2M^2}\lambda^4{
(\bar{\Phi}_1 e^{2gV}\Phi_1-\bar{\phi}e^{-gV}\phi)}^2+
\nonumber\\&+&\big[
g^2(4\bar{\Phi}_1 e^{2gV}\Phi_1+4\bar{\Phi}_2 e^{-2gV}\Phi_2+
\bar{\phi} e^{-gV}\phi)\times\nonumber\\&\times&
\big(\log\frac{g^2(4\bar{\Phi}_1 e^{2gV}\Phi_1+4\bar{\Phi}_2
e^{-2gV}\Phi_2
+\bar{\phi} e^{-gV}\phi)}{\mu^2}-
\gamma+1\big)
\big]
\Big\}
\end{eqnarray}
The effective equations of motion for massive superfields  (with
the action
(\ref{actg})
and the  quantum correction $K^{(1)}$ (\ref{kk})) are of the form
\begin{eqnarray}
&-&\frac{\bar{D}^2}{4}\big(e^{2gV}\bar{\Phi}_1+\frac{\partial
K^{(1)}}{\partial\Phi_1}\big)
+M\Phi_2+\beta\phi^2=0\nonumber\\
&-&\frac{\bar{D}^2}{4}\big(e^{-2gV}\bar{\Phi}_2+\frac{\partial
K^{(1)}}{\partial\Phi_2}\big)
+M\Phi_1=0
\end{eqnarray}
Here $\beta=\lambda(1+\frac{\hbar g^2\alpha}{16\pi^2})$ where
$$\alpha=\int_0^1 dx\frac{\log[x(1-x)]}{1-x(1-x)}$$
(see the next subsection)  
is a one-loop corrected coupling for
the chiral interaction. (Quantum corrections to the chiral potential
are considered in details in the next subsection.) Solutions of
these
equations have the form
\begin{eqnarray}
\Phi_1&=&\frac{\bar{D}^2}{4M^2}\Big(\beta\bar{\phi}^2 e^{-2gV}
[1-4\frac{g^2\hbar}{32\pi^2}(1+\log\frac{g^2\bar{\phi}e^{-gV}\phi}{\mu^2})]
\Big)
+O(\frac{1}{M^3})\nonumber\\
\Phi_2&=&-\frac{\beta\phi^2}{M}+O(\frac{1}{M^3})
\end{eqnarray}
As a
result it turns to be that the one-loop contribution to the
k\"{a}hlerian effective potential
can be expressed as
\begin{eqnarray}
\label{klg}
K^{(1)}&=&
-\frac{1}{32\pi^2}\Big\{
-\lambda^2\bar{\phi}e^{-gV}\phi
(1+\log\frac{M^2}{\mu^2})+\nonumber\\&+&
g^2\bar{\phi}e^{-gV}\phi
\log\frac{g^2\bar{\phi}e^{-gV}\phi}{\mu^2}
\Big\}+O(\frac{1}{M^2})
\end{eqnarray}
Here $\gamma+1$ is absorbed into a redefinition of $\mu$.
It is easy to see that this correction is gauge invariant.
We note that the
quantum correction arises in the effective action of light
superfields that depend on
the gauge coupling.
The expression (\ref{klg}) is quite analogous to the minimal model
of  gauge
neutral
chiral superfields  where the
self-interaction of the light chiral superfield is absent.
Namely, the
only
contribution due to the gauge interactions (i.e. terms proportional
to the gauge coupling
$g$) corresponds to the standard   Coleman-Weinberg potential (the
second term,
in the square brackets), which is
 due to the gauge interactions of the light fields only.

The
redefinition $\phi\to \tilde{\phi}=Z^{1/2}\phi$ where
\begin{equation}
Z=1+\lambda^2\frac{\hbar}{32\pi^2}(1+\log\frac{M^2}{\mu^2})
\end{equation}
in turn yields the action  corresponding to the
sector of chiral superfields only (i.e. with the
background gauge superfields put to zero) and  can be written as
\begin{equation}
\Gamma=\int d^8 z |\tilde{\phi}|^2+O(\frac{1}{M^2})
\end{equation}
This result is  again formally consistent with the  decoupling
theorem.
However, since all
parameters (fields, masses, couplings) are determined from string
theory it
turns out to be that the classical action gets an essential
quantum correction given by (\ref{klg}).

\subsection{Chiral potential corrections}
We now turn to the study of possible chiral corrections to the
effective action.
It turns out that unlike in the  model with interacting gauge
neutral chiral
superfields,
 where  the first set of
quantum chiral correction arises at the  two-loop level (see Section
3),
in  $N=1$ supersymmetric gauge theories  the one-loop corrections
to the chiral effective
potential are possible. This situation is quite analogous to
standard
$N=1$ super - Yang - Mills theory with chiral matter \cite{West2}.
However,
we note that in the present theory chiral corrections to effective
action
depend on the massive superfield $\Phi_1$.

In order to study chiral corrections to the effective chiral
potential
 action it is more suitable to use a
Feynman gauge $\xi=1$. In this gauge the propagator for  the gauge
superfield
reads as
\begin{equation}
<vv>=\frac{1}{\Box}\delta^8(z_1-z_2)
\end{equation}

First,
the chiral effective potential depends only on the chiral
background superfields $\phi,\Phi$.
Second, we note that diagrams containing massive propagators cannot
contribute
to the chiral effective potential (see Section 3).
Therefore possible one-loop diagrams can contain  only vertices of
the type
$\bar{\phi}v\phi$ where  $\phi$ is a background superfield, and
those of the
type
$\lambda\Phi_1\phi^2$ where $\Phi_1$ is a background
superfield. Let the number of vertices
$\bar{\phi}v\phi$ be $N_1$, and the number of $\lambda\Phi_1\phi^2$
be $N_2$.
The number of quantum superfields $\phi$ and $\bar\phi$  is equal to
$2N_2$,
and $N_1$, respectively.
Since the only Green function for
massless superfields is $<\phi\bar{\phi}>$, these two numbers should
be equal,
$2N_2=N_1$. Each vertex $\lambda\Phi_1\phi^2$ corresponds to one
factor
$\bar{D}^2$ since it is purely chiral, and each vertex
$\bar{\phi}v\phi$ with
$\phi$ as the external field corresponds to one
factor $D^2$ (see e.g., \cite{BK0}). Then,
a diagram  that can contribute to the  chiral effective potential
has to have
a number of factors $D^2$  that is one larger   than  the
number of factors $\bar{D}^2$
\cite{Buch5}, i.e. $N_1-N_2=1$. Therefore we conclude that $N_1=2$,
$N_2=1$.
The only supergraph  with such a structure is given  below:

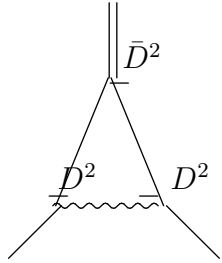
\vspace*{2mm}

\Lengthunit=1cm
\hspace{6cm}
\GRAPH(hsize=3){
\Linewidth{.5pt}
\mov(-.7,-.7){\wavelin(1.4,0)}
\mov(-.1,1){\lin(-.7,-1.7)}
\mov(-.2,1){\lin(.7,-1.7)}
\mov(-1,-.7){\lin(-.7,-.7)}\mov(.4,-.7){\lin(.7,-.7)}
\mov(-0.4,1){\lin(0,1)}\mov(-0.5,1){\lin(0,1)}
\ind(-5,9){-}\ind(-2,12){\bar{D}^2}
\ind(3,-4){D^2}\ind(-14.5,-6){-}\ind(-12,-4){D^2}\ind(-2.5,-6){-}
}

\vspace*{1mm}

This supergraph yields the following contribution
\begin{eqnarray}
J&=&\lambda g^2\int d^4\theta_1 d^4\theta_2 d^4\theta_3
\int\frac{d^4k}{{(2\pi)}^4}\frac{1}{k^2(k-p_1)^2(k-p_2)^2}
\Phi(p_1+p_2,\theta_1)
\phi(-p_1,\theta_2)\phi(-p_2,\theta_3)\times\nonumber\\&\times&
\delta_{12}\frac{\bar{D}^2_1 D^2_3}{12}
\delta_{13}\frac{D^2_2}{4}\delta_{23}
\end{eqnarray}
After $D$-algebra transformation, integration over momenta and
Fourier
transformation
this contribution  has in the  infrared limit $p_1,p_2\to 0$ the
form
(cf. \cite{West2})
\begin{eqnarray}
J&=&\frac{1}{16\pi^2}\lambda g^2\beta\int d^6 z \Phi_1(z)\phi^2(z)
\end{eqnarray}
where
$$\alpha=\int_0^1dx \frac{\log[x(1-x)]}{1-x(1-x)}$$ is a constant
encountered
in \cite{West2}. (Note the same constant is encountered in the
previous
subsection.)
This contribution coincides  in
form with the classical chiral potential.  The one-loop corrected
chiral effective potential is then of the form:
\begin{equation}
W^{(1)}=(\lambda+\frac{\hbar}{16\pi^2}\lambda
g^2\alpha)\Phi_1(z)\phi^2(z)
\end{equation}
However, since the heavy superfield $\Phi$ is at least of first
order in  the inverse
mass (as a solution of the effective equations of motion) this
term is
also of
first order in the  inverse mass. Nevertheless,
this result demonstrates  that quantum corrections to  the
chiral effective potential can depend
not only on massless superfields but also on massive ones.
 (It was commonly
believed that quantum corrections to  the chiral effective potential
 can depend only
on massless superfields.)

\subsection{Strength depending contributions in the effective
action}
In this subsection, we are going to consider  quantum corrections to
the
gauge function, i.e. a holomorphic function that multiplies $W^2$ in
the
 leading  low-energy approximation to the effective action.
To simplify the consideration we again use the Langau gauge $\xi=0$.
 The loops can
contain  only either  propagators for  chiral superfields or
propagators for
vector superfields. However, since the  external vector superfield
can connect  only
to the vertex of the form $v\Phi_i\bar{\Phi}_i$ (or
$v\phi\bar{\phi}$)
the possible
supergraphs must  necessarily contain propagators for the  chiral
superfields
and
therefore  at the one-loop level,
only propagators for  chiral superfields contribute.
(We  also note that in principle it is possible to study
corrections proportional to
$W^2$ in a general gauge.)

Now, we turn to the study of possible supergraphs with two external
gauge lines.
First, we note that the vertex $g\bar{\Phi}_2 v\Phi_2$
can be  associated either with the propagator $<vv>$ or with  the
external vector superfield line.
However, as noted above,  in the  Landau gauge
the loops of the form under
consideration cannot contain propagators $<vv>$. Therefore   the
only
contributing  supergraph containing the vertices
 $v\Phi_2\bar{\Phi}_2$ is of the form

\vspace*{2mm}

\hspace{4cm}
\Linewidth{.5pt}
\GRAPH(hsize=3){\wavelin(1,0)\mov(2,0){\wavelin(1,0)}
\mov(1.5,0){\dashcirc(1)}
\ind(10,3){-}\ind(18,3){-}\ind(10,-3){-}\ind(18,-3){-}
\ind(7,4){D^2}\ind(20,4){\bar{D}^2}\ind(20,-4){D^2}\ind(6,-4){\bar{D}^2}}

\vspace*{2mm}

\noindent Recall that the
dashed line denotes the propagator $<\Phi_2\bar{\Phi}_2>$
(the same on as
for  $<\Phi_1\bar{\Phi}_1>$), and the wavy line denotes the
propagator $<vv>$.  The contribution of the diagram has the form
\begin{eqnarray}
I&=&4g^2\int
\frac{d^4 k}{{(2\pi)}^4}\int d^4\theta_1 d^4\theta_2
V(\theta_1)V(\theta_2)
\frac{1}{(k^2+M^2)^2}\frac{D^2_1\bar{D}^2_2}{16}\delta_{12}
\frac{\bar{D}^2_1 D^2_2}{16}\delta_{12}
\end{eqnarray}
After $D$-algebra transformations and dimensional regularization
this expression is equal to
\begin{eqnarray}
I&=&\frac{4g^2}{16\pi^2}\int d^2\theta \frac{1}{64} W^2
\int\frac{d^n k}{{(2\pi)}^n}
\frac{1}{(k^2+M^2)^2}=\nonumber\\&=&
\frac{4g^2}{16\pi^2}\int d^2\theta \frac{1}{64} W^2
(\frac{2}{\epsilon}-\gamma+\log\frac{M^2}{\mu^2})
\end{eqnarray}
where  $\epsilon$ is the usual dimensional regularization parameter.
The divergent part can be cancelled via a  suitable
counterterm, and the  resulting  correction is
\begin{eqnarray}
\label{i1}
I&=&\frac{4g^2}{16\pi^2}\frac{1}{64}\int d^2\theta  W^2
\log\frac{M^2}{\mu^2}
\end{eqnarray}
The Euler constant $\gamma$
is here absorbed into  a redefinition of the
normalization
parameter $\mu$. As a result we see that this correction leads to a
term proportional to $\log\frac{M^2}{\mu^2}$ which evidently
increases logarithmically with $M$ as  $M\to\infty$.

Now we turn to the study of corrections proportional to $W^2$ which
do not contain
propagators of the field $\Phi_2$. First  of all, let us consider
 corrections at the  zeroth
order in the inverse mass expansion. Since the heavy superfields
$\Phi_1$, $\Phi_2$ (after solving
the effective equations of motion) correspond to  expressions
that are of
the  first and/or
higher orders in the  inverse mass expansion (see the discussions
above), the
corrections that are  of the  zeroth order in $1\over M$ expansion
therefore correspond to
diagrams with massless external chiral superfields only.
Such corrections are described by the supergraphs

\vspace{2mm}

\hspace*{3cm}\GRAPH(hsize=3){\Linewidth{.5pt}\mov(.4,0){\wavelin(1,0)}
\mov(2.2,0){\wavelin(1,0)}
\Linewidth{1.5pt}
\mov(1.5,0){\wavecirc(1)}
\ind(10,3){-}\ind(18,3){-}\ind(10,-3){-}\ind(18,-3){-}
\ind(5,3){D^2}\ind(20,3){\bar{D}^2}\ind(20,-4){D^2}\ind(5,-4){\bar{D}^2}
\ind(10,-10){Fig.2}
\hspace{4cm}\GRAPH(hsize=3){\Linewidth{.5pt}
\mov(.4,0){\wavelin(1,0)}\mov(2.2,0){\wavelin(1,0)}
\Linewidth{1.5pt}
\mov(1.5,0){\dashcirc(1)}
\ind(10,3){-}\ind(18,3){-}\ind(10,-3){-}\ind(18,-3){-}
\ind(5,3){D^2}\ind(20,3){\bar{D}^2}\ind(20,-4){D^2}\ind(5,-4){\bar{D}^2}
\ind(10,-10){Fig.3}
}
}

\vspace{2mm}


Here the bold dashed line (in the following we will denote it  as
$G_1$) is a
propagator of the form

\vspace{2mm}

\hspace*{3cm}\GRAPH(hsize=3){\Linewidth{1.5pt}\dashlin(1,0)\ind(14,0){=}
{\Linewidth{.5pt}}\mov(1.6,0){\dashlin(1,0)}\ind(29,0){+}\mov(3.3,0)
{\dashlin(1,0)}
\mov(5.3,0){\dashlin(1,0)}\mov(4.3,0){\lin(1,0)}\Linewidth{.5pt}
\mov(5.3,0){\lin(0,1)}\mov(4.2,0){\lin(0,1)}\ind(64,0){+}\ind(67,0){\ldots}
}

\vspace{2mm}

and the bold wavy line (in the following we will denote it
 as $G_2$) is a propagator of the form

\vspace{2mm}

\hspace*{3cm}\GRAPH(hsize=3){\Linewidth{1.5pt}\wavelin(1,0)\ind(12,0){=}
{\Linewidth{.5pt}}\mov(1.3,0){\lin(1,0)}\ind(27,0){+}\mov(3,0){\lin(1,0)}
\mov(4,0){\dashlin(1,0)}\mov(5,0){\lin(1,0)}\Linewidth{.5pt}
\mov(5,0){\lin(0,1)}\mov(3.9,0){\lin(0,1)}\ind(61,0){+}\ind(64,0){\ldots}
}

\vspace{2mm}

External lines denote superfields $\phi$.

Consequently, the contributions are of the form
\begin{eqnarray}
G_1&=&<\Phi_1\bar{\Phi}_1>+
\lambda^2|\phi|^2<\Phi_1\bar{\Phi}_1><\phi\bar{\phi}><\Phi_1\bar{\Phi}_1>+
\ldots+\nonumber\\
&+&<\Phi_1\bar{\Phi}_1>(\lambda^2|\phi|^2)^n(<\phi\bar{\phi}>
<\Phi_1\bar{\Phi}_1>)^n+
\ldots\nonumber\\
G_2&=&<\phi\bar{\phi}>+
\lambda^2|\phi|^2<\phi\bar{\phi}><\Phi_1\bar{\Phi}_1><\phi\bar{\phi}>
+\ldots+
\nonumber\\
&+&<\phi\bar{\phi}>(\lambda^2|\phi|^2)^n(<\Phi_1\bar{\Phi}_1>
<\phi\bar{\phi}>)^n+
\ldots
\end{eqnarray}
The superpropagators $<\phi\bar{\phi}>$, and
$<\Phi_1\bar{\Phi}_1>$ are given by
(\ref{prq}) and (\ref{prt}), respectively.

A straightforward calculation yields
\begin{eqnarray}
G_1&=&-\frac{D^2\bar{D}^2}{16
k^2}\frac{k^2+M^2}{k^2+M^2+|\phi|^2}\nonumber\\
G_2&=&-\frac{D^2\bar{D}^2}{16 (k^2+M^2+|\phi|^2)}
\end{eqnarray}

As a result, contributions from the diagrams shown in  Fig.2 and
Fig.3
correspond to
\begin{eqnarray}
I_2&=&g^2\int
\frac{d^4 k}{{(2\pi)}^4}\int d^4\theta_1 d^4\theta_2
V(\theta_1)V(\theta_2)
\frac{{(k^2+M^2)}^2}{(k^2(k^2+M^2+|\phi|^2))^2}\frac{D^2_1\bar{D}^2_2}{16}
\delta_{12}
\frac{\bar{D}^2_1 D^2_2}{16}\delta_{12}\nonumber\\
I_3&=&4g^2\int
\frac{d^4 k}{{(2\pi)}^4}\int d^4\theta_1 d^4\theta_2
V(\theta_1)V(\theta_2)
\frac{1}{(k^2+M^2+|\phi|^2)^2}\frac{D^2_1\bar{D}^2_2}{16}\delta_{12}
\frac{\bar{D}^2_1 D^2_2}{16}\delta_{12}
\end{eqnarray}
respectively. After $D$-algebra transformations, integration
 over momenta, and a subtraction of
divergences we find
\begin{eqnarray}
\label{i56}
I_2&=&\frac{g^2}{16\pi^2}\frac{1}{64}\int d^2\theta W^2
(\log\frac{M^2+|\phi|^2}{\mu^2}+A)\nonumber\\
I_3&=&\frac{4g^2}{16\pi^2}\frac{1}{64}\int d^2\theta W^2
(\log\frac{M^2+|\phi|^2}{\mu^2})
\end{eqnarray}
Here $A=1+\int_{\alpha}^1 dx \frac{1-x}{x}$ is a constant  that
depends on
the infrared regularization parameter $\alpha$. We note that this
parameter
can  also be  absorbed into  a redefinition of  the renormalization
parameter $\mu$.

The total correction in the gauge sector is a sum of corrections
$I_1$ (\ref{i1}),
$I_2$ and $I_3$ (\ref{i56}). It has the form
\begin{eqnarray}
I&=&I_1+I_2+I_3=\frac{g^2}{64}\frac{1}{16\pi^2}\int d^2\theta
W^2\big(4 \log\frac{M^2}{\mu^2}+ (\log\frac{M^2+|\phi|^2}{\mu^2}+A)+
\nonumber\\&+&
4\log\frac{M^2+|\phi|^2}{\mu^2}
\big)
\end{eqnarray}
The leading term of this correction is
\begin{eqnarray}
I&=&\frac{9 g^2}{64}\frac{1}{16\pi^2}\int d^2\theta W^2
\log\frac{M^2}{\mu^2}
+O\Big(\frac{1}{M^2}\Big)
\end{eqnarray}
Here the constant $A$ is absorbed into a   redefinition of the
normalization
parameter $\mu$.

Let us  now consider contributions  that depend on external heavy
chiral
superfields. We assume that the supergraph can contain $n_1$
vertices of
the form $\Phi_1\phi_q^2$, $\bar{n}_1$ vertices of the form
$\bar{\Phi}_1\bar{\phi}_q^2$, $n_2$ vertices
of the form $\phi\Phi_{1q}\phi_q$,
and $\bar{n}_2$ vertices of the form
$\bar{\phi}\bar{\Phi}_{1q}\bar{\phi}_q$.
Thus, the  number of external
$\Phi_1$ is equal to $n_1$, and the number of external $\bar{\Phi}$
is equal
to $\bar{n}_1$.
We note that there are no propagators of heavy superfields $\Phi_2$
in
the supergraphs under consideration (in the  Landau gauge),
therefore the only
possible supergraphs are $<\phi\bar{\phi}>$ and
$<\Phi_1\bar{\Phi}_1>$.
Hence the number of quantum superfields $\phi$ and $\bar{\phi}$, and
correspondingly, the number of  $\Phi_{1q}$ and $\bar{\Phi}_{1q}$
must be equal,
and we arrive at the relations $n_2=\bar{n}_2$,
$2n_1+n_2=2\bar{n}_1+\bar{n}_2$.
These relations leads us to conclude that $n_1=\bar{n}_1$ and
 therefore contributions
of supergraphs must be proportional to
${(\Phi_1\bar{\Phi}_1)}^{n_1}$. Since
the solution of the effective equations of motion  yields the result
 that  the heavy superfield
$\Phi_1$ is at least of the second order in $\frac{1}{M}$ (see
above)
we conclude that terms with a  non-trivial dependence on  the
background field
$\Phi_1$ must be of the fourth order in $\frac{1}{M}$.

As a result, the one-loop corrected effective action is the gauge
sector
if of the form:
\begin{eqnarray}
\label{acg}
\Gamma^{(1)}[W]=\frac{1}{64}\int d^6 z W^2
(1+9
g^2\frac{\hbar}{16\pi^2}\log\frac{M^2}{\mu^2})+O\Big(\frac{1}{M^4}\Big)
\end{eqnarray}
This expression contains terms logarithmically increasing with  $M$
and thus the gauge function $f$ can get significant corrections due
to the
contributions of the heavy fields. Note, however, that these
corrections can be
reinterpreted as   the  standard one-loop  threshold corrections due
to
 the contributions of the heavy fields.  Namely, these leading
effects are due to the
 coupling of the vector gauge fields to  the heavy chiral
superfields, and are
 therefore proportional to the gauge coupling $g$. On the other hand
  a  contribution due to  the coupling $\lambda$ between light and
heavy chiral
 superfields is absent at this order. (It appears only at  higher
orders in the
 $1\over M$ expansion.)

Again it is straightforward to show that the obtained result is
in compliance with the decoupling theorem.
We redefine  the vector  superfield $V$ (and consequently the
superfield strength
$W^{\alpha}$) by the rule $\tilde{V}=Z_1^{1/2}V$ where $Z_1$ is a
finite
renormalization constant
\begin{equation}
Z_1=1+9 g^2\frac{\hbar}{16\pi^2}\log\frac{M^2}{\mu^2}
\end{equation}
Then the  expression (\ref{acg}) takes the form
\begin{eqnarray}
\label{acg1}
\Gamma^{(1)}[W]=\frac{1}{64}\int d^6 z W^2 +O\Big(\frac{1}{M^2}\Big)
\end{eqnarray}
which is again of the same form as the corresponding term in the
classical
action.
On the other hand the  terms that include the
 interaction between  the  vector superfield and the chiral ones
should stay invariant under such a redefinition of fields.
 Therefore we must redefine the gauge
coupling $g$ by the rule $\tilde{g}=Z_1^{-1/2}g$. After these
redefinitions
the one-loop effective action takes the form
\begin{equation}
\Gamma^{(1)}=\frac{1}{64}\int d^6 z\tilde{W}^2+
\int d^8
z\tilde{\bar{\phi}}e^{-\tilde{g}\tilde{V}}\tilde{\phi}+O(\frac{1}{M})
\end{equation}
As a result the one-loop corrected effective action is of the same
form as
the classical action of light (massless) superfields $\phi , v$,
only, as
expected by the
decoupling theorem.
However, since the values of fields and couplings are determined
from string
theory, they cannot be redefined, and the final result for the
one-loop corrected
effective action in the sector of gauge superfields
is thus given by (\ref{acg}) which allows us to conclude that
the classical action  in this sector gets essential corrections.
couplings

\section{Summary}
\setcounter{equation}{0}
\renewcommand{\theequation}{\arabic{section}.\arabic{equation}}

In this paper we presented a systematic analysis of the decoupling
effects in
$N=1$ supersymmetric theories with chiral superfields.
We developed techniques to evaluate explicitly the one-loop
corrected actions
that involve both the heavy and light chiral superfields for
different types of
classical k\"ahlerian and chiral potentials. By a subsequent
elimination of
the heavy chiral superfields by their equations of motion,
 via an iterative procedure,  the
resulting effective actions of light chiral
superfields include the quantitative  decoupling corrections
and the
one-loop level. We considered different examples of theories
describing dynamics of interacting
light (massless) and heavy superfields.
For these theories the  one-loop
k\"ahlerian effective potential was calculated  and subsequently the
heavy superfields were expressed via light ones up to a certain
order
in  the inverse (heavy) mass parameters, thus resulting in an
explicit
form of the  effective action  (up to a certain order in the inverse
mass
expansion) for
light superfields, only.

As a representative example we  carried out a detailed analysis
 for  the so-called minimal model. This model  contains  two
(gauge
neutral) chiral superfields, one heavy and  one massless,  while the
classical
 k\"ahlerian
potential contains only the renormalizable (canonical) terms and the
classical
chiral
potential is renormalizable  with the self-interaction term
of the light fields and
an interaction term linear in the heavy field. The results of  these
calculations  demonstrate that the leading order decoupling
effects at
the loop-level
grow logarithmically with the heavy mass scale. In addition, we also
analysed the loop corrections
to the chiral potential, and found that at the two-loop level
corrections
to the heavy fields  do appear, which are, however, suppressed  in
the effective
action of light fields by inverse powers
of the heavy mass scales.

In the subsequent section we carried out the analysis of the
non-minimal models
with a more general structure of the k\"ahlerian and chiral
potentials, that
involve one  light and one massive field. While  the analysis of
these models cannot be carried out explicitly for a general choice
of the
classical k\"ahlerian and chiral potential, we developed techniques
to
determine the effective action as a series expansion in the inverse
powers of
the heavy mass parameter.
We have also addressed the  decoupling effects for $N=1$
supersymmetric gauge
theories with heavy and light chiral fields charged  under the gauge
group.
In particular, we analysed the  $U(1)$-Abelian (minimal) model with
charged heavy and  light chiral superfields, a minimal (gauge
invariant)
k\"aherian potential and a  chiral potential with an interaction
term linear
in a heavy field. The leading corrections to the k\"ahlerian
potential are independent of the  gauge
coupling, however higher order terms (both for the  one-loop
corrected
k\"ahlerian and chiral potential) are proportional to the gauge
coupling. In
addition we also demonstrated that the leading correction to the
gauge function
is proportional to the  gauge coupling, which can be reinterpreted
as a
standard threshold effect due to the heavy charged chiral
superfields. (Note
however that in the leading order  the gauge function
corrections do not depend on the  chiral coupling of the heavy
chiral superfields to the light ones.)

While our specific results are, of course, in agreement with the
decoupling
theorem, the actual quantitative form of the decoupling  effects may
have
important physics implications.
As it  was stressed in the introduction,   in an effective
field theory  (without reference  to an underlying  fundamental
theory)
the couplings are free parameters, fixed by (low-energy)
experiments,
and thus the above decoupling   effects can indeed be all absorbed
into  rescalings
 of the effective light fields. However, within string theory
the original
 couplings are calculable at the string scale
  and the  corrections due to  decoupling  can now have sizable
 calculable predictions for the low energy couplings.
 The results of our calculations  explicitly  confirm that in
 $N=1$ supersymmetric theories  the decoupling effects at
 loop-level
are  of order $\log\ M$, and in specific  cases they
 can   significantly
modify  the low energy predictions of the effective couplings.

In particular, within a class of perturbative
 string models with an anomalous
$U(1)$, after the vacuum restabilization, there are in  general
renormalizable  interactions between the light and heavy fields,
with
mass of order
 $M\sim
 M_{string}\sim
(10^{17})$ GeV.
 Thus, the above one-loop decoupling effects
can significantly change
  the low energy predictions for the couplings at low energies,
 e.g.,
the electro-weak scale
 ($ \mu\sim 1$ TeV). Let us repeat the analysis for an example, in a
class of string models discussed in
 \cite{cceel2}--\cite{new},  where  typical values of the
  couplings, calculated at $M_{string}$, are
$\lambda =g=g_{gauge}$. Here
$g_{gauge}\sim 0.8 $ is the value of the gauge coupling at
at $M_{string}$.  Renormalization group equations then determine
the values of $\lambda$ and $g$ couplings at low energies $\mu$.
However, due to the one-loop decoupling effects there is now
also an additional correction in the effective coupling
$g$ for the light fields; it is of the order
 of  $\lambda (\mu )^2/(16\pi^2)\log M^2/\mu^2\sim 0.26$. (We used
the typical values $M\sim M_{string}\sim 10^{17}$ GeV, $\mu\sim 1$
TeV and
$\lambda(\mu)\sim
0.8$.) This specific example
   indicates that  for a class of four-dimensional string vacua
 the actual prediction  for the tri-linear couplings at low energy
could be
 corrected by
  $10-50\%$.

 Let us conclude with a remark about the leading contributions at
higher loop
 levels. These contributions are expected to be
 of the form (cf.\cite{Buch4}):
 the form
$$|\phi|^2\log^n\big(\frac{|\phi|^2}{\mu^2}\big).
$$
When choosing a  renormalization condition of the form
$\mu^2=const\times M^2$ (analogous to the one  used in Section 3)
the
typical contribution of these terms would be of the type
$$|\phi|^2\log^n\big(\frac{|\phi|^2}{M^2}\big)
$$
thus implying that
the decoupling effects at the higher loop  leading decoupling
effects
 could be just as significant,
and  they would further modify low-energy predictions of the
effective theory
of light fields.

\vspace*{1mm}

{\bf Acknowledgments.} M.C. acknowledges discussions with  L.
Everett and P.
Langacker. The work by I.L.B. and A.Yu.P. was supported
in part by INTAS grant, INTAS-96-0308; RFBR grant, project
No 99-02-16617; RFBR-DFG grant, project No 99-02-04022; GRACENAS
grant,
project
No 97-6.2-34.  The work of M.C.\ was supported in part by U.S.
Department of
Energy Grant No.  DOE-EY-76-02-3071. I.L.B. and M.C. would like to
thank the
organizers of the 32nd
International Symposium Ahrenshoop on the Theory of Elementary
Particles,
where the work was initiated,  for
hospitality.

\renewcommand{\thesection}{}
\section{Appendix A}
\setcounter{equation}{0}
\renewcommand{\theequation}{A.\arabic{equation}}
To find  an exact form of the solution of (\ref{syst}) for the model
(\ref{stot}) it is more convenient to
study  a system of equations for  components of matrices $A$ and
$C$.
Therefore we arrive at
\begin{eqnarray}
\frac{1}{i}\dot{A}_{11}&=&-C_{11}(\lambda\Phi+g\phi) - C_{12}
\lambda\Phi\nonumber\\
\frac{1}{i}\dot{C}_{11}&=&(\lambda\bar{\Phi}+g\bar{\phi})-
\frac{1}{i}A_{11} (\lambda\bar{\Phi}+g\bar{\phi})\Box - A_{12}
\lambda\bar{\phi}\Box
\nonumber\\
\frac{1}{i}\dot{A}_{12}&=&-C_{11} \lambda\phi - C_{12} M\nonumber\\
\frac{1}{i}\dot{C}_{12}&=&\lambda\bar{\phi}-A_{11} \bar{\phi}\Box
- A_{12} M\Box\nonumber\\
\frac{1}{i}\dot{A}_{21}&=&-C_{21} (\lambda\Phi+g\phi) - C_{22}
\lambda\phi\nonumber\\
\frac{1}{i}\dot{C}_{21}&=&\lambda\bar{\phi}-A_{21}
(\lambda\Phi+g\phi)\Box -
 A_{22} \lambda\bar{\phi}\Box\nonumber\\
\frac{1}{i}\dot{A}_{22}&=&-C_{21} \lambda\phi - C_{22} M\nonumber\\
\frac{1}{i}\dot{C}_{22}&=&M-A_{21} \lambda\bar{\phi}\Box - A_{22}
M\Box
\end{eqnarray}
with an  analogous system of equations for $\tilde{A},\tilde{C}$.
It is straightforward  to solve  this system of equation
 by the method described described in subsection 2.3. It
allows  one to obtain the following form of the coefficients
$A_{11},A_{22},
\tilde{A}_{11},\tilde{A}_{22}$ (i.e. only  those contribute to the
$tr\log\Delta$):
\begin{eqnarray}
A_{11}&=&\tilde{A}_{11}=\frac{N_3}{(N_3-N_1)\Box} cosh (i\omega_1
s)-
\frac{N_1}{(N_3-N_1)\Box} cosh (i\omega_2
s)-\frac{1}{\Box}\nonumber\\
A_{22}&=&\tilde{A}_{22}=\frac{N_1}{(N_1-N_3)\Box} cosh (i\omega_1
s)-
\frac{N_3}{(N_1-N_3)\Box} cosh (i\omega_2
s)-\frac{1}{\Box}\nonumber\\
\end{eqnarray}
Here $\omega_1=\sqrt{P+ \sqrt{Q}}\sqrt{\Box}$,
$\omega_2=\sqrt{P- \sqrt{Q}}\sqrt{\Box}$,\
$N_1=
\frac{\omega_1^2-(|\lambda\Phi+g\phi|^2+2\lambda^2|\phi|^2)\Box}
{\lambda\bar{\phi}(\lambda\Phi+g\phi)+M \lambda\phi}$,\\
$N_3=
\frac{\omega_1^2-(|\lambda\Phi+g\phi|^2+2\lambda^2|\phi|^2)\Box}
{\lambda\bar{\phi}(\lambda\Phi+g\phi)+M \lambda\phi}$, and
\begin{eqnarray}
\label{rpq}
P&=&|\lambda\Phi+g\phi|^2+2\lambda^2|\phi|^2
+M^2\nonumber\\
Q&=&{(|\lambda\Phi+g\phi|^2-M^2)}^2+
4|(\lambda\Phi+g\phi)\lambda\bar{\phi}+\lambda\phi M|^2
\end{eqnarray}
Let us denote
\begin{equation}
\label{rpq1}
R_1=\sqrt{P+ \sqrt{Q}},\ R_2=\sqrt{P- \sqrt{Q}}
\end{equation}
The trace of $e^{is\tilde{\Delta}}$ is determined by the matrix
trace of A
which reads as
\begin{equation}
tr A= A_{11}+A_{22}=\frac{1}{\Box}(\cosh(is R_1\sqrt{\Box})+\cosh(is
R_2\sqrt{\Box})-2)
\end{equation}
The one-loop k\"{a}hlerian effective potential has the form
\begin{equation}
\label{kahl2}
K^{(1)}=-\frac{i}{2}\int_{0}^{\infty}\frac{ds}{s}(A_{11}(s)+A_{22}(s))
U(x,x';s)|_{x=x'}
\end{equation}
since $A_{11}=\tilde{A}_{11}$, $A_{22}=\tilde{A}_{22}$.
Here $U(x,x';s)=e^{is\Box}\delta^4(x-x')$ (cf. \cite{Buch1,Buch2}).
Expanding $A_{11}(s)+A_{22}(s)$ as  a power series in $\Box$ we
find the
one-loop k\"{a}hlerian effective potential in the form
\begin{eqnarray}
K^{(1)}=-\frac{i}{2}\int_{0}^{\infty}\frac{ds}{s}
\sum_{n=0}^{\infty}
\big(\frac{{(iR_1)}^{2n+2}}{(2n+2)!}+
\frac{{(iR_2)}^{2n+2}}{(2n+2)!}\big)s^{2n+2}
\Box^{n}U(x,x';s)|_{x=x'}\nonumber\\
\end{eqnarray}
As usual (see \cite{Buch1,Buch2})
$$\Box^{n}U(x,x';s)|_{x=x'}=
\frac{{(-1)}^{n}(n+1)!}{16\pi^2{(is)}^{n+2}}$$
Thus
the one-loop k\"ahlerian effective potential can be cast in  the
form
\begin{eqnarray}
K^{(1)}=-\frac{1}{32\pi^2}\int_0^{\infty} \frac{ds}{s}
\sum_{n=0}^{\infty}
\big(\frac{{R_1}^{2n+2}}{(2n+2)!}+
\frac{{R_2}^{2n+2}}{(2n+2)!}\big)(n+1)!{(-1)}^n
\frac{{(is)}^{2n+2}}{{(is)}^{n+2}}
\end{eqnarray}
Since this integral is divergent we  use the  dimensional
regularization
 by introducing the  regularization
parameter $\epsilon$ with the  subsequent changes:
$\int_{0}^{\infty}\frac{ds}{s} \to
\int_{0}^{\infty}\frac{ds}{s^{1-\epsilon}}$. As a result we arrive
at
\begin{eqnarray}
K^{(1)}&=&-\frac{1}{32\pi^2}\Big[
R^{2(1+\epsilon)}_1
\int_0^{\infty} \frac{d (isR^2_1)}{{(isR^2_1)}^{1-\epsilon}}
\sum_{n=0}^{\infty}
\frac{{(is R_1^2)}^n (n+1)!{(-1)}^n}{(2n+2)!}+\nonumber\\&+&
R^{2(1+\epsilon)}_2
\int_0^{\infty} \frac{d (isR^2_2)}{{(isR^2_2)}^{1-\epsilon}}
\sum_{n=0}^{\infty}
\frac{{(isR_2^2)}^n (n+1)!{(-1)}^n}{(2n+2)!}\Big]
\end{eqnarray}
Using (cf. \cite{Buch1,Buch2})
$$
\sum_{n=1}^{\infty}s^n
\frac{{(-1)}^{n-1}n!}{(2n)!}=s\int_0^1 dt e^{-\frac{1}{4}s(1-t^2)}
$$
one obtains
\begin{eqnarray}
K^{(1)}&=&
-\frac{1}{16\pi^2}
\Big\{{(R^2_1)}^{1+\epsilon}
\int_0^{\infty}\frac{ds}{s^{1-\epsilon}}\int_0^1 dt
e^{-\frac{1}{4}s(1-t^2)}
+\nonumber\\&+&
{(R^2_2)}^{1+\epsilon}
\int_0^{\infty}\frac{ds}{s^{1-\epsilon}}\int_0^1 dt
e^{-\frac{1}{4}s(1-t^2)}
\Big\}
\end{eqnarray}
Here, we have  made  a redefinition $s\to sR^2_1$ in  the
first term and $s\to sR^2_2$ in
the second one  along with a  subsequent Wick rotation.
Furthermore
\begin{eqnarray}
& &{(R^2_1)}^{1+\epsilon}\int_0^{\infty}\frac{ds}{s^{1-\epsilon}}
\int_0^1 dt e^{-\frac{1}{4}s(1-t^2)}=
{(R^2_1)}^{1+\epsilon}\Gamma(\epsilon)
\int_0^1 dt (1-\epsilon \log (\frac{1-t^2}{4})+O(\epsilon^2))=
\nonumber\\&=&
{(R^2_1)}^{1+\epsilon}[\frac{1}{\epsilon}+\gamma-
\int_0^1 dt \log (\frac{1-t^2}{4})+O(\epsilon)]
\end{eqnarray}
Here $O(\epsilon)$   denotes  terms of order
one and/or higher in $\epsilon$,
and $O(\epsilon^2)$  denotes terms of  orders two and/or higher
in
$\epsilon$.
Therefore
\begin{eqnarray}
\label{kahlr}
K^{(1)}&=&-\frac{1}{32\pi^2}
\Big\{
\frac{R^2_1+R^2_2}{\epsilon}+R^2_1\log\frac{R^2_1}{\mu^2}+
R^2_2\log\frac{R^2_2}{\mu^2}\Big\}
\end{eqnarray}
Here the first term contains all one-loop divergences of
the theory which can be
cancelled by a counterterm of the form
\begin{equation}
K^{(1)}_{countr}=\frac{1}{32\pi^2\epsilon}(R^2_1+R^2_2)
\end{equation}
The terms containing $\gamma$ and $\int_0^1 dt \log
(\frac{1-t^2}{4})$ are
removed by a suitable choice of the renormalization
parameter $\mu$.
Using the  exact expression for $R_1$ and $R_2$ ((\ref{rpq}) and
(\ref{rpq1}),
respectively)
we arrive at the result (\ref{kahl1}).
\section{Appendix B}
\setcounter{equation}{0}
\renewcommand{\theequation}{B.\arabic{equation}}
We note
that the result (\ref{kahl1}) for the one-loop correction in the
k\"{a}hlerian
effective potential can also be obtained via a  diagram technique.

Let us study possible diagrams contributing to the one-loop
k\"{a}hlerian
effective potential in this model.
The propagators of the superfields have the standard form \cite{BK0}
\begin{eqnarray}
\label{props}
<\phi(z_1)\bar{\phi}(z_2)>&=&-\frac{1}{\Box}\delta^8(z_1-z_2)\nonumber\\
<\Phi(z_1)\bar{\Phi}(z_2)>&=&-\frac{1}{\Box-M^2}\delta^8(z_1-z_2)\nonumber\\
<\Phi(z_1)\Phi(z_2)>&=&-\frac{M\bar{D}^2}{4\Box(\Box-M^2)}\delta^8(z_1-z_2)
\end{eqnarray}
The possible vertices (see (\ref{spot})) are
$\frac{g}{3!}\phi^3$ and $\lambda\Phi\phi^2$.
It is easy to see that there are the following types of possible
supergraphs.
The first type of supergraphs consists of
$<\phi\bar{\phi}>$-propagators with an alternating background $\phi$
and
$\bar{\phi}$ (cf. \cite{PW,GR}).
Here the external legs correspond to alternating $g\phi$ and
$g\bar{\phi}$.

\vspace*{3mm}

\hspace{1.5cm}
\Lengthunit=.5cm
\GRAPH(hsize=3){
\Linewidth{.5pt}
\Circle(2)\mov(-1,0){\lin(-1,0)}\mov(1,0){\lin(1,0)}
\mov(8,0){\GRAPH(hsize=3){
\mov(.5,0){\Circle(2)\mov(-1,0){\lin(-1,0)}
\mov(1,0){\lin(1,0)}
\mov(-.2,1){\lin(0,1)}\mov(-.2,-1){\lin(0,-1)} }}
\mov(4,0){\GRAPH(hsize=3){
\mov(.5,0){\Circle(2)\mov(-1,.3){\lin(-.8,.3)}
\mov(.8,.3){\lin(.8,.3)}
\mov(-.2,1){\lin(0,1)}\mov(-.2,-1){\lin(0,-1)}
\mov(-1.7,-.3){\lin(-.8,-.3)}
\mov(0,-.3){\lin(.8,-.3)}
}
\mov(5,0){\ldots}}}}}

\vspace*{3mm}

\noindent The result represents itself as a sum
of all such supergraphs.
The diagram of such a form with $2n$ external legs corresponds to a
ring
containing  $n$ links of the form

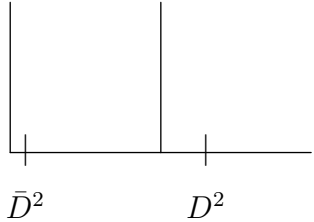
\vspace{2mm}

\Lengthunit=2cm
\hspace{2cm}
\GRAPH(hsize=3){
\mov(.5,0){
\mov(1,0){\lin(1,0)\lin(0,1)}\mov(2,0){\lin(1,0)\lin(0,1)}
}
\ind(16,-4){\bar{D}^2}\ind(16,0){|}\ind(28,-4){D^2}\ind(28,0){|}
}

\vspace{2mm}

\noindent A contribution of such a link is of the form
\begin{equation}
R=\frac{g^2|\phi|^2}{k^4}
\frac{D^2_{l-1}}{4}\delta_{l-1,l}\frac{\bar{D}^2_{l}}{4}\delta_{l,l+1}
\end{equation}
Here, superfields $W'', \bar{W}''$ are treated as  constants, and
external
momenta are put to zero.
Let us denote the contribution of a supergraph
with $2n$ external lines as $I_{2n}$. It is easy to see that
\begin{eqnarray}
I_{2n}&=&\frac{1}{2n}\int\prod_{k=1}^{2n}
d^4\theta_1...d^4\theta_{2n}
\int \frac{d^4 k}{{(2\pi)}^4}\frac{1}{2n} R^n=\nonumber\\
&=&\frac{1}{2n}\int\prod_{k=1}^{2n} d^4\theta_1...d^4\theta_{2n}
\int \frac{d^4 k}{{(2\pi)}^4}
{g^2|\phi|^2}^n {(\frac{1}{k^2})}^{2n}
\times\nonumber\\&\times&
\frac{D^2_1 }{4}\delta_{12} \frac{\bar{D}^2_2}{4}\delta_{23}...
\frac{D^2_{2n-1}}{4}\delta_{2n-1,2n}\frac{\bar{D}^2_{2n}}{4}\delta_{2n,1}
\end{eqnarray}
Here $2n$ is a symmetric factor (see \cite{PW}).

Straightforward $D-$algebra transformations lead to
\begin{equation}
I_{2n}= \int d^4\theta\int \frac{d^4 k}
{{(2\pi)}^4 k^2}
\frac{1}{2n}
{\big(-\frac{g^2|\phi|^2}{k^2}\big)}^n
\end{equation}
Here we used the rule $D^2\bar{D}^2 D^2=-16 k^2 D^2$.

The total contribution of all these diagrams
which is further denoted as $K^{(1)}_0$
is a sum of all $I_{2n}$ (cf. \cite{PW}), i.e.
\begin{equation}
K^{(1)}_0=\sum_{n=1}^{\infty} I_{2n}
=\int d^4 \theta \frac{d^4 k}{{(2\pi)}^4 k^2}
\sum_{n=1}^{\infty}\frac{1}{2n}
{\Big(-\frac{g^2|\phi|^2}{K^2_{\Phi\bar{\Phi}}k^2}\Big)}^n
\end{equation}
Then, since
$$
\sum_{n=1}^{\infty}\frac{1}{2n} {(-a)}^n=-\log(1+a),
$$
$K^{(1)}_0$ is of the form
\begin{equation}
K^{(1)}_0=\sum_{n=1}^{\infty} I_{2n}
=-\int d^4\theta \frac{d^4 k}{{(2\pi)}^4 k^2}
\log\big(1+\frac{g^2|\phi|^2}{k^2}\big)
\end{equation}
We can integrate over angular coordinates which allows us to
change $d^4 k$ for $\pi^2 r dr$ where $r=k^2$. Then, since the
integral over
$r$ is divergent we carry out dimensional regularization
introducing the  regularization parameter $\epsilon$. Namely, we
change $\pi^2 r dr\to\pi^2 r^{1+\epsilon/2} dr$. As a result we
arrive at
\begin{equation}
K^{(1)}_0=-\mu^{-\epsilon}\int d^4 \theta \frac{dr
r^{\epsilon/2}}{32\pi^2}
\log\Big(1+\frac{g^2|\phi|^2}{ K^2_{\Phi\bar{\Phi}}r}\Big)
\end{equation}
Then,
$$
\int_0^{\infty} dr r^{\epsilon/2}
\log(1+\frac{A}{r})=-A^{1+\epsilon}\Gamma(-1-\epsilon/2)=
-\frac{2}{\epsilon}A-A\log\frac{A}{\mu^2}+O(\epsilon)
$$
Inserting  $A=g^2|\phi|^2$ we arrive at the one-loop
contribution to the effective action 
\begin{equation}
\label{cr0}
K^{(1)}_0=-\frac{1}{32\pi^2}g^2|\phi|^2
\big(\frac{2}{\epsilon}+\log\frac{g^2|\phi|^2}{\mu^2}\big)
\end{equation}
(see \cite{West2}).

The most effective method to study  diagrams that contain vertices
proportional to $\Phi\phi^2$   consists
of introducing   the propagator for the  light superfield $\phi_q$
which depends  on the  background light superfield $\phi$
 and  corresponds to the classical action (\ref{stot}).
This propagator is denoted by a bold line as

\vspace{2mm}

\Lengthunit=.8cm
\hspace{1.5cm}
\GRAPH(hsize=3){
\mov(.5,0){\Linewidth{1.5pt}
\mov(.5,0){\lin(1,0)}\ind(21,0){=}\Linewidth{0.5pt}\mov(2.5,0){\lin(1,0)}
\ind(39,0){+}\mov(4.3,0){\lin(3,0)}\mov(5.3,0){\lin(0,1)}
\mov(6.3,0){\lin(0,1)}
\ind(75,0){+}\ind(79,0){\ldots}\ind(92,0){(}
\mov(8.1,0){\lin(1,0)}\mov(9.2,0){\lin(0,1)\lin(2,0)}\mov(10.1,0){\lin(0,1)}
\ind(112,0){)^n}\ind(116,0){+}\ind(120,0){\ldots}
}}

\vspace{2mm}

\noindent A summation  of this chain allows  one to show that the
total
propagator of the light quantum  superfield depends on  the
background light superfield in the following way
\begin{eqnarray}
<\phi_q(z_1)\bar{\phi}_q(z_2)>=-\frac{\bar{D}^2_1
D^2_2}{16(\Box-g^2|\phi|^2)}
\delta_{12}
\end{eqnarray}
The propagator for  heavy superfield $\Phi$ is  given by
(\ref{props}).
Then we can see that aside from the diagrams discussed above
there are two other types
of supergraphs. The first one  consists of  an
arbitrary number of repeating chains
of the form $<\Phi\bar{\Phi}><\phi\bar{\phi}>$, i.e.

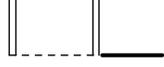
\vspace{2mm}

\Lengthunit=.8cm
\hspace{2cm}
\GRAPH(hsize=3){\Linewidth{0.5pt}
\mov(.5,0){\dashlin(1.4,0)\lin(0,1)\mov(.11,0){\lin(0,1)}
\Linewidth{1.5pt}
\mov(1.2,0){\lin(1,0)}
\Linewidth{0.5pt}
\mov(.7,0){\lin(0,1)\mov(.1,0){\lin(0,1)}}
}}

\vspace{2mm}

\noindent where  double lines 
correspond to
an external alternating $\lambda\Phi$ and $\lambda\bar{\Phi}$, and
bold one to $<\phi\bar{\phi}>$ propagators.  A contribution from a
supergraph consisting of $n$ such chains, because of
(\ref{props}), is
of the form \begin{equation}
K_n=\frac{1}{n}{\Big(\frac{\lambda^2|\Phi|^2\bar{D}^2 D^2}
{{16(\Box-g^2|\phi|^2)(\Box-M^2)}}
\Big)}^n
\end{equation}

The second type  consists of  an arbitrary number of
 repeating chains of the form\\
$
<\phi\bar{\phi}><\Phi\bar{\Phi}><\phi\bar{\phi}><\bar{\Phi}\bar{\Phi}>
<\phi\bar{\phi}><\bar{\Phi}\Phi><\bar{\phi}\phi><\Phi\Phi>
$
which can be written as

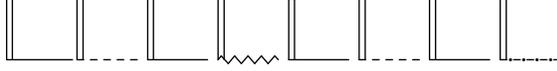
\vspace{2mm}

\Lengthunit=.8cm
\hspace*{2.5cm}
\GRAPH(hsize=3){\lin(0,1)\mov(.1,0){\lin(0,1)}\lin(1.1,0)\Linewidth{0.5pt}
\mov(1,0){\dashlin(1,0)\lin(0,1)\mov(.1,0){\lin(0,1)}\Linewidth{0.5pt}}
\mov(2,0){\lin(1,0)\lin(0,1)\mov(.1,0){\lin(0,1)}\Linewidth{0.5pt}}
\mov(3,0){\trianglin(1,0)\lin(0,1)\mov(.1,0){\lin(0,1)}\Linewidth{0.5pt}}
\mov(4,0){\lin(1,0)\lin(0,1)\mov(.1,0){\lin(0,1)}\Linewidth{0.5pt}}
\mov(5,0){\dashlin(1,0)\lin(0,1)\mov(.1,0){\lin(0,1)}\Linewidth{0.5pt}}
\mov(6,0){\lin(1,0)\mov(.1,0){\lin(0,1)}\lin(0,1)\Linewidth{0.5pt}}
\mov(7,0){\dashdotlin(1,0)\lin(0,1)\mov(.1,0){\lin(0,1)}\Linewidth{0.5pt}}
}

\vspace{2mm}

\noindent Here a dashed-and-dotted line denotes $<\Phi\Phi>$ and a
jagged line denotes $<\bar{\Phi}\bar{\Phi}>$. The external double
line
denote background superfield $\Phi$. A contribution
of a diagram consisting of $n$ such chains is of the form
\begin{equation}
L_n=\frac{1}{n}{\Big(\frac{M^2\lambda^8|\Phi|^8\Box^2\bar{D}^2 D^2}
{{16{(\Box-g^2|\phi|^2)}^4{(\Box-M^2)}^4}}
\Big)}^n
\end{equation}
The total one-loop correction in the k\"{a}hlerian effective
potential is a
sum of $K^{(1)}_0$ (\ref{cr0}), and all the $K_n$ and  $L_n$
contributions.
After a  summation with  a subsequent Fourier transformation
and an integration over
momenta with a  subsequent subtraction of  the one-loop divergences
we arrive at the one-loop correction in the k\"{a}hlerian effective
potential (\ref{kahl1}).

\section{Appendix C}
\setcounter{equation}{0}
\renewcommand{\theequation}{C.\arabic{equation}}
Let us consider a  calculation of the one-loop k\"ahlerian effective
potential
for the case when light superfields are pure background ones and
heavy
superfields are pure quantum ones.

The computation of the one-loop effective action reduces
 to the  problem of evaluating  the operator
\begin{eqnarray}
\Omega(s)=e^{is\tilde{\Delta'}}=\exp\big[i
(H\{D^2,\bar{D}^2\}+
\frac{1}{4}W''\bar{D}^2+\frac{1}{4}\bar{W}''D^2)
\big]
\end{eqnarray}
Representing $\Omega(s)$ in the standard form (see
\cite{Buch1,Buch2})
\begin{eqnarray}
\label{omega2}
 \Omega(s) &=& 1 + \frac{1}{16}A(s)D^2\bar {D}^2 +
 \frac{1}{16}\tilde{A}(s)\bar{D}^2D^2 +
 \frac{1}{8}B^{\alpha}(s)D_{\alpha}\bar{D}^2 + \nonumber\\ &+&
 \frac{1}{8}\tilde{B}_{\dot{\alpha}}(s)\bar{D}^{\dot{\alpha}}D^2
+\frac{1}{4}C(s)D^2 + \frac{1}{4}\tilde{C}(s)\bar{D}^2
\end{eqnarray}
we have the equation
$$
i\frac{\partial \Omega}{\partial s}=\Omega\tilde{\Delta}
$$
and therefore we arrive at the system of equations:
\begin{eqnarray}
\frac{1}{16i}\dot{A}&=&H+HA\Box+\frac{1}{16}W''C\\
\frac{1}{4i}\dot{C}&=&\frac{1}{4}\bar{W}''+4HC\Box+\frac{1}{4}\bar{W}''A\Box
\nonumber\\
\frac{1}{8i}\dot{B}^{\alpha}&=&\frac{1}{8}W''\tilde{B}_{\dot{\alpha}}
\partial^{\alpha\dot{\alpha}}\nonumber
\end{eqnarray}
The initial conditions for
these equations and analogous ones for $\tilde{A}$, $\tilde{B}$,
$\tilde{C}$
are $A=B=C=\tilde{A}=\tilde{B}=\tilde{C}=0$ at $s=0$.
Hence one finds that $B^{\alpha}(s)=\tilde{B}_{\dot{\alpha}}(s)=0$.
Solutions of other equations  are of the form
\begin{eqnarray}
\label{kahlsl}
A(s)=\tilde{A}(s)&=&
\frac{e^{16iH\Box s}}{2\Box}
(\exp(is\sqrt{W''\bar{W}''\Box})+\exp(-is\sqrt{W''\bar{W}''\Box}))-
\frac{1}{\Box}\nonumber\\
C(s)&=&
e^{16iH\Box s}\frac{\bar{W}''}
{\sqrt{W''\bar{W}''\Box}}
{\sh(is\sqrt{W''\bar{W}''\Box})}
\nonumber\\
\tilde{C}(s)&=&
e^{16iH\Box s}\frac{W''}
{\sqrt{W''\bar{W}''\Box}}\sh(is\sqrt{W''\bar{W}''\Box})
\end{eqnarray}
The one-loop k\"{a}hlerian effective potential has the form
\begin{equation}
\label{kahldf}
K^{(1)}=-i\int_{0}^{\infty}\frac{ds}{s}(A(s)+\tilde{A}(s))U(x,x';s)|_{x=x'}
\end{equation}
where $U(x,x';s)=e^{is\Box}\delta^4(x-x')$ (cf. \cite{Buch1,Buch2}).
Using (\ref{kahlsl}), the expression (\ref{kahldf}) can be rewritten
as
\begin{eqnarray}
\label{kahl3}
K^{(1)}&=&-i\int_{0}^{\infty}\frac{ds}{s}
\big(\frac{e^{16iH\Box s}}{2\Box}
(\exp(is\sqrt{W''\bar{W}''\Box})+\exp(-is\sqrt{W''\bar{W}''\Box}))-
\frac{1}{\Box}\big)\times\nonumber\\&\times&
e^{is\Box}\delta^4(x-x')|_{x=x'}
\end{eqnarray}
Let us denote $\sqrt{W''\bar{W}''\Box}$ as $Q$.

Since
\begin{eqnarray}
& &\frac{e^{16iH\Box s}}{2\Box}
(\exp(is\sqrt{W''\bar{W}''\Box})+\exp(-is\sqrt{W''\bar{W}''\Box}))-
\frac{1}{\Box}=\nonumber\\&=&
\frac{(\ch (iQ s)-1)}{\Box}e^{16 isH\Box}+
\frac{e^{16isH\Box}-1}{\Box}
\end{eqnarray}
we can rewrite the expression (\ref{kahl3}) as
\begin{eqnarray}
\label{kahl4}
K^{(1)}&=&-i\int_{0}^{\infty}\frac{ds}{s}
\Big\{\frac{\ch(iQs)-1}{\Box}
e^{16iH\Box s}+\frac{(e^{16iH\Box s}-1)}{\Box}
\Big\}\times\nonumber\\&\times&
e^{is\Box}\delta^4(x-x')|_{x=x'}
\end{eqnarray}
which equals to
\begin{eqnarray}
\label{kp}
K^{(1)}&=&-i\int_{0}^{\infty}\frac{ds}{s}
\Big\{\frac{\ch(iQs)-1}{\Box}
e^{i(16H+1)\Box s}\delta^4(x-x')|_{x=x'}
+\nonumber\\
&+&\frac{(e^{16iH\Box s}-1)}{\Box}
e^{is\Box}\delta^4(x-x')|_{x=x'}\Big\}
\end{eqnarray}

Then,
\begin{eqnarray}
& &\frac{\ch(iQs)-1}{\Box}=\sum_{n=1}^{\infty}
\frac{{(is\sqrt{W''\bar{W}''})}^{2n}}{(2n)!}\Box^{n-1}\nonumber\\
& &\frac{(e^{16iH\Box s}-1)}{\Box}=
\sum_{n=1}^{\infty}
\frac{{(16iHs)}^n}{n!}\Box^{n-1}
\end{eqnarray}

In analogy with \cite{Buch1,Buch2} we conclude that
\begin{eqnarray}
& &\Box^{n-1}e^{is\Box}\delta^4(x-x')|_{x=x'}=
\frac{{(-1)}^{n-1}n!}{16\pi^2{(is)}^{n+1}}\nonumber\\
& &
\Box^{n-1}e^{is(1+16H)\Box}\delta^4(x-x')|_{x=x'}=
\frac{1}{{(1+16H)}^{n+1}}
\frac{{(-1)}^{n-1}n!}{16\pi^2{(is)}^{n+1}}
\end{eqnarray}
The last expression can be obtained from the first one by the change
$s\to s(1+16H)$.

Then,
\begin{eqnarray}
K^{(1)}&=&-i\int_{0}^{\infty}\frac{ds}{s}
\Big\{\sum_{n=1}^{\infty}
\frac{{(2is\sqrt{W''\bar{W}''})}^{2n}}{(2n)!}
\frac{1}{{(1+16H)}^{n+1}}
\frac{{(-1)}^{n-1}n!}{16\pi^2{(is)}^{n+1}}+\nonumber\\&+&
\sum_{n=1}^{\infty}
\frac{{(16iHs)}^n}{n!}
\frac{{(-1)}^{n-1}n!}{16\pi^2{(is)}^{n+1}}
\Big\}
\end{eqnarray}
It is equal to
\begin{eqnarray}
\label{k1}
K^{(1)}&=&
-i\frac{1}{16\pi^2}\int_{0}^{\infty}\frac{ds}{s}
\Big\{\frac{1}{(1+16H)is}\sum_{n=1}^{\infty}
{\Big(\frac{is W''\bar{W}''}{1+16H}\Big)}^n
\frac{{(-1)}^{n-1}n!}{(2n)!}
-\nonumber\\&-&
\frac{16H}{(1+16H)is}
\Big\}
\end{eqnarray}
The integral $\int_{0}^{\infty}\frac{ds}{s}$ is divergent, therefore
we introduce the  regularization parameter $\epsilon$ and change
$\int_{0}^{\infty}\frac{ds}{s}\to
\int_{0}^{\infty}\frac{ds}{s^{1-\epsilon}}$
Then, the last term in the expression (\ref{k1}) can be rewritten as
$$
-\frac{1}{16\pi^2}\int_{0}^{\infty}\frac{ds}{s^{2-\epsilon}}\frac{16H}{(1+16H)}
$$
and
$$\int_{0}^{\infty}\frac{ds}{s^{2-\epsilon}}=
\int_{0}^{\infty}ds s^{\epsilon-2}e^{-s p^2}|_{p^2=0}=
{(p^2)}^{1-\epsilon}\Gamma(-1+\epsilon)|_{p^2=0}
$$
This term vanishes at $\epsilon\neq 0$ and can be defined to be
equal
to zero at $\epsilon=0$.

Therefore the one-loop k\"ahlerian effective potential has the form
\begin{eqnarray}
K^{(1)}&=&
-i\frac{1}{16\pi^2}\frac{1}{(1+16H)}\int_0^{\infty}\frac{ds}{s^{2-\epsilon}}
\Big\{\sum_{n=1}^{\infty}
{\Big(\frac{is W''\bar{W}''}{1+16H}\Big)}^n
\frac{{(-1)}^{n-1}n!}{(2n)!}
\Big\}
\end{eqnarray}
These power series can be summed up.
Namely, after the change  $s\to s\frac{i W''\bar{W}''}{1+16H} $ and
a
subsequent Wick rotation we arrive at
\begin{eqnarray}
K^{(1)}&=&
-i\frac{1}{16\pi^2}\frac{1}{(1+16H)}
\Big\{\frac{W''\bar{W}''}{1+16H}\Big\}^{1+\epsilon}
\int_0^{\infty}\frac{ds}{s^{2-\epsilon}}
\Big\{
\sum_{n=1}^{\infty}s^n
\frac{{(-1)}^{n-1}n!}{(2n)!}
\Big\}
\end{eqnarray}
It is known (cf. \cite{Buch1,Buch2}) that
$$
\sum_{n=1}^{\infty}s^n
\frac{{(-1)}^{n-1}n!}{(2n)!}=s\int_0^1 dt e^{-\frac{1}{4}s(1-t^2)}
$$
Hence
\begin{eqnarray}
K^{(1)}&=&
-\frac{1}{16\pi^2}\frac{1}{(1+16H)}
\Big\{\frac{W''\bar{W}''}{1+16H}\Big\}^{1+\epsilon}
\int_0^{\infty}\frac{ds}{s^{1-\epsilon}}\int_0^1 dt
e^{-\frac{1}{4}s(1-t^2)}
\end{eqnarray}
and
\begin{eqnarray}
& &\int_0^{\infty}\frac{ds}{s^{1-\epsilon}}\int_0^1 dt
e^{-\frac{1}{4}s(1-t^2)}=
\Gamma(\epsilon)\int_0^1 dt (1-\epsilon \log
(\frac{1-t^2}{4})+O(\epsilon^2))=
\nonumber\\&=&
\frac{1}{\epsilon}+\gamma-\int_0^1 dt \log
(\frac{1-t^2}{4})+O(\epsilon)
\nonumber\\
& &\Big\{\frac{W''\bar{W}''}{1+16H}\Big\}^{1+\epsilon}=
\frac{W''\bar{W}''}{1+16H}(1+\epsilon\log\frac{W''\bar{W}''}{1+16H}+O(\epsilon^2))
\end{eqnarray}
Here $O(\epsilon)$ denotes terms of  order one and/or higher in
$\epsilon$, and
$O(\epsilon^2)$ denoted terms of order  two and/or  higher in
$\epsilon$.
Therefore
\begin{eqnarray}
K^{(1)}&=&-\frac{1}{32\pi^2}
\Big\{
\frac{W''\bar{W}''}{{(1+16H)}^2\epsilon}+
\frac{W''\bar{W}''}{{(1+16H)}^2}
\Big[\log\Big\{\frac{W''\bar{W}''}{\mu^2(1+16H)}\Big\}
\Big]
\Big\}
\end{eqnarray}
And $1+16H=F_{\Phi\bar{\Phi}}$.
Here  the first term contains all the one-loop divergences
 of the theory and can be
cancelled by a suitable counterterm.
The terms containing $\gamma$ and $\int_0^1 dt \log
(\frac{1-t^2}{4})$ are
removed by a suitable choice of the renormalization
parameter $\mu$.

As a result we arrive at the expression (\ref{res}) for the one-loop
k\"{a}hlerian effective potential.

\end{document}